\documentclass[twocolumn]{aastex63}

\usepackage{amsmath}
\usepackage{longtable}
\usepackage{amssymb}
\usepackage{tabularx}
\usepackage{xcolor}
\usepackage{scalerel}
\usepackage{fontawesome}
\graphicspath{ {Figures/} }

\defcitealias{M21}{M21}
\defcitealias{K21}{K21}
\defcitealias{P20}{P20}

\submitjournal{AAS Journals}

\shortauthors{Cale et al.}
\shorttitle{Chromatic RVs of AU Mic}

\begin{document}

\title{Diving Beneath the Sea of Stellar Activity: Chromatic Radial Velocities of the Young AU Mic Planetary System}


\correspondingauthor{Bryson L. Cale}
\email{bryson.cale1@gmail.com}

\author[0000-0002-2078-6536]{Bryson L. Cale}
\affiliation{George Mason University, 4400 University Drive, Fairfax, VA 22030, USA}

\author[0000-0003-4701-8497]{Michael Reefe}
\affiliation{George Mason University, 4400 University Drive, Fairfax, VA 22030, USA}

\author[0000-0002-8864-1667]{Peter Plavchan}
\affiliation{George Mason University, 4400 University Drive, Fairfax, VA 22030, USA}

\author[0000-0002-2903-2140]{Angelle Tanner}
\affiliation{Mississippi State University, 75 B. S. Hood Road, Mississippi State, MS 39762}

\author[0000-0002-5258-6846]{Eric Gaidos}
\affiliation{University of Hawai'i at Manoa, 1680 East-West Road, Honolulu, HI 96822}

\author[0000-0002-2592-9612]{Jonathan Gagn\'e}
\affiliation{Plan\'etarium Rio Tinto Alcan, Espace pour la vie, 4801 av. Pierre-De Coubertin, Montr\'eal, QC H1V~3V4, Canada}
\affiliation{Institute for Research on Exoplanets, Universit\'e de Montr\'eal, D\'epartement de Physique, C.P.~6128 Succ. Centre-ville, Montr\'eal, QC H3C~3J7, Canada}

\author[0000-0002-8518-9601]{Peter Gao}
\affiliation{University of California, Santa Cruz, 1156 High St, Santa Cruz, CA 95064}

\author[0000-0002-7084-0529]{Stephen R. Kane}
\affiliation{University of California, Riverside, 900 University Ave. Riverside, CA 92521}


\author[0000-0002-5086-4232]{V\'ictor J.\,S. B\'ejar}
\affiliation{Instituto de Astrof\'isica de Canarias, 38205 La Laguna, Tenerife, Spain}
\affiliation{Departamento de Astrof\'isica, Universidad de La Laguna, 38206 La Laguna, Tenerife, Spain}

\author[0000-0002-3612-8968]{Nicolas Lodieu}
\affiliation{Instituto de Astrof\'isica de Canarias, 38205 La Laguna, Tenerife, Spain}
\affiliation{Departamento de Astrof\'isica, Universidad de La Laguna, 38206 La Laguna, Tenerife, Spain}

\author[0000-0002-3645-5977]{Guillem Anglada-Escud\'e}
\affiliation{Institut de Ci\`encies de l’Espai (ICE, CSIC), Campus UAB, Can Magrans s/n, 08193 Bellaterra, Spain}
\affiliation{Institut d’Estudis Espacials de Catalunya (IEEC), 08034 Barcelona, Spain}

\author[0000-0002-6689-0312]{Ignasi Ribas}
\affiliation{Institut de Ci\`encies de l’Espai (ICE, CSIC), Campus UAB, Can Magrans s/n, 08193 Bellaterra, Spain}
\affiliation{Institut d’Estudis Espacials de Catalunya (IEEC), 08034 Barcelona, Spain}

\author[0000-0003-0987-1593]{Enric Pall\'e}
\affiliation{Instituto de Astrof\'isica de Canarias, 38205 La Laguna, Tenerife, Spain}
\affiliation{Departamento de Astrof\'isica, Universidad de La Laguna, 38206 La Laguna, Tenerife, Spain}

\author[0000-0002-3302-1962]{Andreas Quirrenbach}
\affiliation{Landessternwarte, Zentrum f\"ur Astronomie der Universit\"at Heidelberg, K\"onigstuhl 12, 69117 Heidelberg, Germany}

\author[0000-0002-8388-6040]{Pedro J. Amado}
\affiliation{Instituto de Astrof\'isica de Andaluc\'ia (CSIC), Glorieta de la Astronom\'ia s/n, 18008 Granada, Spain}

\author{Ansgar Reiners}
\affiliation{Institut f\"ur Astrophysik, Georg-August-Universit\"at, Friedrich-Hund-Platz 1, 37077 G\"ottingen, Germany}

\author[0000-0002-7349-1387]{Jos\'e A. Caballero}
\affiliation{Centro de Astrobiolog\'ia (CSIC-INTA), ESAC, Camino bajo del castillo s/n, 28692 Villanueva de la Ca\~nada, Madrid, Spain}

\author[0000-0001-5664-2852]{Mar\'ia Rosa Zapatero Osorio}
\affiliation{Centro de Astrobiolog\'ia (CSIC-INTA), Carretera de Ajalvir km 4, 28850 Torrej\'on de Ardoz, Madrid, Spain}

\author[0000-0001-6187-5941]{Stefan Dreizler}
\affiliation{Institut f\"ur Astrophysik, Georg-August-Universit\"at, Friedrich-Hund-Platz 1, 37077 G\"ottingen, Germany}


\author{Andrew W. Howard}
\affiliation{California Institute of Technology, 1200 E California Blvd, Pasadena, CA 91125}

\author[0000-0003-3504-5316]{Benjamin J. Fulton}
\affiliation{IPAC, 770 South Wilson Avenue, Pasadena, CA 91125}

\author[0000-0002-6937-9034]{Sharon Xuesong Wang}
\affiliation{Department of Astronomy, Tsinghua University, Beijing 100084, People's Republic of China}

\author[0000-0003-2781-3207]{Kevin I. Collins}
\affiliation{George Mason University, 4400 University Drive, Fairfax, VA 22030, USA}

\author[0000-0001-8364-2903]{Mohammed El Mufti}
\affiliation{George Mason University, 4400 University Drive, Fairfax, VA 22030, USA}

\author[0000-0002-7424-9891]{Justin Wittrock}
\affiliation{George Mason University, 4400 University Drive, Fairfax, VA 22030, USA}

\author[0000-0002-0388-8004]{Emily A. Gilbert}
\affiliation{Department of Astronomy and Astrophysics, University of Chicago, 5640 S. Ellis Ave. Chicago, IL 60637, USA}
\affiliation{University of Maryland, Baltimore County, 1000 Hilltop Circle, Baltimore, MD 21250, USA}
\affiliation{The Adler Planetarium, 1300 South Lakeshore Drive, Chicago, IL 60605, USA}
\affiliation{NASA Goddard Space Flight Center, 8800 Greenbelt Road, Greenbelt, MD 20771, USA}
\affiliation{NASA Goddard Space Flight Center Sellers Exoplanet Environments Collaboration}

\author[0000-0001-7139-2724]{Thomas Barclay}
\affiliation{NASA Goddard Space Flight Center, 800 Greenbelt Rd, Greenbelt, MD 20771}
\affiliation{University of Maryland Baltimore County, 1000 Hilltop Cir, Baltimore, MD 21250, USA}

\author[0000-0003-0637-5236 ]{Baptiste Klein}
\affiliation{Sub-department of Astrophysics, Department of Physics, University of Oxford, Oxford OX1 3RH, UK}

\author[0000-0002-5084-168X]{Eder Martioli}
\affiliation{Institut d’Astrophysique de Paris, CNRS, UMR 7095, Sorbonne
Universit\'e,  98 bis bd Arago, 75014 Paris, France}
\affiliation{Laborat\'orio Nacional de Astrof\'isica, Rua Estados Unidos
154, Itajub\'a, MG 37504-364, Brazil}

\author[0000-0001-9957-9304]{Robert Wittenmyer}
\affiliation{University of Southern Queensland, West St, Darling Heights, QLD 4350, Australia}

\author[0000-0001-7294-5386]{Duncan Wright}
\affiliation{University of Southern Queensland, West St, Darling Heights, QLD 4350, Australia}

\author[0000-0003-3216-0626]{Brett Addison}
\affiliation{University of Southern Queensland, West St, Darling Heights, QLD 4350, Australia}


\author[0000-0003-2094-2428]{Teruyuki Hirano}
\affiliation{Astrobiology Center, 2-21-1 Osawa, Mitaka, Tokyo 181-8588, Japan}
\affiliation{National Astronomical Observatory of Japan, NINS, 2-21-1 Osawa, Mitaka, Tokyo 181-8588, Japan}
\affiliation{Department of Astronomical Science, School of Physical Sciences, The Graduate University for Advanced Studies (SOKENDAI), 2-21-1, Osawa, Mitaka, Tokyo, 181-8588, Japan}

\author[0000-0002-6510-0681]{Motohide Tamura}
\affiliation{Department of Astronomy, Graduate School of Science, The University of Tokyo, 7-3-1 Hongo, Bunkyo-ku, Tokyo 113-0033, Japan}
\affiliation{Astrobiology Center, 2-21-1 Osawa, Mitaka, Tokyo 181-8588, Japan}
\affiliation{National Astronomical Observatory of Japan, NINS, 2-21-1 Osawa, Mitaka, Tokyo 181-8588, Japan}

\author[0000-0001-6181-3142]{Takayuki Kotani}
\affiliation{Astrobiology Center, 2-21-1 Osawa, Mitaka, Tokyo 181-8588, Japan}
\affiliation{National Astronomical Observatory of Japan, NINS, 2-21-1 Osawa, Mitaka, Tokyo 181-8588, Japan}
\affiliation{Department of Astronomical Science, School of Physical Sciences, The Graduate University for Advanced Studies (SOKENDAI), 2-21-1, Osawa, Mitaka, Tokyo, 181-8588, Japan}

\author[0000-0001-8511-2981]{Norio Narita}
\affiliation{Komaba Institute for Science, The University of Tokyo, 3-8-1 Komaba, Meguro, Tokyo 153-8902, Japan}
\affiliation{JST, PRESTO, 3-8-1 Komaba, Meguro, Tokyo 153-8902, Japan}
\affiliation{Astrobiology Center, 2-21-1 Osawa, Mitaka, Tokyo 181-8588, Japan}
\affiliation{Instituto de Astrof\'{i}sica de Canarias (IAC), 38205 La Laguna, Tenerife, Spain}

\author[0000-0002-4501-564X]{David Vermilion}
\affiliation{George Mason University, 4400 University Drive, Fairfax, VA 22030, USA}

\author[0000-0001-7058-4134]{Rena A. Lee}
\affiliation{University of Hawai'i at Manoa, 2500 Campus Rd, Honolulu, HI 96822}

\author[0000-0001-9596-8820]{Claire Geneser}
\affiliation{Mississippi State University, 75 B. S. Hood Road, Mississippi State, MS 39762}

\author{Johanna Teske}
\affiliation{Carnegie Earth and Planets Laboratory, 5241 Broad Branch Road, N.W., Washington, DC 20015-1305}

\author[0000-0002-8964-8377]{Samuel N. Quinn}
\affiliation{Center for Astrophysics \textbar Harvard and Smithsonian, 60 Garden St, Cambridge, MA 02138, USA}

\author[0000-0001-9911-7388]{David W. Latham}
\affiliation{Center for Astrophysics \textbar Harvard and Smithsonian, 60 Garden St, Cambridge, MA 02138, USA}

\author[0000-0002-9789-5474]{Gilbert A. Esquerdo}
\affiliation{Center for Astrophysics \textbar Harvard and Smithsonian, 60 Garden St, Cambridge, MA 02138, USA}

\author[0000-0002-2830-5661]{Michael L. Calkins}
\affiliation{Center for Astrophysics \textbar Harvard and Smithsonian, 60 Garden St, Cambridge, MA 02138, USA}

\author{Perry Berlind}
\affiliation{Center for Astrophysics \textbar Harvard and Smithsonian, 60 Garden St, Cambridge, MA 02138, USA}

\author[0000-0003-2872-9883]{Farzaneh Zohrabi}
\affiliation{Louisiana State University, 202 Nicholson Hall, Baton Rouge, LA 70803}

\author[0000-0003-0091-3769]{Caitlin Stibbards}
\affiliation{George Mason University, 4400 University Drive, Fairfax, VA 22030, USA}

\author{Srihan Kotnana}
\affiliation{George Mason University, 4400 University Drive, Fairfax, VA 22030, USA}
\affiliation{Westfield High School, 4700 Stonecroft Blvd, Chantilly, VA 20151, USA}

\author[0000-0002-4715-9460]{Jon Jenkins}
\affiliation{NASA Ames Research Center, Moffett Field, CA 94035}

\author[0000-0002-6778-7552]{Joseph D. Twicken}
\affiliation{SETI Institute, Mountain View, CA 94043}
\affiliation{NASA Ames Research Center, Moffett Field, CA 94035}

\author{Christopher Henze}
\affiliation{NASA Ames Research Center, Moffett Field, CA 94035}

\author[0000-0002-4715-9460]{Richard Kidwell Jr.}
\affiliation{Mikulski Archive for Space Telescopes, 3700 San Martin Dr, Baltimore, MD 21218}

\author[0000-0002-7754-9486]{Christopher Burke}
\affiliation{Department of Physics and Kavli Institute for Astrophysics and Space Research Massachusetts Institute of Technology, 70 Vassar St, Cambridge, MA 02139}

\author{Joel Villase{\~ n}or}
\affiliation{Department of Physics and Kavli Institute for Astrophysics and Space Research Massachusetts Institute of Technology, 70 Vassar St, Cambridge, MA 02139}

\author{Patricia Boyd}
\affiliation{Astrophysics Science Division, NASA Goddard Space Flight Center, 800 Greenbelt Rd, Greenbelt, MD 20771}

\begin{abstract}

We present updated radial-velocity (RV) analyses of the AU Mic system. AU Mic is a young (22 Myr) early M dwarf known to host two transiting planets - $P_{b}\sim8.46$ days, $R_{b}=4.38_{-0.18}^{+0.18}\ R_{\oplus}$, $P_{c}\sim18.86$ days, $R_{c}=3.51_{-0.16}^{+0.16}\ R_{\oplus}$. With visible RVs from CARMENES-VIS, CHIRON, HARPS, HIRES, {\sc {\textsc{Minerva}}}-Australis, and TRES, as well as near-infrared (NIR) RVs from CARMENES-NIR, CSHELL, IRD, iSHELL, NIRSPEC, and SPIRou, we provide a $5\sigma$ upper limit to the mass of AU Mic c of $M_{c}\leq20.13\ M_{\oplus}$ and present a refined mass of AU Mic b of $M_{b}=20.12_{-1.57}^{+1.72}\ M_{\oplus}$. Used in our analyses is a new RV modeling toolkit to exploit the wavelength dependence of stellar activity present in our RVs via wavelength-dependent Gaussian processes. By obtaining near-simultaneous visible and near-infrared RVs, we also compute the temporal evolution of RV-``color'' and introduce a regressional method to aid in isolating Keplerian from stellar activity signals when modeling RVs in future works. Using a multi-wavelength Gaussian process model, we demonstrate the ability to recover injected planets at $5\sigma$ significance with semi-amplitudes down to $\approx$ 10\,m\,s$^{-1}$ with a known ephemeris, more than an order of magnitude below the stellar activity amplitude. However, we find that the accuracy of the recovered semi-amplitudes is $\sim$50\% for such signals with our model.

\end{abstract}


\section{Introduction} \label{sec:intro}

Characterizing young planetary systems is key to improving our understanding of their formation and evolution. Young transiting systems in particular offer a means to directly probe the radii, and together with masses from precise radial-velocity (RV) measurements, the bulk densities of the planets. RV observations are also crucial to constrain the eccentricity of the orbit to understand the kinematic history and stability of the system. A precision of 20\% for the mass determination is further recommended for enabling detailed atmospheric characterization, particularly for terrestrial-mass planets \citep{2019ApJ...885L..25B}.

Unfortunately, searches for planets orbiting young stars have been limited by stellar activity signals comparable in amplitude to that of typical Keplerian signals. Stellar surface inhomogeneities (e.g., cool spots, hot plages) driven by the dynamic stellar magnetic field rotate in and out of view, leading to photometric variations over time. The presence of such active regions breaks the symmetry between the approaching and receding limbs of the star, introducing RV variations over time as well \citep{desort2007etal}. These active regions further affect the integrated convective blue-shift over the stellar disk, and will therefore manifest as an additional net red- or blue-shift \citep{meunier2013, 2014ApJ...796..132D}. Various techniques have been introduced to lift the degeneracy between activity- and planetary-induced signals in RV datasets such as line-by-line analyses \citep{dumusque2018, 2018AJ....156..180W, cretignier2020} and Gaussian process (GP) modeling \citep[e.g.,][]{2014MNRAS.443.2517H, 2015ApJ...808..127G, 2016AJ....152..204L, 2021arXiv210209441T, Robertson_2020, P20, K21}, but such measurements remain challenging due to the sparse cadence of typical RV datasets compared to the activity timescales.

AU Mic is a young \citep[22 Myr;][]{2014MNRAS.445.2169M}, nearby \citep[$\beta$ Pictoris moving group, $\sim$ 10 pc;][]{gaiadr2}, and active pre-main-sequence M1 dwarf \citep[][hereafter referred to as P20]{P20}. AU Mic hosts an edge-on debris disk \citep{2013ApJS..208....9P}, and therefore the probability for planets to transit is greater than for other systems. Using photometric observations from \textit{TESS} \citep{2015JATIS...1a4003R} in Sector 1 (2018-July-25 to 2018-August-22), \citetalias{P20} discovered an $\approx8.46$ day Neptune-size ($R_{b}=4.38_{-0.18}^{+0.18}\ R_{\oplus}$) transiting planet, which was further validated to transit with \textit{Spitzer} observations (hereafter referred to as AU Mic b). \citetalias{P20} also reported the detection of a single-transit event in the \textit{TESS} Sector 1 light curve, but did not constrain the period with only an isolated event. With high cadence RVs from SPIRou, \cite{K21} (hereafter referred to as K21) measured the mass of AU Mic b and confirmed it to be consistent with a Neptune-mass planet ($M_{b}=17.1_{-4.5}^{+4.7}\ M_{\oplus}$). With more observations of AU Mic from the \textit{TESS} extended mission in Sector 27 (2020-July-04 to 2020-July-30), \cite{M21} (hereafter referred to as M21) determined AU Mic c to be a smaller Neptune-sized planet ($R_{c}=3.51_{-0.16}^{+0.16}\ R_{\oplus}$) with a period of $\approx$ 18.86 days.

In this paper, we present and discuss analyses of several years of multi-wavelength RV observations of AU Mic that further elucidates this planetary system. In section \ref{sec:data}, we summarize the visible and near-infrared (NIR) RV observations, as well as photometric observations which are used to inform our RV model. In section \ref{sec:rv_fitting}, we introduce two joint quasi-periodic Gaussian process kernels which are first steps in taking into account the expected wavelength dependence of stellar activity through simple scaling relations between wavelengths. We then apply our model to RVs of AU Mic and present results in section \ref{sec:results}. In section \ref{sec:discussion}, we assess the sensitivity of our RV model through planet injection and recovery tests. We then briefly discuss the utility of ``RV-color'' between two wavelengths in isolating Keplerian from stellar activity induced signals in Section \ref{sec:disc_utility}. We finally note the assumptions and caveats in this work in section \ref{sec:caveats_future_work}. A summary of this work is provided in section \ref{sec:conclusion}.

\section{Observations} \label{sec:data}

\subsection{RVs} \label{sec:data_rvs}

Our analyses make use of new and archival high-resolution echelle spectra from a variety of facilities, which are summarized in Table \ref{tab:spectrographs}. We briefly detail new spectroscopic observations and the corresponding RVs from observing programs primarily intended to characterize the AU Mic planetary system.

\subsubsection{CARMENES} \label{sec:carmenes}

The CARMENES (Calar Alto high-Resolution search for M dwarfs with Exo-earths with Near-infrared and optical echelle Spectrographs) instrument \citep{2018SPIE10702E..0WQ} is a pair of two high-resolution spectrographs installed at the 3.5\,m telescope at the Calar Alto Observatory in Spain. The visual (VIS) and near-infrared (NIR) arms cover a wavelength range of 520--960\,nm and 960--1710\,nm, with resolving powers of R=94,600 and R=80,400, respectively. AU Mic was observed 100 times with CARMENES during two different campaigns between 14 July and 9 October 2019, and between 19 July and 16 November 2020, respectively. This last observing period was partially contemporaneous with \textit{TESS} observations of AU Mic in Sector 27 (04 July -- 30 July 2020). One or two exposures of 295\,s were obtained per epoch with typical $S/N$ larger than 70--100, and at airmasses larger than 2.5, due the low declination of the target at the Calar Alto observatory. CARMENES data were processed by the {\it caracal} pipeline \citep{2016SPIE.9910E..0EC}, which includes bias, flat-field, and dark correction, tracing the echelle orders on the detector, optimal extraction of the one-dimensional spectra, and performance of the initial wavelength calibration using U-Ar, U-Ne, and Th-Ne lamps. The RVs were obtained with the \texttt{serval} pipeline \citep{serval} by cross-correlating the observed spectrum with a reference template constructed from all observed spectra of the same star. In addition, the \texttt{serval} pipeline also computes the correction for barycentric motion, secular acceleration, instrumental drift using simultaneous observations of Fabry-P\'erot etalons, and nightly zero-points using RV standards observed during the night \citep{trifinov2018}.

\subsubsection{CHIRON} \label{sec:chiron}

We obtained 14 nightly observations of AU Mic with the CHIRON spectrometer \citep{2018SPIE10702E..11K} on the SMARTS 1.5\,m telescope at the Cerro Tololo Inter-American Observatory (CTIO) between UT dates 2019-09-14 and 2019-11-10. Observations are recorded in narrow slit mode (R$\sim$136,000) using the iodine cell to simultaneously calibrate for the wavelength scale and instrument profile. Like iSHELL observations \citep[see][]{caleetal2019}, exposure times ($t_{exp}$) were limited to 5 minutes due to the uncertainties of barycenter corrections scaling as $t_{exp}^{2}$ \citep{2019MNRAS.489.2395T}, and the dynamicity of telluric absorption over a single exposure. We initially recorded 22 exposures per-night, and later increased this to 42 as the cumulative $S/N$ within a night was insufficient ($\sim100$).\footnote{Unlike iSHELL (and like many modern echelle spectrographs), CHIRON makes use of an exposure meter in order to calculate the proper (flux-weighted) exposure midpoint, and therefore longer exposure times will be less-impacted by the uncertainty in computing the exposure midpoint. Further, tellurics at visible wavelengths are far more sparse than for iSHELL at K-band wavelengths. We therefore recommend significantly longer exposure times ($\geq$30 minutes) for future observations of AU Mic (or targets of similar brightness) with CHIRON in narrow slit mode.} Raw CHIRON observations are reduced via the \texttt{REDUCE} package \citep{Piskunov2002}, and the corresponding RVs are computed using \texttt{pychell}. Unfortunately, a significant fraction of the extracted 1-dimensional spectra are too noisy to robustly measure the precise RVs from (peak $S/N\approx 20-30$ per spectral pixel). We therefore flag clear outliers in the RV measurements, and re-compute the nightly (binned) RVs resulting in 12 epochs to be included in our analyses.

\subsubsection{HIRES} \label{sec:hires}

We include 60 Keck-HIRES \citep{1994SPIE.2198..362V} observations of AU Mic in in our analyses. The majority of these observations took place in the second half of 2020 with several nights yielding contemporaneous observations with other facilities. Exposure times range from 204--500 seconds, yielding a median $S/N\approx234$ at 550 nm per spectral pixel. HIRES spectra are processed and RVs computed via methods described in \cite{2010ApJ...721.1467H}.

\subsubsection{{\textsc{Minerva}}-Australis} \label{sec:minerva_aus}

Spectroscopic observations of AU Mic were carried out using the {\textsc{Minerva}}-Australis facility situated at the Mount Kent Observatory in Queensland, Australia \citep{2018arXiv180609282W, 2019PASP..131k5003A, 2021MNRAS.502.3704A} between 2019 July 18 and 2019 November 5. {\sc {\textsc{Minerva}}}-Australis consists of an array of four independently operated 0.7\,m CDK700 telescopes, three of which were used in observing AU Mic. Each telescope simultaneously feeds stellar light via fiber optic cables to a single KiwiSpec R4-100 high-resolution ($R\sim80,000$) spectrograph \citep{2012SPIE.8446E..88B} with wavelength coverage from 480 to 620\,nm. In total, we obtained 31 observations with telescope 3 (M-A Tel3), 35 observations with telescope 4 (M-A Tel4), and 33 observations with telescope 6 (M-A Tel6). Exposure times for these observations were set to 1800\,s, providing a signal-to-noise ratio between 15 and 35 per spectral pixel. RVs are derived for each telescope by using the least-squares shift and fit technique \citep{2012ApJS..200...15A}, where the template being matched is the mean spectrum of each telescope. Spectrograph drifts are corrected for using simultaneous thorium-argon (ThAr) arc lamp observations.

\subsubsection{TRES}

We include 85 observations (archival and new) of AU Mic observed with the Tillinghast Reflector Echelle Spectrograph \citep[TRES;][]{tres_paper,furesz:2008} in our analyses. The majority of these observations took place in the second half of 2019 with several nights yielding contemporaneous observations with other facilities. Typical exposure times range from 600--1200 seconds, with a median $S/N\approx60$\ per resolution element. Spectra are processed using methods outlined in \cite{buchhave:2010} and \cite{quinn:2014}, with the exception of the cross-correlation template, for which we use the high-$S/N$ median observed spectrum.

\subsubsection{IRD} \label{sec:ird}

We obtained near infrared, high resolution spectra of AU Mic using the InfraRed Doppler (IRD) instrument \citep[e.g.,][]{2018SPIE10702E..11K} on the Subaru 8.2\,m telescope. The observations were carried out between June -- October 2019, and we obtained a total of 430 frames with integration times of 30--60 seconds. Half of these frames were taken on the transit night (UT 2019 June 17) with the goal of measuring the stellar obliquity for AU Mic b, whose RVs were already presented in \citet{2020ApJ...899L..13H}. The raw data are reduced in a standard manner using our custom code as well as \texttt{IRAF} \citep{1993ASPC...52..173T}, and the extracted one-dimensional spectra are analyzed by the RV-analysis pipeline for IRD as described in \citet{2020PASJ...72...93H}. The typical precision of the derived RVs is 9--13\,m\,s$^{-1}$.

\subsubsection{iSHELL} \label{sec:ishell}

We obtained 46 out-of-transit observations of AU Mic with iSHELL on the NASA Infrared Telescope Facility \citep{2016SPIE.9908E..84R} from October 2016 to October 2020. The exposure times varied from 20--300 seconds, and the exposures were repeated 2--23 times within a night to reach a cumulative $S/N$ per spectral pixel $>$ 200 (the approximate center of the blaze for the middle order, 2.35 $\mu \mathrm{m}$) for most nights. Raw iSHELL spectra are processed in \texttt{pychell} using methods outlined in \cite{caleetal2019}.

The corresponding iSHELL RVs are computed in \texttt{pychell} using updated methods to those described in \cite{caleetal2019}. Instead of starting from an unknown (flat) stellar template, we start with a BT-Settl \citep{2012RSPTA.370.2765A} stellar template with $T_{eff}=3700\ \mathrm{K}$, and with solar values for $\log g$ and Fe/H. We further Doppler-broaden the template using the \texttt{rotBroad} routine from \texttt{PyAstronomy} \citep{pya} with $v \sin i = 8.8\ \mathrm{km\,s^{-1}}$. Qualitatively, this broadened template matches the iSHELL observations well. We also ``iterate'' the template by co-adding residuals in a quasi-inertial reference frame with respect to the star according to the bary-center velocities ($v_{BC}$), however the stellar RVs for subsequent iterations tend to be highly correlated with $v_{BC}$ and exhibit significantly larger scatter than the first iteration suggests. We therefore use RVs from the first iteration only and leave the cause of this correlation as a subject for future work.

\subsection{Photometry from \textit{TESS}} \label{sec:data_photometry}

The NASA \textit{TESS} mission \citep{2015JATIS...1a4003R} observed AU Mic in Sectors 1 (2018-July-25 to 2018-August-22) and 27 (2020-July-04 to 2020-July-30). We download the light-curves from the Mikulski Archive for Space Telescopes \citep[][MAST]{2018SPIE10704E..15S}. We use the Science Processing Operations Center \citep[SPOC;][]{jenkinsSPOC2016} ``Presearch Data Conditioning'' light curves utilizing ``Simple Aperture Photometry'' \citep[PDCSAP;][]{2012PASP..124..985S, 2014PASP..126..100S, 2012PASP..124.1000S} to inform our model in section \ref{sec:kernel_parameter_estimation}.

\begin{table*}
\caption{A summary of the RV datasets used in this work. The nightly-binned measurements are provided in appendix \ref{app:rvs_all}. $\mathrm{N_{tot}}$ and $\mathrm{N_{nights}}$ refers to the number of individual and per-night epochs, respectively. The median intrinsic error bars $\sigma_{\mathrm{RV}}$ consider all observations.}
\begin{center}
    \begin{tabular}{ | p{2.8cm} | p{1cm} | p{.8cm} | p{.7cm} | p{1.5cm} | p{1.2cm} | p{2.8cm} | p{2.8cm} | }
    \hline
    Spectrograph/ \newline Facility & $\lambda/\Delta \lambda$ \newline [$\times 10^{3}$] & $\mathrm{N_{nights}}$ & $\mathrm{N_{used}}$ & Median \newline $\sigma_{\scaleto{RV}{2pt}}$ [m\,s$^{-1}$] & Adopted \newline $\lambda$ [nm] & Pipeline \newline  & Comm. Paper \\
    \hline
    HIRES/Keck & 85 & 60 & 41 & 2.6 & 565 & -- & \cite{1994SPIE.2198..362V} \\
    Tillinghast/TRES & 44 & 85 & 55 & 24.2 & 650 & -- & \cite{tres_paper} \\
    CARMENES-VIS/ \newline Calar Alto 3.5\,m & 94.6 & 63 & 60 & 11.4 & 750 & \texttt{caracal} \citep{2016SPIE.9910E..0EC} \newline \texttt{serval} \citep{serval} & \cite{carmcomm} \\
    CARMENES-NIR/ \newline Calar Alto 3.5m & 80.4 & 62 & 49 & 32.6 & 1350 & -- & -- \\
    SPIRou/CFHT & 75 & 27 & 27 & 5.0 & 1650 & \citetalias{K21} & \cite{2018haex.bookE.107D} \\
    iSHELL/IRTF & 85 & 46 & 31 & 5.0 & 2350 & \texttt{pychell} \newline \cite{caleetal2019} & \cite{2016SPIE.9908E..84R} \\
    HARPS-S/ \newline La Silla 3.6m & 115 & 34 & 0 & 2.2 & 565 & ESO DRS \citep{2010Msngr.142...42C} \newline \texttt{HARPS-TERRA} \citep{2012ApJS..200...15A} & \cite{2003Msngr.114...20M} \\
    {\sc {\textsc{Minerva}}}- \newline Australis-T3 & 80 & 13 & 0 & 9.5 & 565 & \citep{2012ApJS..200...15A} & \cite{2018arXiv180609282W} \newline \cite{2019PASP..131k5003A} \newline \cite{2021MNRAS.502.3704A} \\
    {\sc {\textsc{Minerva}}}-\newline Australis-T4 & 80 & 13 & 0 & 9.5 & 565 & -- & -- \\
    {\sc {\textsc{Minerva}}}-\newline Australis-T6 & 80 & 13 & 0 & 9.5 & 565 & -- & -- \\
    CHIRON/CTIO & 136 & 12 & 0 & 46 & 565 & \cite{reduce2002} \newline \cite{caleetal2019} & \cite{chiron2013} \\
    IRD/Subaru & 70 & 6 & 0 & 3.0 & 1350 & \texttt{IRAF}; \citep{1993ASPC...52..173T} \newline \cite{2020PASJ...72...93H} & \cite{2018SPIE10702E..11K} \\
    NIRSPEC/Keck & 25 & 14 & 0 & 50 & 2350 & \cite{2012ApJ...749...16B} & \cite{1998SPIE.3354..566M} \\
    CSHELL/IRTF & 36 & 21 & 0 & 26 & 2350 & \cite{gagne2016}, \cite{Gao2016} & \cite{1993SPIE.1946..313G} \\
    \hline
    \end{tabular}
\end{center}
\label{tab:spectrographs}
\end{table*}

\section{Radial Velocity Fitting} \label{sec:rv_fitting}

\subsection{Bayesian Inference for Radial-Velocities} \label{sec:bayesian_inference}

We primarily seek to utilize a global (joint) Gaussian process model with multiple realizations that give rise to the data we observe with all of the above instruments simultaneously. To implement our desired framework, we have developed two \textit{Python} packages. We leave the description of \texttt{optimize} - a high-level Bayesian inference framework to appendix \ref{app:optimize}.

To provide RV-specific routines, we extend the \texttt{optimize} package within the \texttt{orbits} sub-module of the \texttt{pychell} \citep{caleetal2019} package\footnote{Documentation: https://pychell.readthedocs.io/en/latest/}. We define classes specific for RV-data, models, and likelihoods, with much of the ``boiler-plate'' code handled through \texttt{optimize}. A top-level ``RVProblem'' further defines a pool of RV-specific methods for pre- and post-optimization routines, such as plotting phased RVs, periodogram tools, model comparison tests, and propagation of MCMC chains for deterministic Keplerian parameters (e.g, planet masses, semi-major axes, and densities).

\subsection{Two Chromatic Gaussian Processes} \label{sec:gp_kernel}

A Gaussian process kernel is defined through a square matrix, $\mathbf{K}$ (also called the covariance matrix), where each entry describes the covariance between two measurements\footnote{See \cite{haywoodthesis} for a thorough discussion of Gaussian processes.}. We introduce two GP kernels as extensions of the quasi-periodic (QP) kernel, which has been demonstrated in numerous cases to model rotationally modulated stellar activity in both photometric and RV observations (see Section \ref{sec:intro}) \footnote{Other parameterizations are also common.}.

\begin{gather}
    \mathbf{K_{QP}}(t_{i},t_{j}) = \eta_{\sigma}^{2} \exp \bigg[-\frac{\Delta t^{2}}{2 \eta_{\tau}^{2}} - \frac{1}{2 \eta_{\ell}^{2}} \sin^{2} \bigg( \pi \frac{\Delta t}{\eta_{p}} \bigg) \bigg] \label{eq:gp_qp} \\ \nonumber \\
    \mathrm{where}\ \ \Delta t = |t_{i} - t_{j}| \nonumber
\end{gather}

\noindent Here, $\eta_{P}$ typically represents the stellar-rotation period, $\eta_{\tau}$ the mean spot lifetime, and $\eta_{\ell}$ is the relative contribution of the periodic term, which may be interpreted as a smoothing parameter (larger is smoother). $\eta_{\sigma}$ is the amplitude of the auto-correlation of the activity signal.

We seek to use a fully-inclusive QP-like kernel that accounts for the wavelength-dependence of the stellar activity present in our multi-wavelength dataset. In this work, we only modify the amplitude parameter, $\eta_{\sigma}$; we leave further chromatic modifications (namely convective blue-shift and limb-darkening, see Section \ref{sec:intro}), as subjects for future work. To first order, we expect the amplitude from activity to be linearly proportional to frequency (or inversely proportional to wavelength). This approximation is a direct result of the spot-contrast scaling with the photon frequency (or inversely with wavelength) from the ratio of two black-body functions with different effective temperatures \citep{2010ApJ...710..432R}.

We first re-parametrize the amplitude through a linear kernel as follows:

\begin{gather}
    \mathbf{K_{J1}}(t_{i}, t_{j}) = \eta_{\sigma,\mathrm{s}(i)} \eta_{\sigma,\mathrm{s}(j)} \times \exp[...] \label{eq:gp_j1}
\end{gather}

\noindent Here,  $\eta_{\sigma,s(i)} \eta_{\sigma,s(j)}$ are the effective amplitudes for the spectrographs at times $t_{i}$ and $t_{j}$, respectively, where $s(i)$ represents an indexing set between the observations at time $t_{i}$ and spectrograph $s$.\footnote{Truly simultaneous measurements (i.e., $t_{i} = t_{j}$) would necessitate a more sophisticated indexing set.} Each amplitude is a free parameter.

We also consider a variation of this kernel which further enforces the expected inverse relationship between the amplitude with wavelength. We rewrite the kernel to become:

\begin{gather}
    \mathbf{K_{J2}}(t_{i}, t_{j}, \lambda_{i}, \lambda_{j}) = \eta_{\sigma,0}^{2} \Bigg( \frac{\lambda_{0}}{\sqrt{\lambda_{i} \lambda_{j}}} \Bigg) ^{2 \eta_{\lambda}}
    \times \exp[...] \label{eq:gp_j2}
\end{gather}

\noindent Here, $\eta_{\sigma,0}$ is the effective amplitude at $\lambda = \lambda_{0}$, and $\eta_{\lambda}$ is an additional power-law scaling parameter with wavelength to allow for a more flexible non-linear (with frequency) relation. $\lambda_{i}$ and $\lambda_{j}$ are the ``effective'' wavelengths for observations at times $t_{i}$ and $t_{j}$, respectively. For both eqs. \ref{eq:gp_j1} and \ref{eq:gp_j2}, the expression within square brackets is identical to that in eq. \ref{eq:gp_qp}.

To make predictions from $\mathbf{K_{J2}}$ (eq. \ref{eq:gp_j2}), we follow \cite{2006gpml.book.....R} (eqs. 2.23 and 2.24). We construct the matrix $\mathbf{K_{J2}}(t_{i,*}, t_{j}, \lambda_{*}, \lambda_{j})$, which denotes the $n_{*} \times n$ matrix of the covariances evaluated at all pairs of test points and training points (the data). Wavelengths in the * dimension are identical, and therefore each realization corresponds to a unique wavelength. This formulation allows us to realize the GP with high accuracy for all wavelengths so long as at least one wavelength is sampled near $t_{i,*}$. Predictions with kernel $\mathbf{K_{J1}}$ (eq. \ref{eq:gp_j1}) are found in a similar fashion, where each realization corresponds to a particular spectrograph.

\subsection{Primary RV Analyses} \label{sec:primary_analyses}

We first bin out-of-transit RV observations from each night (per-spectrograph). While not negligible, we expect changes from rotationally modulated activity to be small within a night, so we choose to mitigate activity on shorter timescales our model is not intended to capture (e.g, p-mode oscillations, granulation). The median RV for each spectrograph is also subtracted. We choose to ignore poorly-sampled regions with respect to our adopted mean spot lifetime $\eta_{\tau}$ (100 days, see Section \ref{sec:kernel_parameter_estimation}); each instance of a covariance matrix represents a \textit{family} of functions, and therefore the GP regression may be too flexible (and thus poorly constrained) in regions of low-cadence observations. We also ignore regions with only low precision measurements (median errors $\gtrsim$ 10 m\,s$^{-1}$). This limits our analyses to all observations between September 2019 -- December 2020, and the spectrographs HIRES, TRES, CARMENES-VIS and NIR, SPIRou, and iSHELL. We do not include six binned IRD or thirteen binned {\sc {\textsc{Minerva}}}-Australis observations in our primary analyses as we expect the offsets to be poorly constrained in the presence of stellar activity. Finally, we discard 3 CARMENES-VIS and 13 CARMENES-NIR measurements from our analyses primarily near the beginning of each season due to residuals $>$ 100 m\,s$^{-1}$ that are inconsistent with our other datasets. We suspect that telluric contamination which is further exacerbated by the high airmass of the observations may have degraded the CARMENES observations. For completeness, we present fit results including all spectrographs in appendix \ref{app:full_fit}. A summary of measurements is provided in Table \ref{tab:spectrographs}.

Our RV model first consists of two Keplerian components for the known transiting planets, a GP model for stellar activity, and per-instrument zero points. The zero points are each assigned to 1 m\,s$^{-1}$ with a uniform prior of $\pm$ 300 m\,s$^{-1}$. We further adopt a normal prior of $\mathcal{N}(0, 100)$ to make each offset well-behaved. When using multiple priors, the composite prior probability for such a parameter will not integrate to unity. For the combination of a uniform $+$ normal prior, this is not a concern; the normal prior is properly normalized and takes on a continuous range of values, whereas the uniform prior will either result in a constant term added to the likelihood function if the parameter is in bounds, or $-\infty$ if out of bounds.

Analyses of the \textit{TESS} transits in \citetalias{M21} found $P_{b}=8.4629991 \pm 0.0000024$ days, $TC_{b}=2458330.39046 \pm 0.00016$, $P_{c}=18.858991 \pm 0.00001$ days, and $TC_{c}=2458342.2243 \pm 0.0003$. For all of our analyses, we fix $P$ and $TC$ for planets b and c; the uncertainties in these measurements are insignificant even for our full baseline of $\approx$ 17 years. The semi-amplitudes of each planet start at $K_{b}=8.5$ m\,s$^{-1}$ and $K_{c}=5$ m\,s$^{-1}$, and are only enforced to be positive. Preliminary analyses of a secondary eclipse observed in \textit{Spitzer} observations support a moderately eccentric orbit for AU Mic b, with $e_{b}=0.189 \pm 0.04$ (Collins et al., in prep.), which is somewhat larger than the eccentricity determined from the duration of the primary transits observed with \textit{TESS} ($e_{b}=0.12 \pm 0.04$, Gilbert et al., submitted to \textit{Astrophysical Journal}). We assume a circular orbit for AU Mic c, and further examine eccentric cases in section \ref{sec:disc_eccentricity}. The Keplerian component of our RV model in \texttt{pychell} is nearly identical to that used in \texttt{RadVel} \citep{radvel}. Kepler's equation is written in \textit{Python} and makes use of the \texttt{numba.@njit} decorator \citep{numba} for optimal performance. We exclusively use the orbit basis $\{P,$ $TC,$ $e,$ $\omega,$ $K\}$.

Our optimizer seeks to maximize the natural-logarithm of the a posteriori probability (MAP) under the assumption of normally distributed errors:

\begin{equation} \label{eq:map}
    \ln \mathcal{L} = -\frac{1}{2} \bigg[\vec{r}^{\mathsf{T}} \mathbf{K_{o}}^{-1} \vec{r} + \ln |\mathbf{K_{o}}| + N \ln(2 \pi) \bigg] + \sum_{i} \ln \mathcal{P}_{i}
\end{equation}

\noindent Here, $\vec{r}$ is the vector of residuals between the observations and model. $\mathbf{K_{o}}$ is the covariance matrix sampled at the same observations, $N$ is the number of data points, and $\{\mathcal{P}_{i}\}$ is the set of prior knowledge. We maximize eq. \ref{eq:map} using the iterative Nelder-Mead algorithm described in \cite{caleetal2019}, which is included as part of the \texttt{optimize} package. We also sample the posterior distributions using the \texttt{emcee} package \citep{emcee} for a subset of models to determine parameter uncertainties, always starting from the MAP-derived parameters. In all cases, we use twice the number of chains as varied parameters. We perform a burn-in phase of 1000 steps followed by a full MCMC analysis for $\approx$ 50$\times$ the median auto-correlation time (steps) of all chains.

\subsubsection{Estimation of Kernel Parameters} \label{sec:kernel_parameter_estimation}

\begin{figure*}
    \centering
    \includegraphics[width=0.98\textwidth]{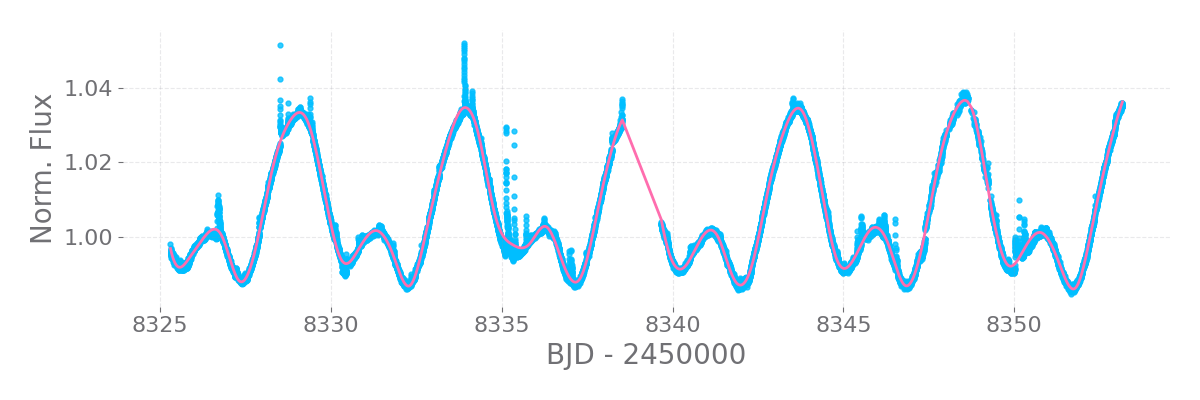}
    \includegraphics[width=0.68\textwidth]{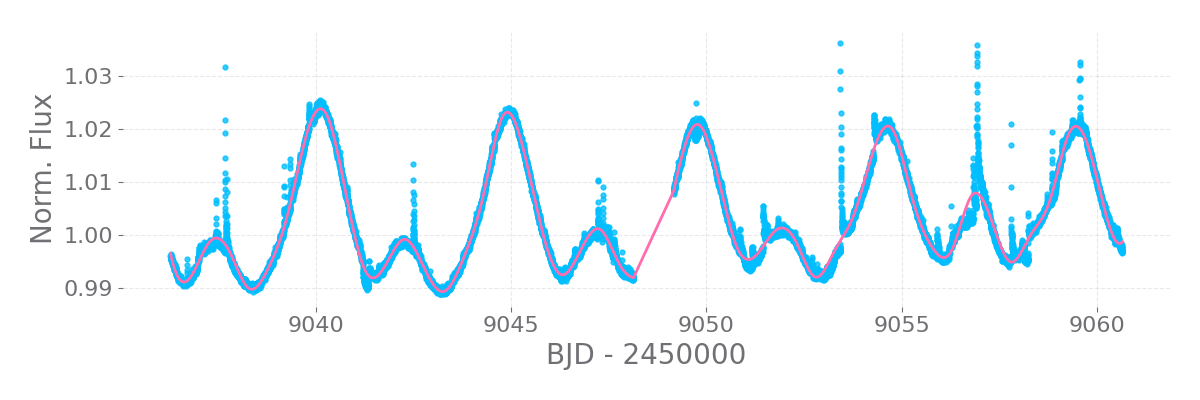}
    \includegraphics[width=0.3\textwidth]{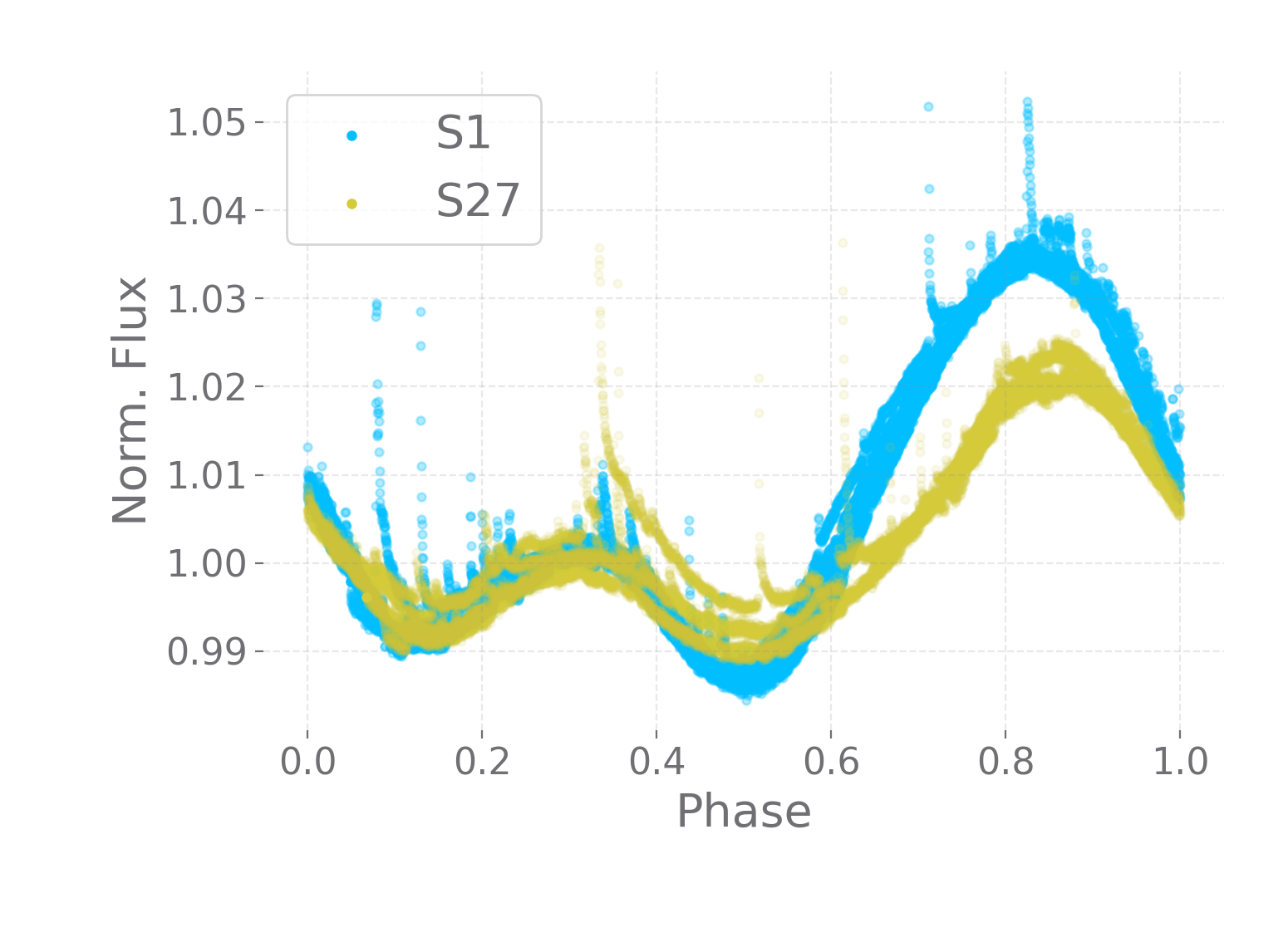}
    \caption{The \textit{TESS} PDCSAP light curves of AU Mic from Sectors 1 (top) and 27 (bottom). The lower right plot shows both Sectors phased to 4.836 days. Although the two seasons exhibit nearly identical periodic signals, Sector 27 exhibits moderate evolution. The least-squares cubic spline fit for each Sector is shown in pink.}
    \label{fig:light_curves}
\end{figure*}

We briefly analyze both Sectors of \textit{TESS} photometry in order to estimate the GP kernel parameters $\eta_{\tau}, \eta_{\ell}$, and $\eta_{P}$. We note that the rotationally modulated structure in both Sectors is consistent (fig. \ref{fig:light_curves}). If we assume spots are spatially static in the rest-frame of the stellar-surface (i.e., spots do not migrate), this suggests a similar spot configuration and contrast for each Sector. We first determine $\eta_{P}$ by qualitatively analyzing both \textit{TESS} Sectors phased up to periods close to that used in \citetalias{M21} ($4.862\pm0.032$ days) with a step size of 0.001 days (see fig. \ref{fig:light_curves}). We find $\eta_{P} \approx 4.836$ \textit{or} $\eta_{P} \approx 4.869$ days from our range of periods tested; no periods between these two values are consistent with our assumption of an identical spot configuration. The difference in these two periods further corresponds to one additional period between the two sectors (i.e., $|1/\eta_{P,1} - 1/\eta_{P,2}| \approx 1/700$ days$^{-1}$). The smaller of these two values implies AU Mic b is in a 7:4 resonance with the stellar rotation period \citep{2021arXiv210802149S}, potentially indicating tidal interactions between the planet and star. We adopt $\eta_{P} \sim \mathcal{N}(4.836, 0.001)$ in all our analyses where the uncertainty is a conservative estimate determined by our step size.

Although the \textit{TESS} light curve itself can provide insight into $\eta_{\tau}$ and $\eta_{\ell}$, we instead try to estimate these values directly from the predicted spot-induced RV variability via the $FF'$ technique \citep{2012MNRAS.419.3147A}:

\begin{equation} \label{eq:ffp}
    \Delta RV_{\mathrm{spots}}(t) = - F(t) F'(t) R_{\star} / f
\end{equation}

\begin{figure}
    \centering
    \includegraphics[width=0.45\textwidth]{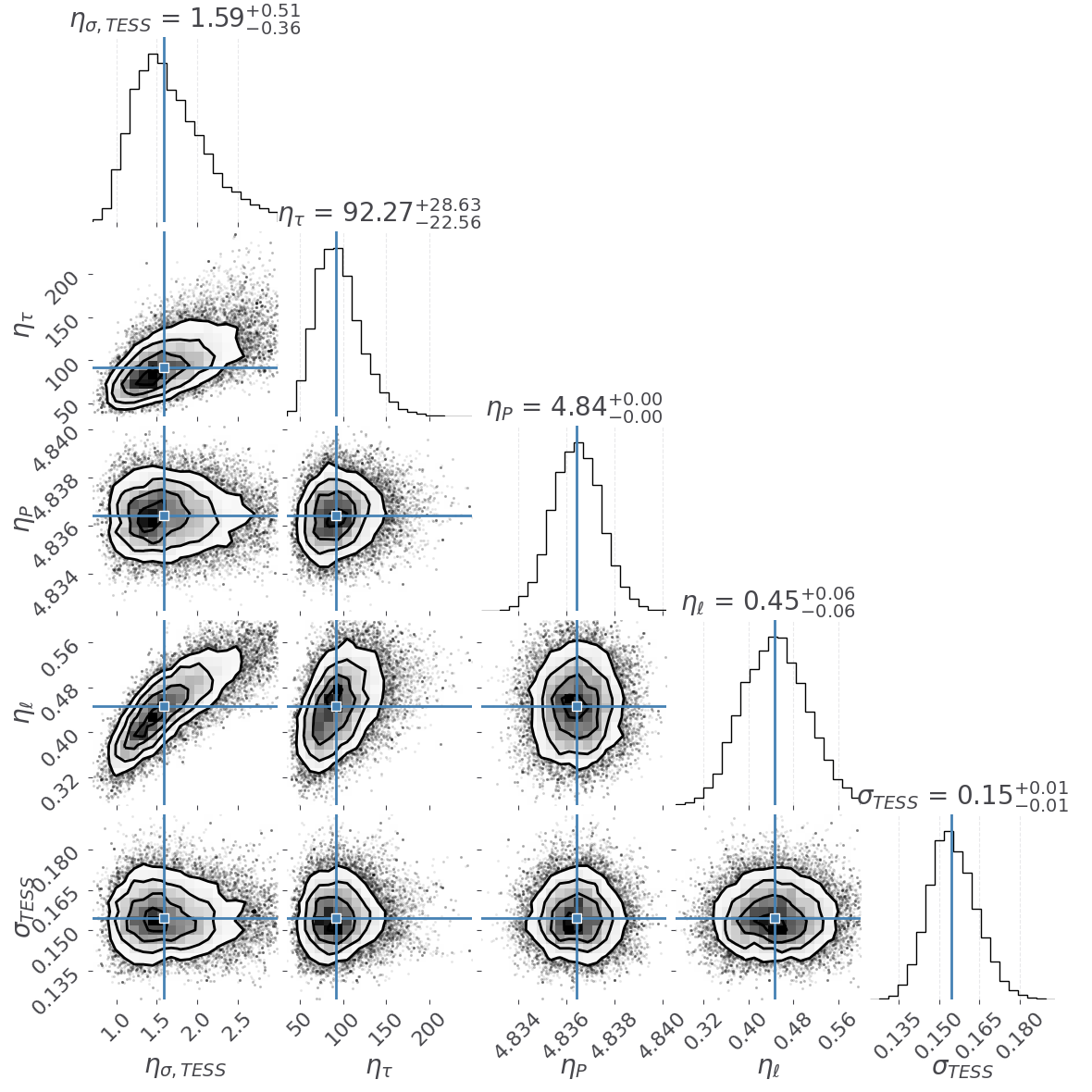}
    \caption{Posterior distributions from fits to the predicted RV variability from the $FF'$ technique (eq. \ref{eq:ffp}).}
    \label{fig:ffprime_corner}
\end{figure}

\noindent Here, $F$ is the photometric flux and $f$ represents the relative flux drop for a spot at the center of the stellar disk. To compute $F$ and $F'$ (the derivative of $F$ with respect to time), we first fit the \textit{TESS} light curve via cubic spline regression \citep[\texttt{scipy.interpolate.LSQUnivariateSpline};][]{scipy} for each Sector individually with knots sampled in units of 0.5 days ($\approx$ 10\% of one rotation period) to average over transits and the majority of flare events (fig. \ref{fig:light_curves}). The nominal cubic splines are then used to directly compute both $F$ and $F'$ on a down-sampled grid of 100 evenly-spaced points for each Sector. We then divide the resulting joint-Sector curve by its standard deviation for normalization; we do not care to directly fit for the chromatic parameter $f(TESS, ...)$. We further assume $f$ to be constant in time (i.e., spots are well-dispersed on the stellar surface). We then perform both MAP and MCMC analyses for this curve using a standard QP kernel (eq. \ref{eq:gp_qp}) with loose uniform priors of $\eta_{\tau} \sim \mathcal{U}(10, 2000)$ (days) and $\eta_{\ell} \sim \mathcal{U}(0.05, 0.6)$. We set the intrinsic error bars of the curve to zero but include an additional ``jitter'' (white noise) term in the model with a Jeffrey's prior \citep{1946RSPSA.186..453J} distribution with the knee at zero to help keep the jitter well-behaved by discouraging larger values unless it significantly improves the fit-quality through an inversely proportional penalty term. The amplitude of the model is drawn from a wide uniform distribution of $\mathcal{U}(0.3, 3.0)$. The posterior distributions are provided in fig. \ref{fig:ffprime_corner}.

A fit to the $FF'$ curve suggests the mean activity timescale $\eta_{\tau}\approx92_{-23}^{+29}$ days. Although our interpretation implies $\eta_{\tau}$ should be comparable to the gap between the two Sectors, ($\sim$ 700 days) we do not have photometric measurements between the two Sectors, and therefore cannot speak to evolution which will be important for our 2019 observations. We further note that the \textit{TESS} Sector 27 light curve exhibits moderate evolution whereas Sector 1 appears more stable (fig. \ref{fig:light_curves}). The posterior distributions are also consistent with a relatively smooth GP with the period length scale $\eta_{\ell}\approx 0.45 \pm 0.06$.

Before making use of our joint kernels, we first assess the performance of the standard QP kernel (eq. \ref{eq:gp_qp}) for each instrument individually. Here, each spectrograph makes use of a unique QP kernel and amplitude term, but the remaining three GP parameters are shared across all kernels. Each amplitude is drawn from a normal prior with mean equal to the standard deviation of the dataset, and a conservative width of 30 m\,s$^{-1}$. The expected semi-amplitudes for AU Mic b and c ($\lesssim 10$ m\,s$^{-1}$) will negligibly affect this estimation. We also apply a Jeffrey's prior with the knee at zero to help keep the amplitude well-behaved\footnote{Although the composite prior for the GP amplitudes is not proper (i.e., does not integrate to unity, see Section \ref{sec:primary_analyses}), we find nearly identical results with only the normal prior.}. For $\eta_{\tau}$ and $\eta_{\ell}$, we first make use of the same priors used to model the $FF'$ curve. We further include a fixed jitter term at 3 m\,s$^{-1}$ added in quadrature along the diagonal of the covariance matrix $\mathbf{K_{o}}$ for the HIRES observations only; HIRES observations provide the smallest intrinsic uncertainties, but are most impacted by activity (largest in amplitude), so we choose to moderately down-weight the HIRES observations. Given the flexibility of GP regression with a nightly-cadence, we choose not to fit for (nor include) jitter-terms for other spectrographs, and further discuss this decision in Section \ref{sec:caveats_future_work}. This is the most flexible model we employ to the RVs, and we therefore use these results to flag the aforementioned CARMENES-VIS and CARMENES-NIR measurements.

We find normally distributed posteriors for $\eta_{\tau}$ and $\eta_{\ell}$ (fig. \ref{fig:corner_2planets_disjoint_float}), but the reduced $\chi^{2}$ statistic of $0.32$ indicates the model over fits the data. The per-spectrograph amplitudes are reasonably consistent with their respective priors, so we assert this is a result of $\eta_{\tau}$ ($\approx43$ days) and/or $\eta_{\ell}$ ($\approx0.23$) taking on too small of values, indicating our RV model is insufficient to constrain these values from the RV observations, either due to insufficient cadence and/or an inadequate model. We therefore again fix $\eta_{\tau}=100$ days to let each season have mostly distinct activity models, while minimizing the flexibility within each season, which is consistent with what the $FF'$ curve suggests. As a compromise between the $FF'$ and RV analyses, we also fix $\eta_{\ell}=0.28$. Our adopted value of $\eta_{\tau}$ is larger than that used in \citetalias{K21} ($\approx$70 days\footnote{In \citetalias{K21}, the hyperparameters $\eta_{\tau}$ and $\eta_{\ell}$ absorb the factors of two present in the formulation used in this work (eq. \ref{eq:gp_qp}).}), while $\eta_{\ell}$ is nearly identical. We further explore these decisions and its impact on our derived semi-amplitudes in section \ref{sec:disc_kernel_parameters}. With fixed value for $\eta_{\tau}$, we re-run MAP and MCMC fits with disjoint kernels, yielding a reduced $\chi^{2}$ of 0.86, indicating the model is now only slightly over-fit.

\begin{figure*}
    \centering
    \includegraphics[width=0.95\textwidth]{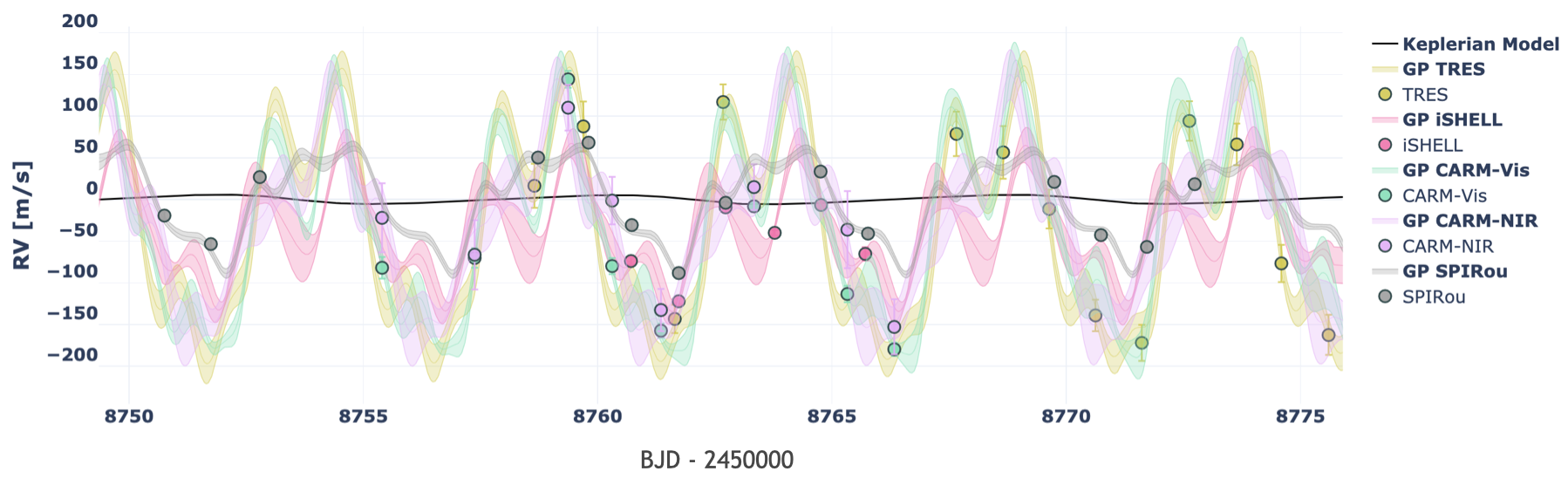}
    \caption{RVs of AU Mic zoomed in on a window with high-cadence, multi-wavelength observations from 2019. Here, we use a disjoint QP GP kernel (eq. \ref{eq:gp_qp}) to model the stellar activity. Each plotted dataset is only corrected according to the best-fit zero points. Data errors are computed by adding the intrinsic errors in quadrature with any uncorrelated noise terms (i.e., 3 m\,s$^{-1}$ for HIRES, see Table \ref{tab:pars_priors}). Although each GP makes use of the same parameters, each still exhibits unique features. This indicates either an insufficient activity model with our cadence or yet-to-be characterized chromatic effects of activity from different wavelength regimes not consistent with a simple scaling relation.}
    \label{fig:rvs_zoom_disjoint}
\end{figure*}

\begin{figure*}
    \centering
    \includegraphics[width=0.95\textwidth]{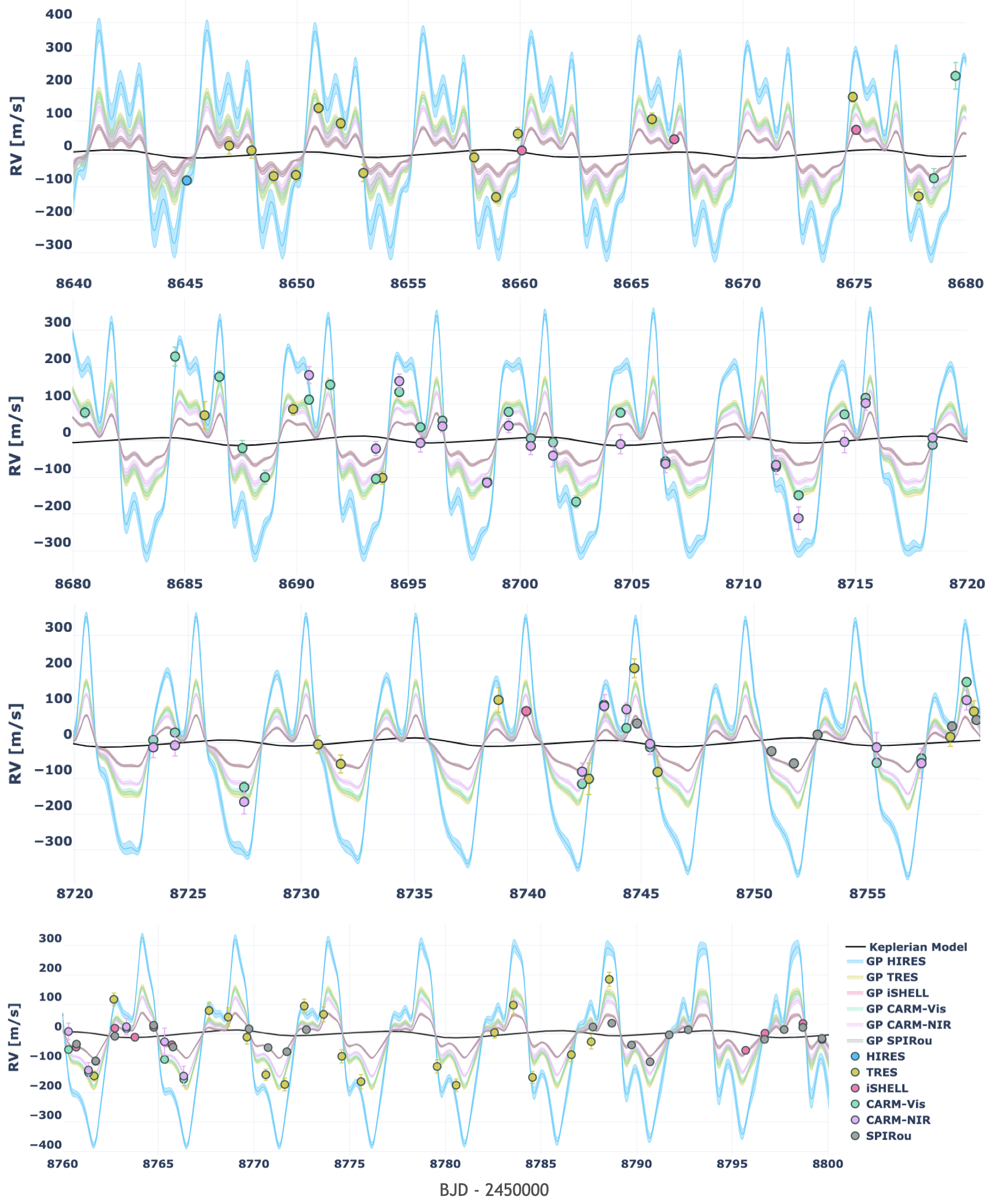}
    \caption{The 2019 RVs using kernel $\mathbf{K_{J1}}$ (eq. \ref{eq:gp_j1}) to model the stellar activity. Although there is only one HIRES observation in early 2019, we are still able to make predictions for the HIRES GP for the entire baseline by using joint kernels.}
    \label{fig:rvs_2019}
\end{figure*}

\begin{figure*}
    \centering
    \includegraphics[width=0.95\textwidth]{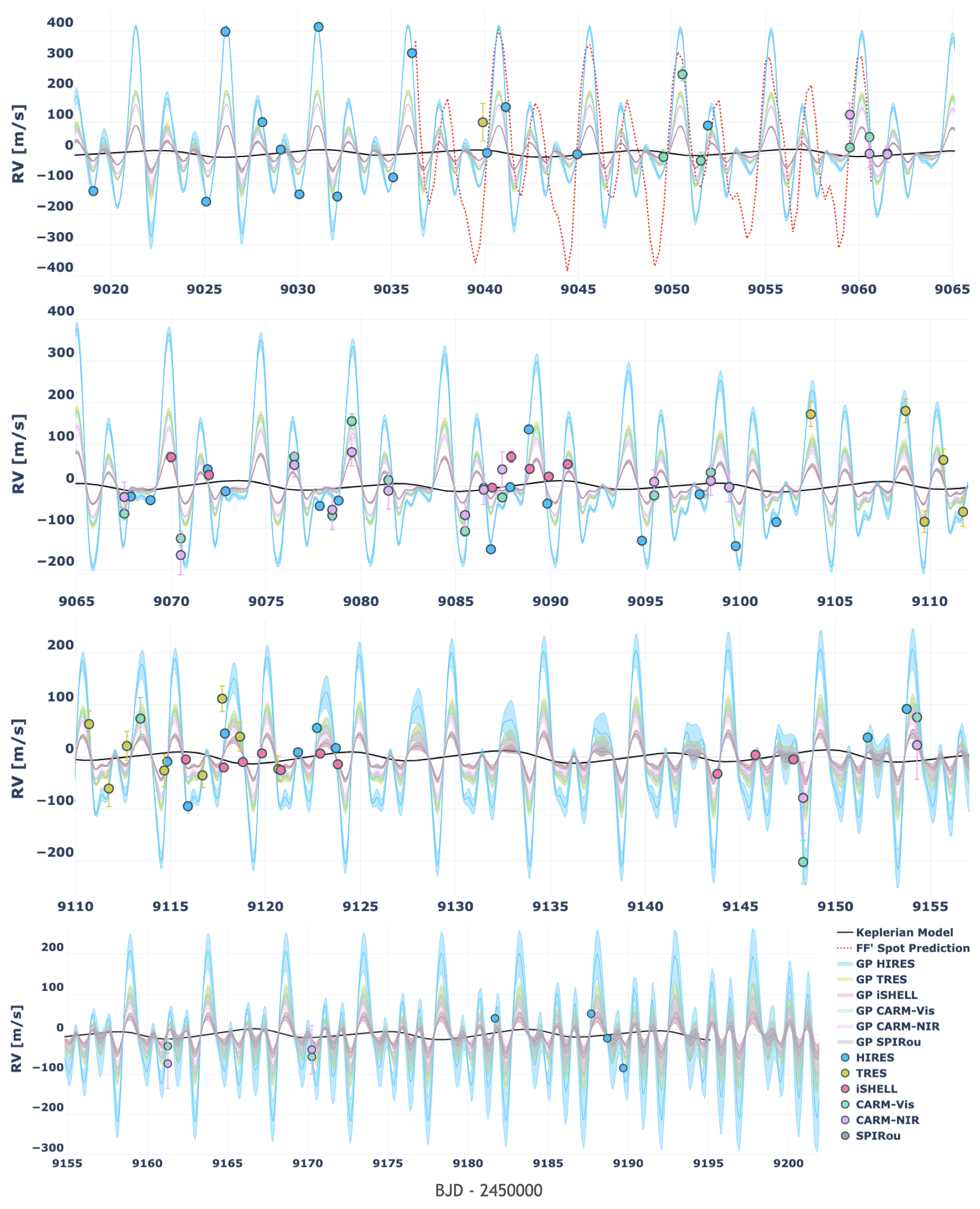}
    \caption{Same as fig. \ref{fig:rvs_2019}, but for our 2020 observations which overlap with the \textit{TESS} Sector 27 photometry. In red we show the generated $FF'$ curve for spot-induced activity signals (eq. \ref{eq:ffp}, arbitrarily scaled) generated from the \textit{TESS} light-curve (section \ref{sec:kernel_parameter_estimation}).}
    \label{fig:rvs_2020}
\end{figure*}

\subsubsection{Joint Kernel RV Fitting}

We use results from the disjoint case to inform our primary joint-kernel models. Although the different GPs appear similar (fig. \ref{fig:rvs_zoom_disjoint}), each still exhibits unique features, suggesting a simple scaling is not valid, and/or insufficient sampling for each kernel individually. Regardless, our two joint kernels will enforce a perfect scaling between any two spectrographs.

\begin{figure}
    \centering
    \includegraphics[width=0.5\textwidth]{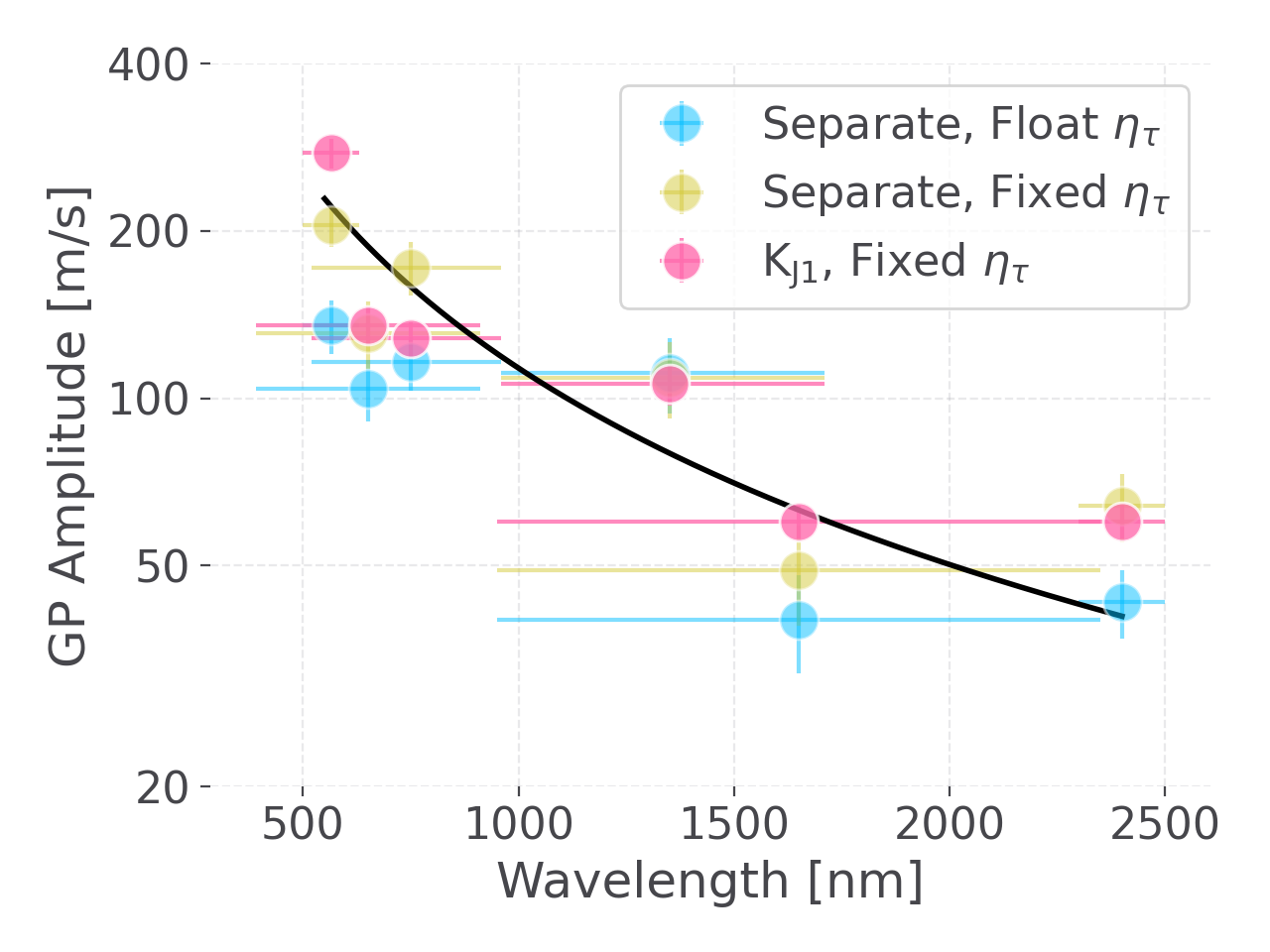}
    \caption{The best-fit GP amplitudes and uncertainties from kernels without enforcing any dependence with wavelength. We consider cases which let $\eta_{\tau}$ and $\eta_{\ell}$ float as well as and our fixed values (see Table \ref{tab:pars_priors}). The solid line is a least-squares solution to the amplitudes for kernel $\mathbf{K_{J2}}$ (eq. \ref{eq:gp_j2}) for the joint-kernel fixed case (pink markers). Horizontal bars correspond to the adopted spectral range for each instrument.}
    \label{fig:gp_amps_wavelength}
\end{figure}

We run MAP and MCMC fits using the joint kernel $\mathbf{K_{J1}}$ (eq. \ref{eq:gp_j1}) again making use of the same normal and Jeffrey's priors for each amplitude. We then fit the resulting set of best-fit amplitudes using our proposed power law relation (see eq. \ref{eq:gp_j2}): $\eta_{\sigma}(\lambda) = \eta_{\sigma,0} (\lambda_{0} / \lambda)^{\eta_{\lambda}}$ with \texttt{scipy.optimize.curve\_fit} \citep{jones2001scipy} (fig. \ref{fig:gp_amps_wavelength}). We arbitrarily anchor $\lambda_{0}$ at $\lambda=565$ nm. The effective mean wavelength of each spectrograph should consider the RV information content (stellar and calibration), and ignore regions with dense telluric features. For gas-cell calibrated spectrographs (HIRES, CHIRON, and iSHELL), we do limit the the range to regions with gas cell features. For all other spectrographs, we take the effective RV information content to be uniform over the full spectral range as a ``zeroth-order'' approximation \citep{2020ApJS..247...11R}. We further do not consider regions of tellurics which may have been masked (e.g., CARMENES RVs generated with \texttt{serval}). Although these estimations are imperfect, they are only relevant to kernel $\mathbf{K_{J2}}$ (eq. \ref{eq:gp_j2}). The adopted wavelengths for each spectrograph are listed in Table \ref{tab:spectrographs}.

We find $\eta_{\sigma,0} \approx 221$ m\,s$^{-1}$, and $\eta_{\lambda} \approx 1.17$. This amplitude is significantly larger than the intrinsic scatter of our observations (namely HIRES) suggests, so we adopt a tight normal prior of $\mathcal{N}(221, 10)$ to restrict it from getting any larger. We only apply a loose uniform prior for $\eta_{\lambda}\sim\mathcal{U}(0.2, 2)$. We then run corresponding MAP and MCMC fits with kernel $\mathbf{K_{J2}}$ (eq. \ref{eq:gp_j2}). A summary of all parameters is provided in Table \ref{tab:pars_priors}. We present and discuss fit results from both joint kernels in Section \ref{sec:results}.

\begin{table*}
\caption{The model parameters and prior distributions used in our primary fitting routines. \faLock\ indicates the parameter is fixed. We run models utilizing $\mathbf{K_{J1}}$ and $\mathbf{K_{J2}}$. We list the radii of AU Mic b and c measured in \citetalias{M21} which we use to compute the corresponding densities of each planet.}
\begin{center}
    \begin{tabular}{ | c | c | c | c | }
    \hline
    Parameter [units] & Initial Value ($P_{0}$) & Priors & Citation\\
    \hline
    \hline
    $P_{b}$ [days] & 8.4629991 & \faLock & Primary transit; \citetalias{M21} \\
    $TC_{b}$ [days] & 2458330.39046 & \faLock & Primary transit;  \citetalias{M21} \\
    $e_{b}$ & 0.189 & $\mathcal{N}(P_{0}, 0.04)$ & Secondary eclipse; Collins et al. in prep \\
    $\omega_{b}$ [rad] & 1.5449655 & $\mathcal{N}(P_{0}, 0.004)$ & Secondary eclipse; Collins et al. in prep \\
    $K_{b}$ [$\mathrm{ms}^{-1}$] & 8.5 & Positive & \citetalias{K21} \\
    \hline
    $P_{c}$ [days] & 18.858991 & \faLock & Primary transit; \citetalias{M21} \\
    $TC_{c}$ [days] & 2458342.2243 & \faLock & Primary transit; \citetalias{M21} \\
    $e_{c}$ & 0 & \faLock & -- \\
    $\omega_{c}$ [rad] & $\pi$ & \faLock & -- \\
    $K_{c}$ [$\mathrm{ms}^{-1}$] & 5 & Positive & \citetalias{M21} \\
    \hline
    $\eta_{\sigma,0}$ [m\,s$^{-1}$] & 216 & $\mathcal{J}(1, 600)$, $\mathcal{N}(P_{0}, 10)$ & RVs; this work \\
    $\eta_{\lambda}$ & 1.18 & $\mathcal{U}(0.3, 2)$ & RVs; this work \\
    $\eta_{\sigma,HIRES}$ [m\,s$^{-1}$] & 130 & $\mathcal{J}(1, 600)$, $\mathcal{N}(P_{0}, 30)$ & RVs; this work \\
    $\eta_{\sigma,TRES}$ [m\,s$^{-1}$] & 103 & $\mathcal{J}(1, 600)$, $\mathcal{N}(P_{0}, 30)$ & RVs; this work \\
    $\eta_{\sigma,CARM-VIS}$ [m\,s$^{-1}$] & 98 & $\mathcal{J}(1, 600)$, $\mathcal{N}(P_{0}, 30)$ & RVs; this work \\
    $\eta_{\sigma,CARM-NIR}$ [m\,s$^{-1}$] & 80 & $\mathcal{J}(1, 600)$, $\mathcal{N}(P_{0}, 30)$ & RVs; this work \\
    $\eta_{\sigma,SPIRou}$ [m\,s$^{-1}$] & 42 & $\mathcal{J}(1, 600)$, $\mathcal{N}(P_{0}, 30)$ & RVs; this work \\
    $\eta_{\sigma,iSHELL}$ [m\,s$^{-1}$] & 40 & $\mathcal{J}(1, 600)$, $\mathcal{N}(P_{0}, 30)$ & RVs; this work \\
    $\eta_{\tau}$ [days] & 100 & \faLock & \textit{TESS} light curve and RVs; this work \\
    $\eta_{\ell}$ & 0.28 & -- & \textit{TESS} light curve and RVs; this work \\
    $\eta_{p}$ [days] & 4.836 & $\mathcal{N}(P_{0}, 0.001)$ & \textit{TESS} light curve; this work \\
    $\gamma$ (per-spectrograph) [m\,s$^{-1}$] & 1 & $\mathcal{U}(-300, 300)$, $\mathcal{N}(0, 100)$ & RVs; this work \\
    \hline
    $\sigma_{HIRES}$ [m\,s$^{-1}$] & 3 & \faLock & -- \\
    $\sigma_{TRES}$ [m\,s$^{-1}$] & 0 & \faLock & -- \\
    $\sigma_{CARM-VIS}$ [m\,s$^{-1}$] & 0 & \faLock & -- \\
    $\sigma_{CARM-NIR}$ [m\,s$^{-1}$] & 0 & \faLock & -- \\
    $\sigma_{SPIRou}$ [m\,s$^{-1}$] & 0 & \faLock & -- \\
    $\sigma_{iSHELL}$ [m\,s$^{-1}$] & 0 & \faLock & -- \\
    \hline
    $M_{\star}$ [$M_{\odot}$] & $0.5_{-0.03}^{+0.03}$ & -- & \citetalias{P20} \\
    $R_{b}$ [$R_{\oplus}$] & $4.38_{-0.18}^{+0.18}$ & -- & \citetalias{P20} \\
    $R_{c}$ [$R_{\oplus}$] & $3.51_{-0.16}^{+0.16}$ & -- & \citetalias{M21} \\
    \hline
    \end{tabular}
\end{center}
\label{tab:pars_priors}
\end{table*}

\section{Results} \label{sec:results}

The best-fit parameters and corresponding uncertainties from the MAP and MCMC analyses with a two-planet model using joint kernels $\mathbf{K_{J1}}$ (eq. \ref{eq:gp_j1}) and $\mathbf{K_{J2}}$ (eq. \ref{eq:gp_j2}) are provided in Table \ref{tab:pars_results_2planets}. We compute planet masses, densities, and orbital semi-major axes by propagating the appropriate MCMC chains. The uncertainties in $M_{\star}$ and the planetary radii from Table \ref{tab:pars_priors} are added in quadrature where appropriate. A corner plot presenting the posterior distributions of each varied parameter are provided in figs. \ref{fig:corner_2planets_j1} and \ref{fig:corner_2planets_j2} for kernels $\mathbf{K_{J1}}$ and $\mathbf{K_{J2}}$, respectively. All chains are well-converged, with posteriors resembling Gaussian distributions. We find the offsets for each spectrograph are highly correlated with one-another; we note this is unique to the cases leveraging a joint-kernel, and strongest when datasets overlap, but do not further explore this result.

\begin{figure*}
    \centering
    \includegraphics[width=0.98\textwidth]{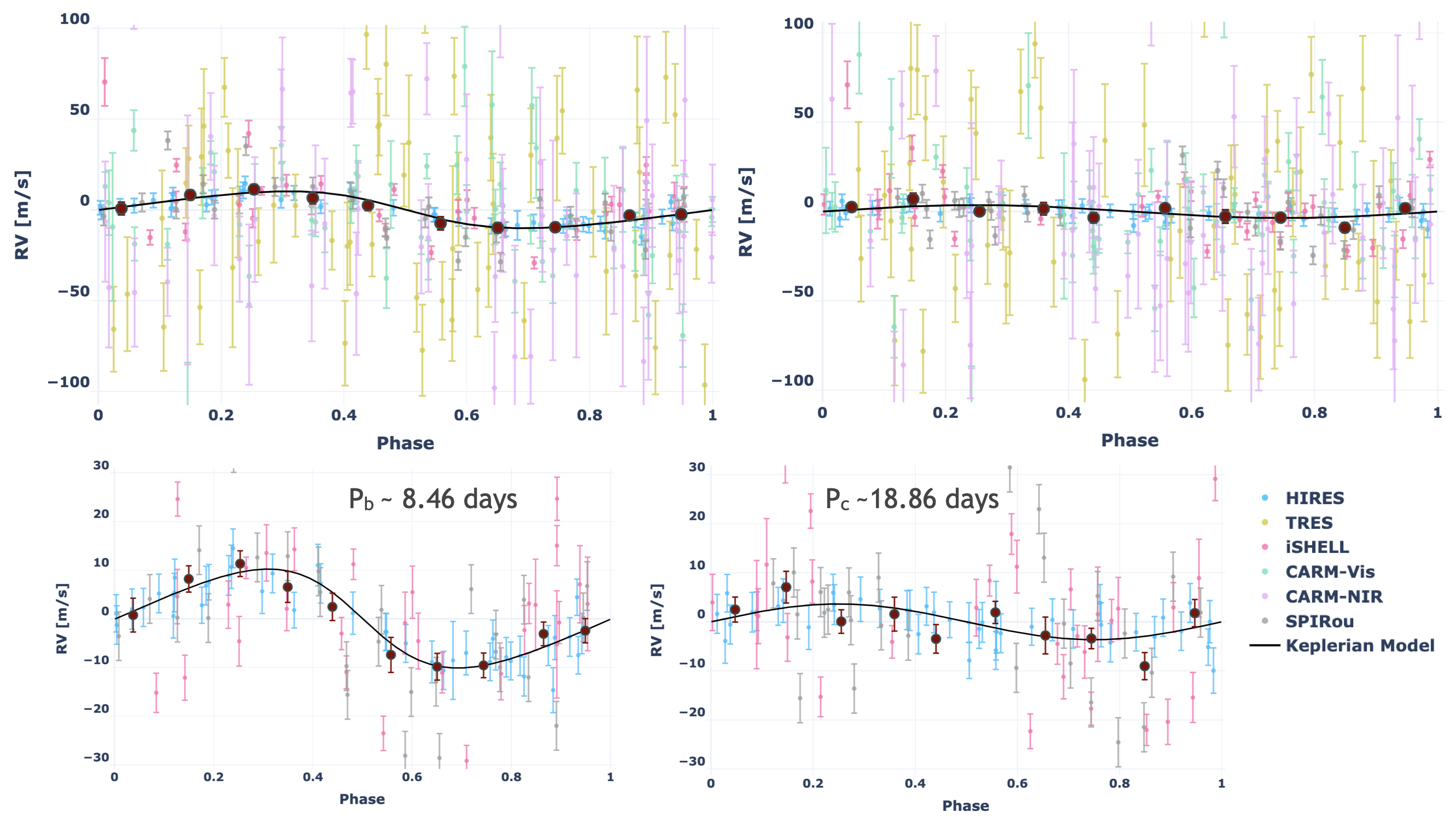}
    \caption{The phased RVs for AU Mic b (left column), and c (right column), and the corresponding best fit Keplerian models, generated from our nominal two-planet model. For each spectrograph, we subtract the unique zero-points, all other planet signals, and the appropriate GP. Corresponding data errors are computed by adding the intrinsic error in quadrature with additional uncorrelated noise (i.e., 3 m\,s$^{-1}$ for HIRES, see Table \ref{tab:pars_priors}). The dark red points are generated by binning the phased RVs using a window size of 0.1, weighted by $1/\sigma_{RV}^{2}$ where $\sigma_{\mathrm{RV}}$ are the data errors. In the top row we plot all data used in the fit. In the bottom row, we only show HIRES, iSHELL, and SPIRou. Although the HIRES cadence in 2020 was relatively dense with respect to the activity timescales $\eta_{\tau}$ and $\eta_{P}$, the data still appears to be over-fit.}
    \label{fig:2planets_phased}
\end{figure*}

Unlike kernel $\mathbf{K_{J2}}$, $\mathbf{K_{J1}}$ only enforces a scaling relation between the different spectrographs but no correlation with wavelength, so we adopt results from $\mathbf{K_{J1}}$ for our primary results, although the results for $K_{b}$ and $K_{c}$ are moderately consistent between kernels. With kernel $\mathbf{K_{J1}}$ (eq. \ref{eq:gp_j1}), we report the median semi-amplitudes of AU Mic b and c to be $10.23_{-0.91}^{+0.88}$ m\,s$^{-1}$ and $3.68_{-0.86}^{+0.87}$ m\,s$^{-1}$, corresponding to masses of $20.12_{-1.72}^{+1.57}\ M_{\oplus}$ and $9.60_{-2.31}^{+2.07}\ M_{\oplus}$, respectively. The phased-up RVs for AU Mic b and c are shown in fig. \ref{fig:2planets_phased}. With kernel $\mathbf{K_{J2}}$ (eq. \ref{eq:gp_j2}), we find $K_{b}=8.92_{-0.85}^{+0.85}$ m\,s$^{-1}$ and $K_{c}=5.21_{-0.87}^{+0.90}$ m\,s$^{-1}$. Both our findings for $K_{b}$ are larger but within $1\sigma$ of the semi-amplitude reported in \citetalias{K21} ($8.5_{-2.2}^{+2.3}$ m\,s$^{-1}$). The mass of AU Mic c is also consistent with a Chen-Kipping mass-radius relation \citep[$\approx 12.1\ M_{\oplus}$;][]{chen_kipping}. The posterior distributions for $e_{b}$ and $\omega_{b}$ are also consistent with their respective priors. Our finding for $K_{b}$ is nearly twice as large as that obtained when using disjoint QP kernels (5.58 m\,s$^{-1}$, fig. \ref{fig:corner_2planets_disjoint_locked}), although the uncertainties are similar. With disjoint kernels, we find no evidence in the RVs for AU Mic c.

We further validate our results by computing the Bayesian information criterion (BIC) and the small-sample Akaike information criterion \citep{akaike1974, aicc2002}. We compute the relevant quantities for a power-set of planet models. We are not trying to independently detect the eccentricity of AU Mic b and therefore do not include cases with $e_{b}=0$. Prior probabilities are not included in the calculation of the corresponding $\ln \mathcal{L}$ (eq. \ref{eq:map}) to maintain normalization between different models. The results are summarized in Table \ref{tab:model_selection} and are consistent with the relative precisions for each derived semi-amplitude.

Lastly, we compute and present the reduced chi-squared statistic ($\chi_{red}^{2}$) for each spectrograph individually to assess their respective goodness of fit (Table \ref{tab:redchi2s_per_spectrograph}). We add in quadrature the intrinsic error bars with any additional uncorrelated noise (i.e., 3 m\,s$^{-1}$ for HIRES, see Table \ref{tab:pars_priors}). We find the HIRES observations are moderately over-fit ($\chi_{red}^{2}$=0.64), whereas the other spectrographs are under-fit. We suspect this is due to the activity amplitude for HIRES being significantly larger than the other spectrographs despite exhibiting a similar overall dispersion. Although we include an additional 3 m\,s$^{-1}$ white noise term for the HIRES observations, they still yield the smallest overall error bars and therefore are given the most weight in the GP regression. Although a more flexible uncorrelated noise model may yield a more accurate weighting scheme for the different spectrographs (i.e., a varied ``jitter'' parameter for each spectrograph), we favor the model without them for the variety of reasons discussed in Section \ref{sec:caveats_future_work}. Lastly, we note that we find moderately similar results when not using the HIRES RVs altogether ($K_{b}=12.95 \pm 1.1$ m\,s$^{-1}$, $K_{c}=3.5 \pm 1.0$ m\,s$^{-1}$).

\begin{table*}
\caption{The best-fit parameters and corresponding Keplerian variables for our primary two-planet fits using joint-kernels $\mathbf{K_{J1}}$ (eq. \ref{eq:gp_j1}) and $\mathbf{K_{J2}}$ (eq. \ref{eq:gp_j2}). The MCMC values correspond to the 15.9$^{\textrm{th}}$, 50$^{\textrm{th}}$, and 84.1$^{\textrm{th}}$ percentiles. Planet masses, densities, and semi-major axes are computed by propagating the appropriate MCMC chains. We also add in quadrature the uncertainties in $M_{\star}$ and planetary radii from Table \ref{tab:pars_priors} where relevant.}
\begin{center}
    \begin{tabular}{ | c | c | c | c | c | }
    \hline
    Name [units] & MAP (J1) & MCMC (J1) & MAP (J2) & MCMC (J2) \\
    \hline
    \hline
    $P_{b}$ [days] & 8.4629991 & -- & -- & -- \\
    $TC_{b}$ [days; BJD] & 2458330.39046 & -- & -- & -- \\
    $e_{b}$ & 0.187 & $0.186_{-0.035}^{+0.036}$ & 0.182 & $0.181_{-0.035}^{+0.035}$ \\
    $\omega_{b}$ [radians] & 1.5452 & $1.5451_{-0.0038}^{+0.0038}$ & 1.5453 & $1.5454_{-0.0041}^{+0.0041}$ \\
    $K_{b}$ [m\,s$^{-1}$] & 10.21 & $10.23_{-0.91}^{+0.88}$ & 8.94 & $8.92_{-0.85}^{+0.85}$ \\
    $M_{b}$ [$M_{\oplus}$] & 20.14 & $20.12_{-1.72}^{+1.57}$ & 17.66 & $17.73_{-1.62}^{+1.68}$ \\
    $a_{b}$ [AU] & 0.0645 & $0.0645_{-0.0013}^{+0.0013}$ & -- & -- \\
    $\rho_{b}$ [g/cm$^{3}$] & 1.32 & $1.32_{-0.20}^{+0.19}$ & 1.16 & $1.16_{-0.18}^{+0.18}$ \\
    \hline
    $P_{c}$ [days] & 18.858991 & -- & -- & -- \\
    $TC_{c}$ [days; BJD] & 2458342.2243 & -- & -- & -- \\
    $e_{c}$ & 0 & -- & -- & -- \\
    $\omega_{c}$ [radians] & $\pi$ & -- & -- & -- \\
    $K_{c}$ [m\,s$^{-1}$] & 3.62 & $3.68_{-0.86}^{+0.87}$ & 5.23 & $5.21_{-0.87}^{+0.90}$ \\
    $M_{c}$ [$M_{\oplus}$] & 9.50 & $9.60_{-2.31}^{+2.07}$ & 13.71 & $14.12_{-2.71}^{+2.48}$ \\
    $a_{c}$ [AU] & 0.1101 & $0.1101_{-0.002}^{+0.002}$ & -- & -- \\
    $\rho_{c}$ [g/cm$^{3}$] & 1.21 & $1.22_{-0.29}^{+0.26}$ & 1.75 & $1.80_{-0.34}^{+0.31}$ \\
    \hline
    $\gamma_{HIRES}$ [m\,s$^{-1}$] & 2.9 & $4.1_{-57.0}^{+55.6}$ & -19.4 & $-8.7_{-42.9}^{+43.3}$ \\
    $\gamma_{TRES}$ [m\,s$^{-1}$] & 11.4 & $12.1_{-27.8}^{+27.4}$ & -0.2 & $9.3_{-38.0}^{+38.4}$ \\
    $\gamma_{CARM-VIS}$ [m\,s$^{-1}$] & 3.7 & $4.3_{-26.6}^{+26.0}$ & -12.1 & $-3.4_{-33.9}^{+34.5}$ \\
    $\gamma_{CARM-NIR}$ [m\,s$^{-1}$] & 2.6 & $2.9_{-21.8}^{+21.7}$ & -6.8 & $-1.5_{-21.2}^{+21.2}$ \\
    $\gamma_{SPIRou}$ [m\,s$^{-1}$] & 5.5 & $5.6_{-12.4}^{+12.3}$ & 0.72 & $5.1_{-17.3}^{+17.8}$ \\
    $\gamma_{iSHELL}$ [m\,s$^{-1}$] & -2.8 & $-2.4_{-12.4}^{+12.1}$ & -7.5 & $-4.3_{-12.9}^{+13.0}$ \\
    \hline
    $\eta_{\sigma,0}$ [m\,s$^{-1}$] & -- & -- & 242.4 & $243.1_{-9.1}^{+8.8}$ \\
    $\eta_{\lambda}$ & -- & -- & 0.843 & $0.845_{-0.024}^{+0.024}$ \\
    $\eta_{\sigma,HIRES}$ [m\,s$^{-1}$] & 269.4 & $275.7_{-16.4}^{+17.4}$ & -- & -- \\
    $\eta_{\sigma,TRES}$ [m\,s$^{-1}$] & 132.3 & $135.4_{-9.5}^{+10.6}$ & -- & -- \\
    $\eta_{\sigma,CARM-VIS}$ [m\,s$^{-1}$] & 125.1 & $128.2_{-8.2}^{+8.8}$ & -- & -- \\
    $\eta_{\sigma,CARM-NIR}$ [m\,s$^{-1}$] & 103.0 & $105.5_{-8.7}^{+9.1}$ & -- & -- \\
    $\eta_{\sigma,SPIRou}$ [m\,s$^{-1}$] & 58.5 & $60.1_{-4.0}^{+4.3}$ & -- & -- \\
    $\eta_{\sigma,iSHELL}$ [m\,s$^{-1}$] & 58.5 & $60.0_{-3.9}^{+4.2}$ & -- & -- \\
    $\eta_{P}$ [days] & 4.8384 & $4.8384_{-0.0009}^{+0.0008}$ & 4.8376 & $4.8376_{-0.0009}^{+0.0009}$ \\
    \hline
    \end{tabular}
\end{center}
\label{tab:pars_results_2planets}
\end{table*}

\begin{table}
    \centering
    \begin{tabular}{|c|c|}
    \hline
    Spectrograph & $\chi_{red}^{2}$ \\
    \hline
    HIRES & 1.09 \\
    TRES & 6.86 \\
    CARMENES-VIS & 5.89 \\
    CARMENES-NIR & 4.39 \\
    SPIRou & 16.70 \\
    iSHELL & 22.43 \\
    \hline
    \end{tabular}
    \caption{The reduced chi-squared for each spectrograph from our nominal two-planet model using kernel $\mathbf{K_{J1}}$. Unlike when using quasi-disjoint kernels (Section \ref{sec:kernel_parameter_estimation}), we find the model is overall under-fit with the joint kernel. We suspect this is primarily due to an inadequate stellar-activity (i.e., a scaling relation is insufficient between spectrographs) and/or the exclusion of per-spectrograph jitter terms, and discuss these details further in Section \ref{sec:caveats_future_work}.}
    \label{tab:redchi2s_per_spectrograph}
\end{table}

\subsection{Evidence For Additional Candidates?}

We compute periodograms to further assess the relative statistical confidence of the two transiting planets and to search for other planets in the system. We first compute a series of generalized Lomb-Scargle \citep[GLS;][]{2018ascl.soft07019Z, pya} periodograms out to 500 days after removing the nominal zero-points, appropriate GPs, and the two planets, all generated using parameters from our nominal two-planet model (Table \ref{tab:pars_results_2planets}) with kernel $\mathbf{K_{J1}}$ (eq. \ref{eq:gp_j1}) to model the stellar activity. We also compute an activity-filtered periodogram from a planet-free model to assess how much the GP model will absorb planetary signals, and inform our interpretation of other peaks present in the periodogram. We further plot the normalized power-levels for false alarm probabilities (FAPs) of 10\%, 1\%, and 0.1\%.

We also compute ``brute-force'' periodograms by performing MAP fits for a wide range of fixed orbital periods for a user-defined ``test-planet'' with various assumptions for other model parameters \citep[see][]{2021MNRAS.502.3704A}. Given the time complexity of GP regression, we only consider periods out to 100 days. We first run two searches with no other planets in the model, first allowing for the test-planet's $TC$ to float, and second fixing $TC$ to the nominal value for AU Mic b (Table \ref{tab:pars_results_2planets}). We then run searches for a second-planet, this time including a planetary model to account for the orbit of AU Mic b, with $K_{b}\sim\mathcal{N}(8.5, 2.5)$, consistent with the semi-amplitude found in \citetalias{K21}. We again consider the case of letting the test-planet's $TC$ float, then run three cases with fixing the test-planet's TC to each time of transit for AU Mic c from the \textit{TESS} Sector 1 and 27 light curves (\citetalias{M21}). Lastly, we perform a search for a third planet letting its $TC$ float, and including models for AU Mic b and c ($K_{b}\sim\mathcal{N}(8.5, 2.5)$, $K_{c}>0$).

\begin{figure*}
    \centering
    \includegraphics[width=0.98\textwidth]{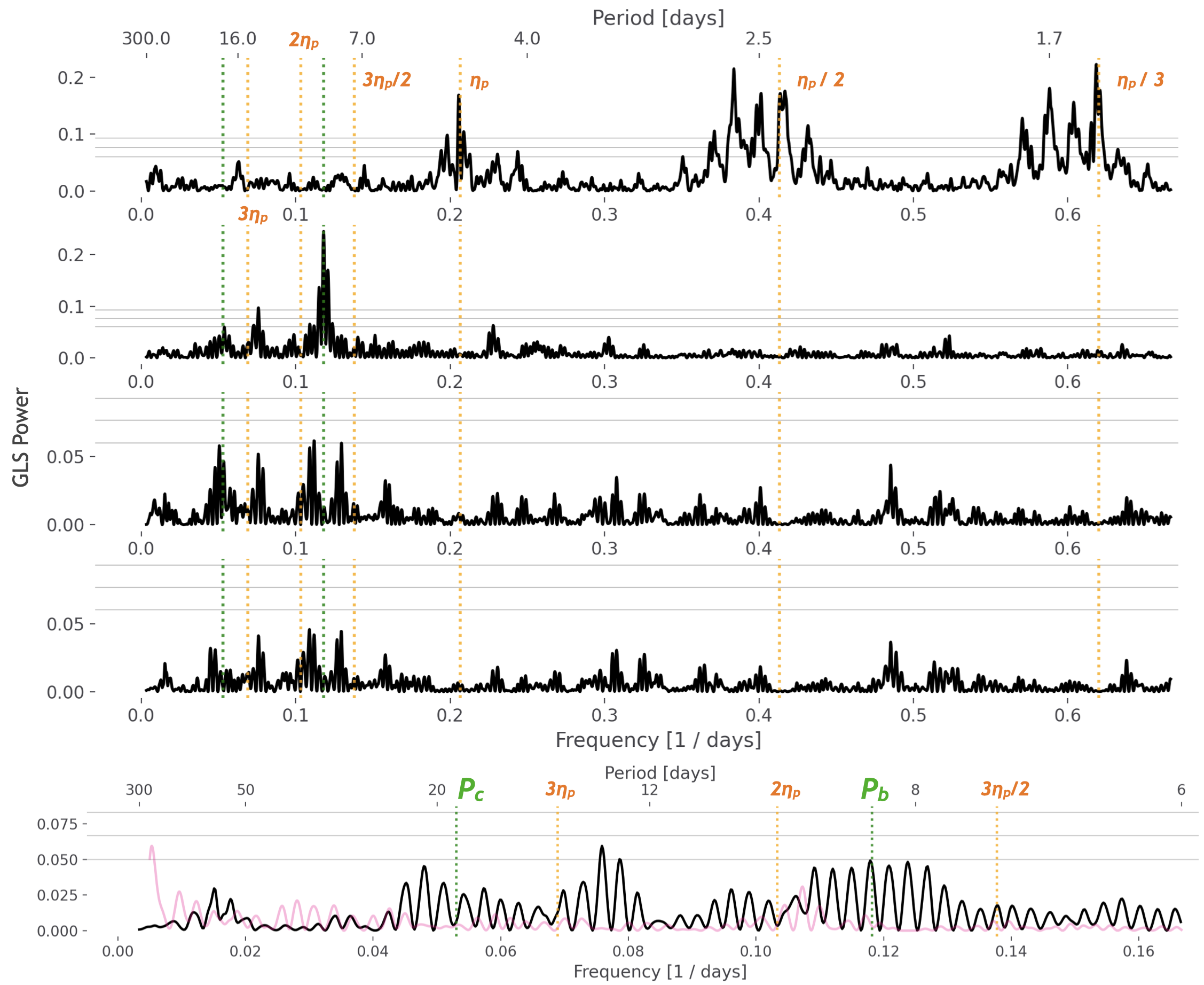}
    \caption{GLS periodograms for AU Mic. Rows 1--4 are generated from our nominal two-planet MAP fit result using $\mathbf{K_{J1}}$ (eq. \ref{eq:gp_j1}) to model the stellar activity. From top to bottom, with each step applying an additional ``correction'': 1. zero-point corrected RVs, 2. activity-filtered RVs, 3. planet b-filtered RVs, 4. planet c-filtered RVs. Annotated from left to right in green are the periods for AU Mic c and b. In the top row, we also annotate in orange (from left to right) potential aliases of the stellar rotation period $3\eta_{P}$, $2\eta_{P}$, and $3\eta_{P}/2$, followed by the first three harmonics. In the bottom row, we compute a periodogram from an activity-filtered and trend-corrected zero-planet model to indicate how power from planets is absorbed by the GP. In each periodogram, we also identify the false alarm probability (FAP) power levels corresponding to 0.1\% (highest), 1\%, and 10\% (lowest). The clear alias present in all periodograms is caused from the large gap between the two seasons of observations. In the bottom panel, we also plot in pink a Lomb-Scargle periodogram (arbitrarily scaled) of our window function (i.e, identical yet arbitrary RVs at each observation).}
    \label{fig:gls_periodograms}
\end{figure*}

\begin{figure*}
    \centering
    \includegraphics[width=0.98\textwidth]{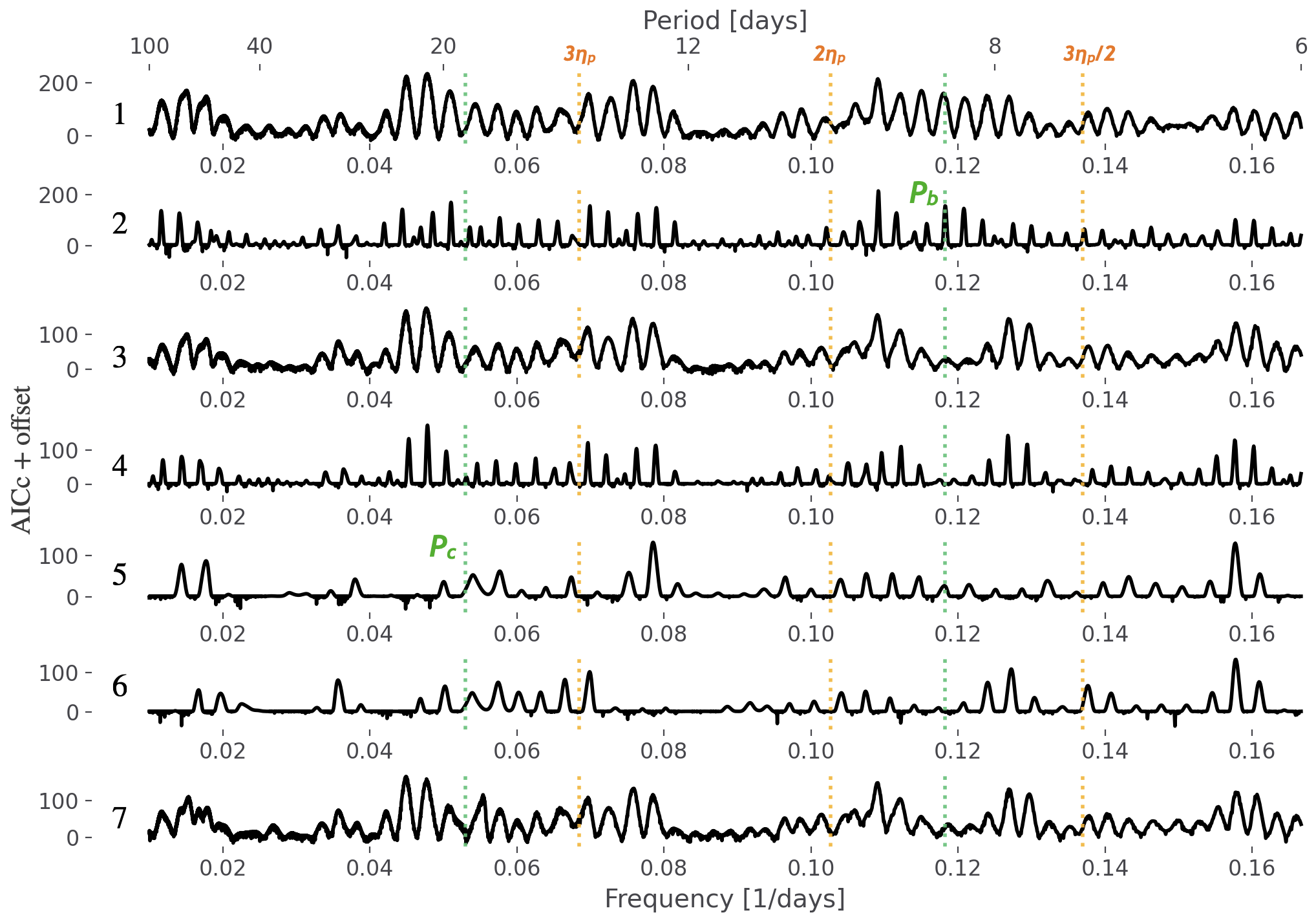}
    \caption{``Brute-force'' periodograms for AU Mic with different assumptions for planetary models, but all making use of kernel $\mathbf{K_{J1}}$ (eq. \ref{eq:gp_j1}) to model the stellar activity. In each row, we perform a MAP fit for a wide range of fixed periods for a particular ``test''-planet. In row 1, we include no other planets in our model, and allow for the test-planet's $TC$ to float. In row 2, we perform the same search but fixing $TC$ to the nominal value for AU Mic b (Table \ref{tab:pars_priors}). In row 3, we include a model for AU Mic b (with $K_{b}\sim\mathcal{N}(8.5, 2.5)$, see \citetalias{K21}), and search for a second planet again letting $TC$ float. In rows 4--6, we perform the same search but fix the test-planet's $TC$ to one of the three times of transit for AU Mic c from \textit{TESS} (in chronological order). In the bottom row, we include nominal models for AU Mic b and c ($K_{b}\sim\mathcal{N}(8.5, 2.5)$, $K_{c}>0$). We also annotate the same potential aliases with the stellar-rotation period (orange) and planetary periods (green) as in fig. \ref{fig:gls_periodograms}.}
    \label{fig:brute_force_periodograms}
\end{figure*}

\begin{figure}
    \centering
    \includegraphics[width=0.45\textwidth]{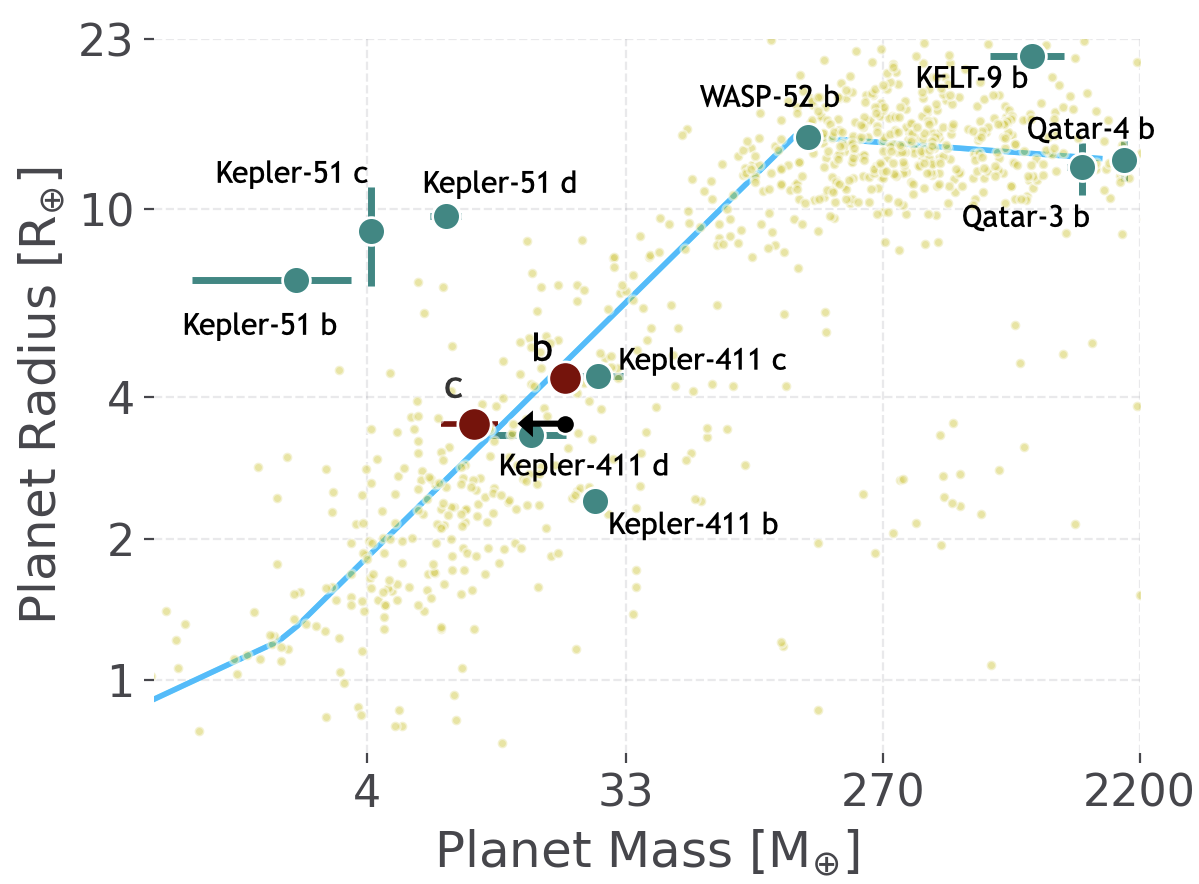}
    \caption{The mass vs. radius for all exoplanets with provided radii and masses from the NASA Exoplanet Archive \citep{nasa_exoplanet_archive}. For AU Mic b and c, we plot (maroon markers) the masses determined from our two-planet model with kernel $\mathbf{K_{J1}}$. We also indicate with an arrow the $5\sigma$ upper limit to the mass of AU Mic c determined from the posterior of $K_{c}$. The radii for b and c are those reported in \citetalias{M21}. In blue, we plot a piece-wise Chen-Kipping mass-radius relation \citep{chen_kipping}. We also annotate (cyan markers) the masses and radii for a sample of young planets (stellar-age estimated $\lesssim 400$ Myr).}
    \label{fig:mass_radius}
\end{figure}

Both the GLS (fig. \ref{fig:gls_periodograms}) and brute-force (fig. \ref{fig:brute_force_periodograms}) periodograms exhibit clear aliasing with a frequency of $\approx 0.00281$ days$^{-1}$ (or 356 days) which we attribute to having two seasons of observations separated by $\approx 200$ days. Given the respective power of AU Mic b in both the GLS and brute-force planet-free periodograms, we briefly explore other peaks with similar power, even though all other peaks are below all three FAPs after removing the nominal two-planet model (fig. \ref{fig:gls_periodograms}, row 3; fig. \ref{fig:brute_force_periodograms}, row 7). Both two- and zero-planet periodograms (as well as GLS and brute-force) show power between AU Mic b and c's orbits near 12.72 and 13.19 days, as well as power near 66.7 days for the residual RVs. Although these peaks are comparable in power to AU Mic b in both planet-free periodograms, they may be spurious. We further discuss the confirmation of AU Mic b and c as well as the validation of such additional potential candidates in Section \ref{sec:injection_recovery}. A mass-radius diagram is shown in fig. \ref{fig:mass_radius} to place the mass and radius of all AU Mic b and c in context with other known exoplanets, including a subset of young sample of exoplanets shown in \citetalias{P20}. The plotted masses for AU Mic b and c are from our nominal two-planet model using kernel $\mathbf{K_{J1}}$ (eq. \ref{eq:gp_j1}).

\begin{table}
\caption{Model information criterion for AU Mic b and c using kernel $\mathbf{K_{J1}}$ (eq. \ref{eq:gp_j1}) to model the stellar activity.}
\begin{center}
    \begin{tabular}{ | c | c | c | c | c | c | }
    \hline
    Planets & $\ln\mathcal{L}$ & $\chi_{red}^{2}$ & N free &  $\Delta$AICc & $\Delta$BIC \\
    \hline
    b, c & -1753.1 & 4.73 & 17 & 0 & 0 \\
    b & -1762.0 & 4.77 & 16 & 15.5 & 12.2 \\
    c & -1816.4 & 5.14 & 14 & 119.8 & 109.8 \\
    None & -1828.8 & 5.23 & 13 & 142.4 & 129.13 \\
    \hline
    \end{tabular}
\end{center}
\label{tab:model_selection}
\end{table}

\section{Discussion} \label{sec:discussion}


\subsection{Constraints on Eccentricity} \label{sec:disc_eccentricity}

Here we briefly explore eccentric orbits for the two-transiting planets b and c. For each planet, we take $e \sim \mathcal{U}(0, 0.7)$ and $\omega\sim \mathcal{U}(0, 2\pi)$. We only use kernel $\mathbf{K_{J1}}$ (eq. \ref{eq:gp_j1}) to mode the stellar activity. Posterior distributions are presented in fig. \ref{fig:corner_2planets_j1_ecc}. We find $e_{b}=0.30\pm0.04$, which is $\approx$ 50\% larger than our prior informed by a secondary eclipse event indicates. The corresponding finding of $\omega_{b}=3.01\pm0.27$ is also inconsistent with our adopted prior for $\omega_{b}$. The posterior distribution for $e_{c}$ is concentrated at the upper bound (0.7), implying an overlapping orbit with AU Mic b. Orbital stability calculations presented in \citetalias{M21} indicate $e_{c}<0.2$, so we assert our model is unable to accurately constrain its eccentricity. The behavior of $e_{c}$ further indicates our detection of $K_{c}$ may not be significant.

\subsection{Sensitivity to Kernel Hyperparameters} \label{sec:disc_kernel_parameters}

Our analyses in section \ref{sec:kernel_parameter_estimation} make use of a fixed mean spot lifetime $\eta_{\tau}=100$ days and smoothing parameter $\eta_{\ell}=0.28$. Here we determine how sensitive the recovered semi-amplitudes of AU Mic b and c are to these two parameters. We consider $\eta_{\tau} \in$ \{40, 70, 100, 200, 300\} (days), and $\eta_{\ell} \in$ \{0.15, 0.2, 0.25, 0.3, 0.35\}. We perform MAP and MCMC fits for all pairs of these two fixed parameters using $\mathbf{K_{J1}}$ (eq. \ref{eq:gp_j1}) for a two-planet model. All other parameters adopt initial values and priors from Table \ref{tab:pars_priors}. Results are summarized in Table \ref{tab:kernel_param_search}.

We find $K_{b}$ is only moderately sensitive to the values of each hyper-parameter, ranging from $\sim$ 7--11 m\,s$^{-1}$. With a larger spot lifetime, $K_{b}$ tends towards larger values, indicating the GP is likely absorbing power from planet b with a more flexible model (smaller $\eta_{\tau}$). However, $K_{b}$ is relatively insensitive to the value of $\eta_{\ell}$. The range of values for $K_{c}$ is larger, changing by nearly a factor of three. Unlike $K_{b}$, $K_{c}$ is more unstable and tends towards larger values when using a more flexible (smaller) spot lifetime. The reduced chi-squared statistic indicates the model is not over-fit in any of the cases performed, but is also larger than unity by a several factors in most cases indicating our modeling is inadequate.

\begin{table}[]
    \centering
    \begin{tabular}{| c | c | c | c | c | c |}
    \hline
     $\eta_{\tau}$ [days] & $\eta_{\ell}$ & $K_{b}$ [m\,s$^{-1}$] & $K_{c}$ [m\,s$^{-1}$] & $\chi^{2}_{red}$ \\
     \hline
     40 & 0.15 & $8.79\pm1.47$ & $7.38\pm1.65$ & 1.58 \\
     40 & 0.2 & $8.84\pm1.30$ & $8.51\pm1.33$ & 2.10 \\
     40 & 0.25 & $8.23\pm1.17$ & $9.05\pm1.17$ & 2.45 \\
     40 & 0.3 & $7.41\pm1.08$ & $9.13\pm1.13$ & 2.68 \\
     40 & 0.35 & $6.95\pm0.98$ & $9.23\pm1.05$ & 2.85 \\
     \hline
     70 & 0.15 & $8.74\pm1.24$ & $6.88\pm1.30$ & 1.99 \\
     70 & 0.2 & $10.32\pm1.13$ & $5.90\pm1.07$ & 2.69 \\
     70 & 0.25 & $10.45\pm1.04$ & $4.76\pm0.95$ & 3.25 \\
     70 & 0.3 & $9.61\pm0.91$ & $4.16\pm0.89$ & 3.65 \\
     70 & 0.35 & $9.18\pm0.82$ & $3.94\pm0.83$ & 3.90 \\
     \hline
     100 & 0.15 & $9.28\pm1.17$ & $5.88\pm1.08$ & 2.46 \\
     100 & 0.2 & $10.85\pm1.00$ & $4.73\pm0.98$ & 3.20 \\
     100 & 0.25 & $10.78\pm0.95$ & $3.78\pm0.87$ & 3.76 \\
     100 & 0.3 & $9.81\pm0.85$ & $3.63\pm0.80$ & 4.12 \\
     100 & 0.35 & $9.22\pm0.80$ & $3.60\pm0.77$ & 4.39 \\
     \hline
     200 & 0.15 & $9.38\pm0.98$ & $4.01\pm0.96$ & 3.44 \\
     200 & 0.2 & $11.04\pm0.89$ & $3.35\pm0.84$ & 4.16 \\
     200 & 0.25 & $11.09\pm0.87$ & $3.38\pm0.77$ & 4.73 \\
     200 & 0.3 & $10.14\pm0.84$ & $4.51\pm0.76$ & 5.28 \\
     200 & 0.35 & $9.06\pm0.73$ & $4.77\pm0.66$ & 5.59 \\
     \hline
     300 & 0.15 & $9.32\pm0.92$ & $3.84\pm0.87$ & 3.82 \\
     300 & 0.2 & $10.60\pm0.88$ & $3.78\pm0.81$ & 4.68 \\
     300 & 0.25 & $10.42\pm0.80$ & $4.99\pm0.74$ & 5.59 \\
     300 & 0.3 & $10.51\pm0.76$ & $4.52\pm0.68$ & 5.96 \\
     300 & 0.35 & $9.89\pm0.75$ & $4.41\pm0.68$ & 6.27 \\
     \hline
    \end{tabular}
    \caption{MCMC results with different assumptions for the mean spot lifetime $\eta_{\tau}$ and $\eta_{\ell}$ using kernel $\mathbf{K_{J1}}$ (eq. \ref{eq:gp_j1}). For each row, we fix the values of $\eta_{\tau}$ and $\eta_{\ell}$. All other model parameters take on the initial values and priors from Table \ref{tab:pars_priors} for a two-planet model. We perform a MAP fit followed by MCMC sampling for each case. We report the nominal values and uncertainties for the semi-amplitudes of AU Mic b and c from the MCMC fitting, as well as the reduced chi-square statistic, $\chi^{2}_{red}$ using the MAP-derived parameters. Uncertainties reported here for $K_{b}$ and $K_{c}$ are the average of the upper and lower uncertainties.}
    \label{tab:kernel_param_search}
\end{table}

\subsection{Planet Injection and Recovery} \label{sec:injection_recovery}

\begin{figure*}
    \centering
    \includegraphics[width=0.98\textwidth]{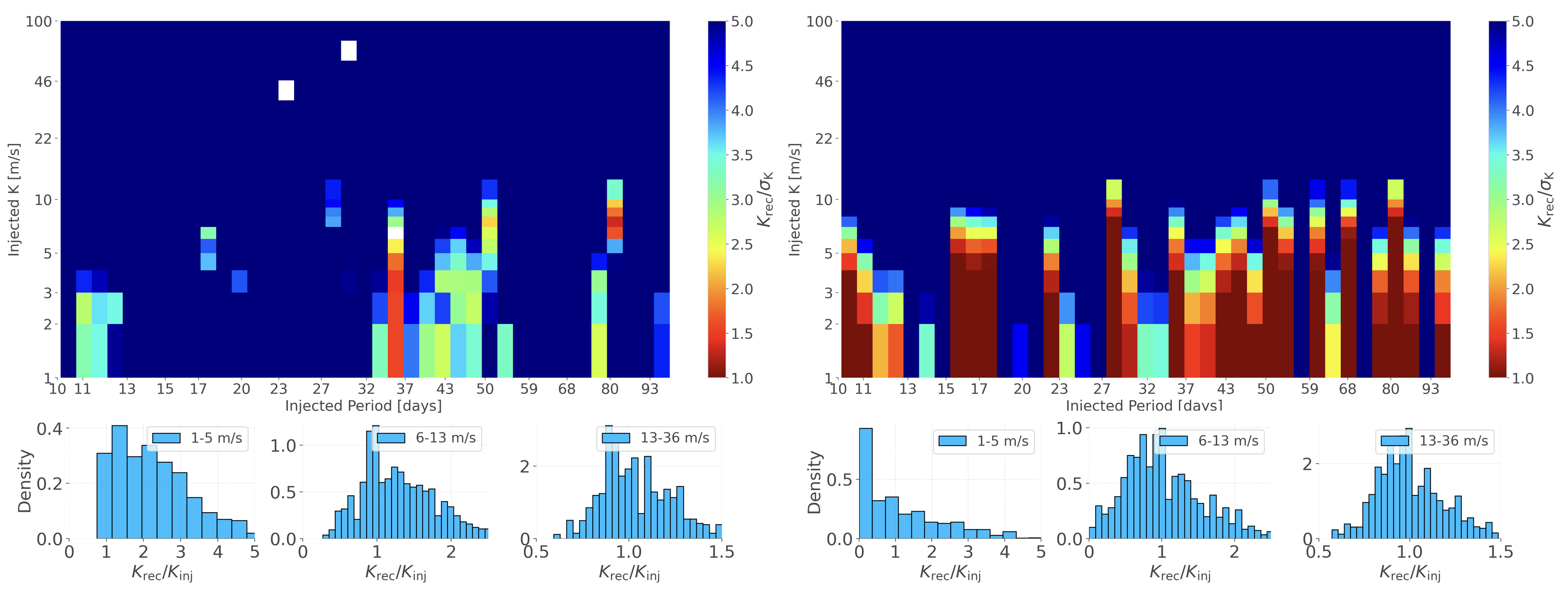}
    \caption{Histograms depicting our injection and recovery-test results. In the top row, we show the relative confidence interval of the recovered semi-amplitudes ($K_{\mathrm{rec}}$) derived from the MCMC analysis in the case of letting the ephemeris ($P$, $TC$) float (left) and fixing the ephemeris to the injected values (right). In the bottom row, we compare the recovered semi-amplitude to the injected value ($K_{\mathrm{inj}}$).}
    \label{fig:injrec_real}
\end{figure*}

\begin{figure*}
    \centering
    \includegraphics[width=0.98\textwidth]{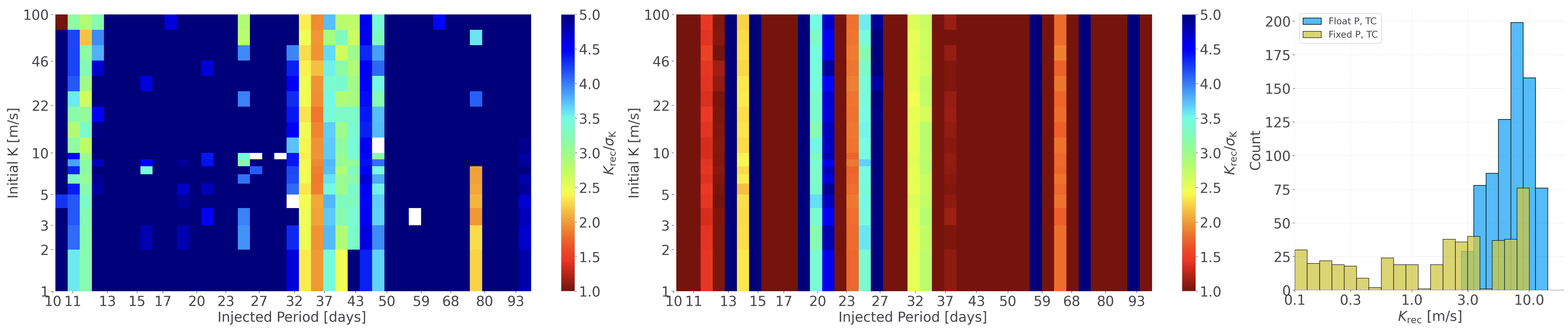}
    \caption{Histograms depicting the recovery of planetary signals without having injected any into the data. In panels 1 and 2, we show the relative confidence interval of the recovered semi-amplitudes ($K_{rec}$) derived from the MCMC analysis in the case of letting the ephemeris ($P$, $TC$) float (left) and fixing the ephemeris to arbitrary the arbitrary TC=2457147.36589 (middle). On the right, we show the recovered semi-amplitudes for each case.}
    \label{fig:injrec_fake}
\end{figure*}

Here we assess the fidelity of our RV model applied to the AU Mic system through planetary injection and recovery tests. We first inject planetary signals into the RV data with well-defined semi-amplitudes, periods, and ephemerides ($TC$). We arbitrarily choose $TC=2457147.36589$ for all injected cases. We consider 40 unique periods between 5.12345--100.12345 days, uniformly distributed in log-space. For the semi-amplitude $K$, we consider values from 1--10 m\,s$^{-1}$ with a step size of 1 m\,s$^{-1}$, as well as values between 10--100 m\,s$^{-1}$ which are uniformly distributed in log-space (20 total values). In all cases, we include a model for AU Mic b with fixed $P$ and $TC$ such that $K_{b}\sim\mathcal{N}(8.5, 2.5)$. We first assess our recovery capabilities using a Gaussian prior for $P$ such that $P\sim\mathcal{N}(P_{\mathrm{inj}}, P_{\mathrm{inj}}/50)$ and a uniform prior for $TC$ such that $TC \sim \mathcal{U}(TC_\mathrm{inj} \pm P_{\mathrm{inj}}/2)$. For each injected planet (one at a time), we run our MAP and MCMC analyses to determine the recovered $K$ and corresponding uncertainty. The starting value for $K$ and $TC$ are always the injected values. We also consider the same injection and recovery test but with fixing $P$ and $TC$ to the injected value. We finally determine how susceptible our RV model is to pick out ``fake''-planets by running these same two trials with no injected planets. Although there are no injected planets, we still run the same trials as the injected case with different initial values for $K$. A two-dimensional histogram of the recovered $K$ as a fraction of the injected $K$, as well as the associated uncertainty (also as a fraction of the injected $K$) for each case are shown in figs. \ref{fig:injrec_real} and \ref{fig:injrec_fake} for the injected and non-injected cases, respectively.

In the case of injected planets, we find our RV-model is able to confidently recover semi-amplitudes down to a few m\,s$^{-1}$ in this data set with a relative precision of $\gtrsim 4\sigma$. However, a closer inspection reveals the recovered semi-amplitudes are typically larger than the injected K, particularly for smaller injected values (1--5 m\,s$^{-1}$) that includes our measured semi-amplitude AU Mic c. When the ephemeris is known, we tend to poorly measure the smallest values of K, indicating the recovered $TC$ in the non-fixed case is unlikely what we have injected. In the 6--13 m\,s$^{-1}$ range, which covers the recovered semi-amplitude of AU Mic b, we find that the accuracy of the recovered semi-amplitudes are $\sim$50\%. So, while we quote a formal precision on the mass of AU Mic b to be $M_{b}=20.12_{-1.57}^{+1.72}\ M_{\oplus}$ ($\sim$9\% precision), our injection and recovery tests indicate that the accuracy on the mass of AU Mic b is only known to a factor of two.

Unfortunately, attempts to recover non-injected planets are ``unsuccessful'', in that our modeling finds strong evidence for planets we did not inject (fig. \ref{fig:injrec_fake}) in the case of allowing $P$ and $TC$ to float. A deeper investigation into the posteriors of such fits indicates certain parameters (primarily $P$ and $TC$) are typically not well-behaved and yield non-Gaussian distributions. When fixing $P$ and $TC$ to ``nominal'' values, our modeling does not tend to find such non-existent planets (fig. \ref{fig:injrec_fake}).

The confident recoveries of ``fake''-planets in our tests indicate our GP model is flexible enough to find relatively (quasi)-stable islands in probability space with high confidence for $K$ specifically. Although several peaks stand out in our periodogram analyses (figs. \ref{fig:gls_periodograms} and \ref{fig:brute_force_periodograms}), more observations and/or more sophisticated modeling are needed to robustly claim these periods as statistically validated planets. We further note that the recovered values of $K$ for the smallest injected values are inaccurate, indicating our measurement of $K_{c}=3.68$ m\,s$^{-1}$ is also moderately unconvincing, and is likely an overestimate given the behavior of all recoveries at this level of $K$. We finally note this analysis is limited by planets we do not account for in the model, which may impact our ability to recover certain combinations of $P$ and $TC$. Further tests using several values for the injected $TC$ may also yield different results. With these limitations in mind, we also provide an estimation of the upper-limit to the mass of AU Mic c. We find a $5\sigma$ upper-limit to the semi-amplitude of AU Mic c of $\leq7.68$ m\,s$^{-1}$, corresponding to a mass of $\leq20.13\ M_{\oplus}$.

\subsection{Utility of RV-Color} \label{sec:disc_utility}

Our chromatic kernel used in this work is an initial step to exploit the expected correlation of stellar activity versus wavelength by introducing a scaling relation between wavelengths (eqs. \ref{eq:gp_j1} and \ref{eq:gp_j2}). Here we examine the ``RV-color'' for our multi-wavelength dataset in order to further assess the correlation between our RVs with expected activity:

\begin{equation}
    RV_{\mathrm{color}}(t, \lambda, \lambda') = \mathrm{RV}(t, \lambda) - \mathrm{RV}(t, \lambda')
\end{equation} \label{eq:rv_color}

\begin{figure}
    \centering
    \includegraphics[width=0.48\textwidth]{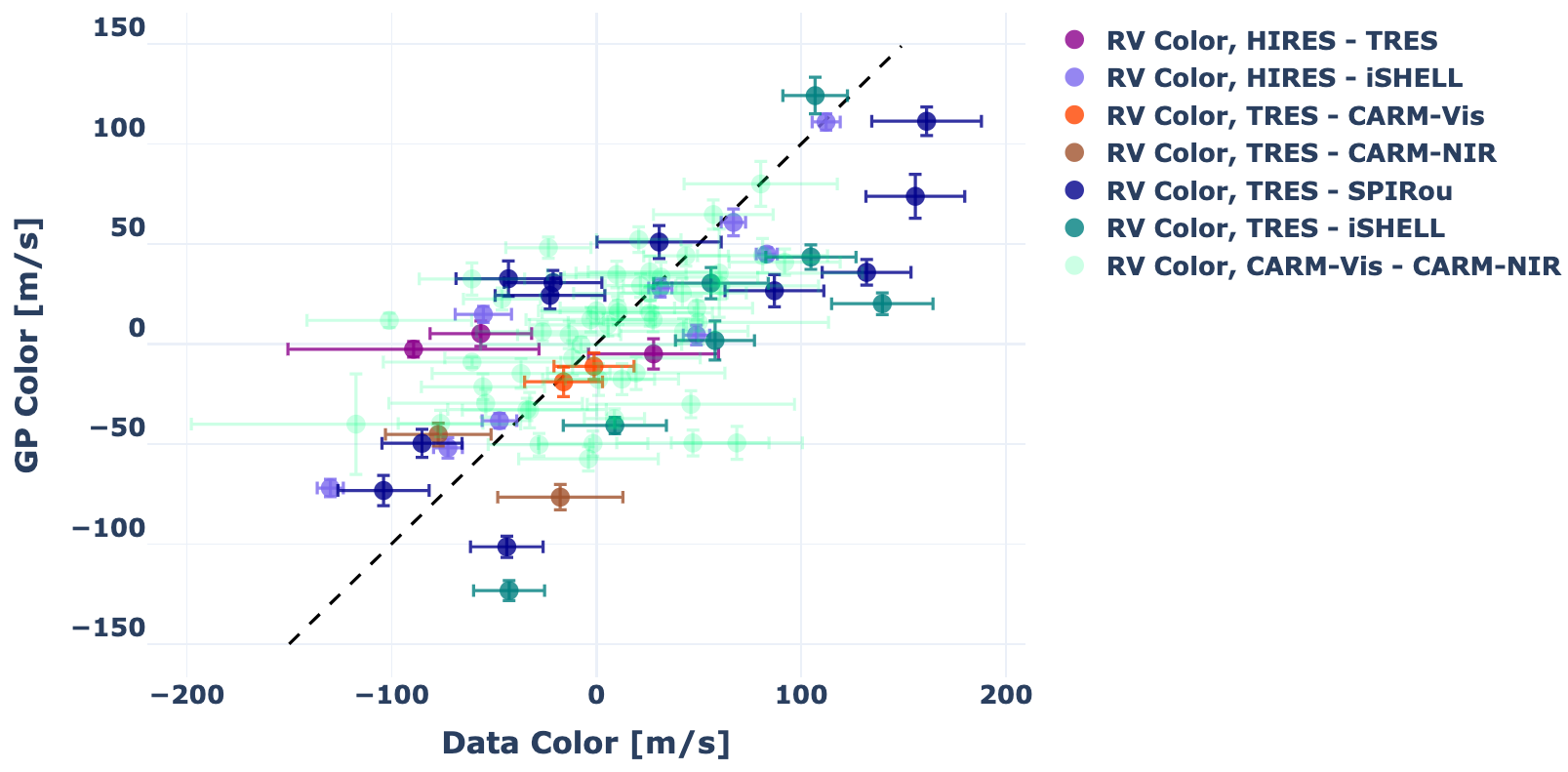}
    \caption{The observed ``RV-color'' = RV(t, $\lambda$) - RV(t, $\lambda'$) ($\lambda' > \lambda$) from our 2019 and 2020 nights with nearly simultaneous measurements at unique wavelengths. These are plotted against the same RV-color difference predicted by our chromatic GP model using kernel $\mathbf{K_{J2}}$. Pairs consisting of CARMENES-VIS and CARMENES-NIR measurements are nearly transparent to make other pairs more visible. We do not plot pairs of SPIRou and iSHELL because they are tightly centered near zero. A dashed one-to-one line is also shown. The weighted coefficient of determination ( R$^{2}$) is $\approx 0.68$.}
    \label{fig:rv_color_1_to_1}
\end{figure}

We first determine which nights contain nearly simultaneous measurements at unique wavelengths. We require observations to be within 0.3 days ($\approx 6\%$ of one rotation period) of each other to minimize differences from rotationally modulated activity but to increase the number of pairs for our brief use. For each nearly simultaneous chromatic pair, we compute the ``data-color'' directly from the measured RVs as well as the ``GP-color'' by computing the differences between the two measurements and two GPs sampled at the identical times, respectively (such that $\lambda' > \lambda$). This calculation requires knowledge of the parameters in order to remove the per-instrument zero points and realize each appropriate GP, so we make use of the MAP-derived parameters in Table \ref{tab:pars_results_2planets} with kernel $\mathbf{K_{J2}}$ (eq. \ref{eq:gp_j2}). The correlation between the data- and GP-color is shown in fig. \ref{fig:rv_color_1_to_1}. The agreement between the data and the model (weighted $R^{2}\approx0.71$) indicates that our chromatic GP technique is doing a good job of reproducing the RV-color phenomenon for multiple wavelength pairs.

With a sufficient model for stellar activity, we expect the data and GP RV-color to match (up to white noise). Therefore, the ``RV-color'' between the data and GP may be used to further constrain (in future analyses) the model (and therefore prevent over-fitting) by including an effective L2 regularization penalty as follows:

\begin{equation}
    \ln \mathcal{L}\ \textrm{+=}\ -\Lambda \sum_{t} r_{col}(t)^{2}
\end{equation}

\noindent Here, $\vec{r}_{col}$ is the vector of residuals between the GP and data RV-color. $\Lambda>0$ is a tunable hyperparameter whose value is directly correlated with the relative importance and confidence of the stellar activity model. $+=$ represents the standard ``addition assignment'' operator. The vector $\vec{r}_{col}$ may be computed for all pairs of wavelengths with (nearly)-simultaneous measurements, and each pair can make use of an identical or unique values of $\Lambda$. We finally note this regularization term is not limited to our assumption of a simple-scaling relation, and could also be used in the case of disjoint kernels.

\subsection{Additional Caveats and Future Work} \label{sec:caveats_future_work}

Kernels $\mathbf{K_{J1}}$ (\ref{eq:gp_j1}) and $\mathbf{K_{J2}}$ (eq. \ref{eq:gp_j2}) make use of a scaling relation for stellar activity models at different wavelengths (spectrographs) where each activity model is drawn from a Gaussian process characterized by a covariance matrix utilizing all observations. Using such joint kernels yield fits with larger scatter than cases using disjoint QP kernels (one per-spectrograph, eq. \ref{eq:gp_qp}). In the latter case, we find that although each activity model appears to be ``in-phase'' with one another, each GP exhibits unique features that are inconsistent with a simple scaling relation (fig. \ref{fig:rvs_zoom_disjoint}). With nightly sampling, it is difficult to determine whether the observed differences between disjoint GPs is indicative of inadequate sampling or an inadequate RV model (activity + planets). Further, all activity models used in this work make use of identical kernel hyperparameters (excluding the amplitude) which may further be an inadequate assumption. We expect the stellar rotation period ($\eta_{P}$) to be identical across wavelengths (or nearly so), however it is not clear whether the mean activity timescale ($\eta_{\tau}$) or period length scale ($\eta_{\ell}$) in particular should be achromatic hyperparameters.

Our work further excluded per-spectrograph uncorrelated ``jitter'' terms. We suspect this may be the source of our model's ability to find planets we did not inject into the model (Section \ref{sec:injection_recovery}), which we defer to future work. The reduced-$\chi^2$ values in Tables \ref{tab:redchi2s_per_spectrograph} and \ref{tab:kernel_param_search} quantify the degree to which our models do not capture signals from possible additional planets, incorrect values for eccentricity and/or $\omega$, per-spectrograph systematics not included in the formal measurement uncertainties, stellar activity such as p-mode oscillations, convection noise, or longer time-scale variations. Therefore, although our specific likelihood function (eq. \ref{eq:map}) assumes normally distributed errors, we choose not to combine any remaining (i.e., unaccounted for by the provided error bars) potentially correlated noise into an additional uncorrelated jitter term to keep our model simple.

More accurately characterizing the masses and orbits of AU Mic b and c may require a more sophisticated stellar activity model and more intensive multi-wavelength cadence. Our work further does not make use of activity indicators (e.g., Ca \textsc{II} H and K, H$\alpha$) or asymmetries in the cross-correlation function (e.g., the bisector inverse slope (BIS) or differential line width dLW; \cite{serval}) to help constrain the activity model \citep[see][]{rajpaul2015a}. The \texttt{serval} pipeline in particular provides a measure of the chromaticity (CRX) for both the CARMENES-VIS and NIR datasets which we do not use in our modeling. For AU Mic, we expect that each spectrograph is precise enough to resolve first-order chromatic effects within their respective spectral grasp's which will unfortunately make the formal uncertainties of each spectrograph larger. Further, our QP-based kernels are primarily intended to capture rotationally modulated activity induced from temperature inhomogeneities on the stellar surface. Although the flexibility of disjoint GPs likely captures other rotationally modulated effects such as convective blueshift and limb-darkening, it will not capture short-term activity such as flares. We finally note that more seasons with high-cadence RVs will help mitigate the strong 1 year alias present in our dataset, and will help determine the correct periods for potential non-transiting planets.

\section{Conclusion} \label{sec:conclusion}

In this work, we have developed two joint-Gaussian process kernels which begin to take into account the expected wavelength dependence of stellar activity through a simple-scaling relation. We apply our kernels to a dataset of AU Mic, which is composed of RVs from multiple facilities, and wavelengths ranging from visible to K-band. With our analyses, we report a refined mass of AU Mic b of $M_{b}=20.12_{-1.72}^{+1.57}\ M_{\oplus}$, and provide a $4.2\sigma$ mass estimate of the recently validated transiting planet AU Mic c to be $M_{c}=9.60_{-2.31}^{+2.07}\ M_{\oplus}$, corresponding to a $5\sigma$ upper limit of $M_{c}\leq20.13\ M_{\oplus}$. We also identify additional peaks present in the activity-filtered RVs, but such periods require more evidence for a robust validation given the overall flexibility of our RV model with an unknown ephemeris.

In Section \ref{sec:disc_eccentricity}, we find our model is unable to robustly constrain the eccentricity for AU Mic b or c. In section \ref{sec:disc_kernel_parameters}, we find the derived planetary semi-amplitudes for AU Mic b and c are moderately sensitive to the choice of kernel-parameters, indicating careful attention must be made when interpreting planetary masses with such a flexible model. Through injection and recovery tests in section \ref{sec:injection_recovery}, we further validate our RV-model by demonstrating our ability to recover planets down to $\approx$ 10 m\,s$^{-1}$ when the orbit's ephemeris is known. However, we find that the accuracy in the recovered semi-amplitudes is $\sim$50\% at 10 m\,s$^{-1}$. In section \ref{sec:disc_utility}, we introduce a method to further leverage the ``RV-color'' correlation between the observations and activity model through penalizing the objective function by including an effective L2 regularization term.

\section{Acknowledgments} \label{sec:acknowledgements}

All data processed with \texttt{pychell} (iSHELL and CHIRON) were run on ARGO, a research computing cluster provided by the Office of Research Computing, and the exo computer cluster, both at George Mason University, VA.

We thank all support astronomers, observers, and engineers from all facilities in helping enable the collection of the data presented in this paper.

The authors wish to recognize and acknowledge the very significant cultural role and reverence that the summit of Maunakea has always had within the indigenous Hawaiian community, where the iSHELL, HIRES, IRD, and SPIRou observations were recorded. We are most fortunate to have the opportunity to conduct observations from this mountain.

This work is supported by grants to Peter Plavchan from NASA (awards 80NSSC20K0251 and 80NSSC21K0349), the National Science Foundation (Astronomy and Astrophysics grants 1716202 and 2006517), and the Mount Cuba Astronomical Foundation.

Emily A. Gilbert also wishes to thank the LSSTC Data Science Fellowship Program, which is funded by LSSTC, NSF Cybertraining Grant \#1829740, the Brinson Foundation, and the Moore Foundation; her participation in the program has benefited this work. Emily is thankful for support from GSFC Sellers Exoplanet Environments Collaboration (SEEC), which is funded by the NASA Planetary Science Division’s Internal Scientist Funding Model. The material is based upon work supported by NASA under award number 80GSFC21M0002.

This work is partly supported by JSPS KAKENHI Grant Number JP18H05439, JST PRESTO Grant Number JPMJPR1775, the Astrobiology Center of National Institutes of Natural Sciences (NINS) (Grant Number AB031010).

M.T. is supported by JSPS KAKENHI grant Nos. 18H05442, 15H02063, and 22000005.

The authors also with to acknowledge funding from the Agencia Estatal de Investigaci\'on del Ministerio  de Ciencia e Innovaci\'on (AEI-MCINN) under grant PID2019-109522GB-C53.

The authors also wish to thank the California Planet Search (CPS) collaboration for carrying out the HIRES observations recorded in 2020 presented in this work.

{\textsc{Minerva}}-Australis is supported by Australian Research Council LIEF Grant LE160100001, Discovery Grant DP180100972, Mount Cuba Astronomical Foundation, and institutional partners University of Southern Queensland, UNSW Australia, MIT, Nanjing University, George Mason University, University of Louisville, University of California Riverside, University of Florida, and The University of Texas at Austin.

CARMENES is an instrument at the Centro Astron\'omico Hispano-Alem\'an de Calar Alto (CAHA, Almer\'{\i}a, Spain).
CARMENES is funded by the German Max-Planck-Gesellschaft (MPG), the Spanish Consejo Superior de Investigaciones Cient\'{\i}ficas (CSIC), the European Union through FEDER/ERF FICTS-2011-02 funds, and the members of the CARMENES Consortium (Max-Planck-Institut f\"ur Astronomie, Instituto de Astrof\'{\i}sica de Andaluc\'{\i}a, Landessternwarte K\"{o}onigstuhl, Institut de Ci\`encies de l’Espai, Institut f\"ur Astrophysik G\"{o}ttingen, Universidad Complutense de Madrid, Th\"uringer Landessternwarte Tautenburg, Instituto de Astrof\'{\i}sica de Canarias, Hamburger Sternwarte, Centro de Astrobiolog\'{\i}a and Centro Astron\'omico Hispano-Alem\'an), with additional contributions by the Spanish Ministry of Economy, the German Science Foundation through the Major Research Instrumentation Programme and DFG Research Unit FOR2544 ``Blue Planets around Red Stars'', the Klaus Tschira Stiftung, the states of Baden-W\"urttemberg and Niedersachsen, and by the Junta de Andaluc\'{\i}a.

We acknowledge financial support from the Agencia Estatal de Investigaci\'on of the Ministerio de Ciencia, Innovaci\'on y Universidades and the ERDF through projects PID2019-109522GB-C5[1:4]/AEI/10.13039/501100011033, PGC2018-098153-B-C33, and the Centre of Excellence ``Severo Ochoa'' and ``Mar\'ia de Maeztu'' awards to the Instituto de Astrof\'isica de Canarias (CEX2019-000920-S), Instituto de Astrof\'isica de Andaluc\'ia (SEV-2017-0709), and Centro de Astrobiolog\'ia (MDM-2017-0737), and the Generalitat de Catalunya/CERCA programme.

This paper includes data collected by the NASA \textit{TESS} mission that are publicly available from the Mikulski Archive for Space Telescopes (MAST). Funding for the \textit{TESS} mission is provided by NASA's Science Mission Directorate. We acknowledge the use of public \textit{TESS} data from pipelines at the \textit{TESS} Science Office and at the TESS Science Processing Operations Center \citep{jenkinsSPOC2016}.

Baptiste Klein acknowledges funding from the European Research Council under the European Union’s Horizon 2020 research and innovation programme (grant agreement No 865624, GPRV).

Eder Martioli acknowledges funding from the French National Research Agency (ANR)
under contract number ANR-18-CE31-0019 (SPlaSH).

\textit{Software:} pychell \citep{caleetal2019}, optimize\footnote{\url{https://optimize.readthedocs.io/en/latest/}}, Matplotlib \citep{matplotlib}, SciPy \citep{scipy}, NumPy \citep{numpy}, Numba \citep{numba}, corner \citep{2020zndo...3937526F}, plotly \citep{plotly}, Gadfly Matplotlib theme \url{https://gist.github.com/JonnyCBB/c464d302fefce4722fe6cf5f461114ea}, emcee, \citep{emcee}

\bibliography{bib_master}{}
\bibliographystyle{aasjournal}

\appendix

\section{Posterior Distributions}

Here we show the posterior distributions for the relevant RV models employed in this work. In each corner plot, blue lines correspond to the 50$^{\textrm{th}}$ percentile of the distribution. Upper and lower uncertainties correspond to the $84.1^{st}$ and $15.9^{th}$ percentiles, respectively.

\begin{figure}
    \centering
    \includegraphics[width=0.98\textwidth]{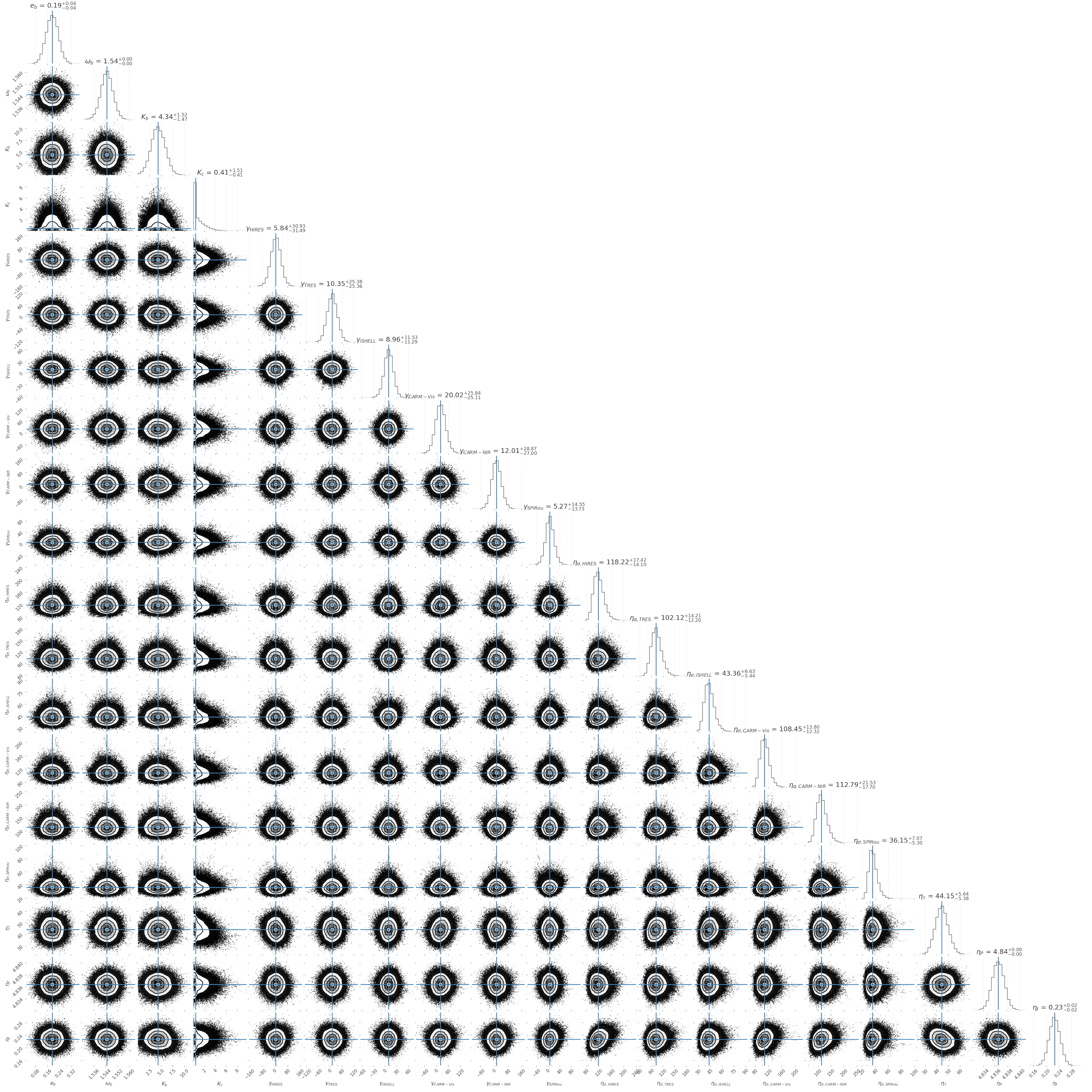}
    \caption{Posterior distributions using disjoint QP kernels (eq. \ref{eq:gp_qp}) for each spectrograph to model the stellar activity, including a two-planet model for the transiting planets b and c. The derived values for $\eta_{\tau}$ and $\eta_{\ell}$ suggests a more dynamic activity model than the $FF'$ curve prediction suggests.}
    \label{fig:corner_2planets_disjoint_float}
\end{figure}

\clearpage

\begin{figure}
    \centering
    \includegraphics[width=0.95\textwidth]{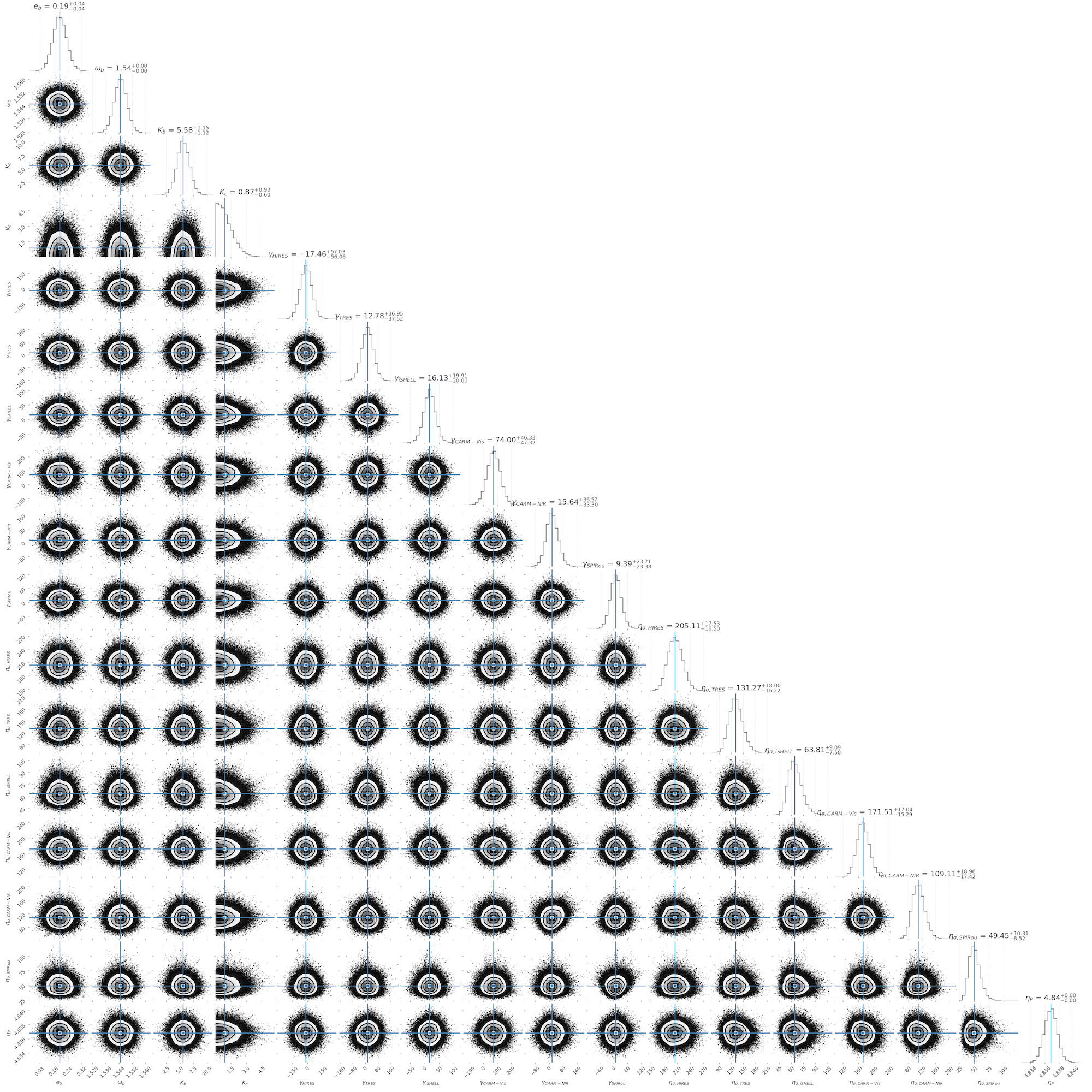}
    \caption{Same as fig. \ref{fig:corner_2planets_disjoint_float} but fixing $\eta_{\tau}=100$ days and $\eta_{\ell}=0.28$.}
    \label{fig:corner_2planets_disjoint_locked}
\end{figure}

\clearpage

\begin{figure}
    \centering
    \includegraphics[width=0.98\textwidth]{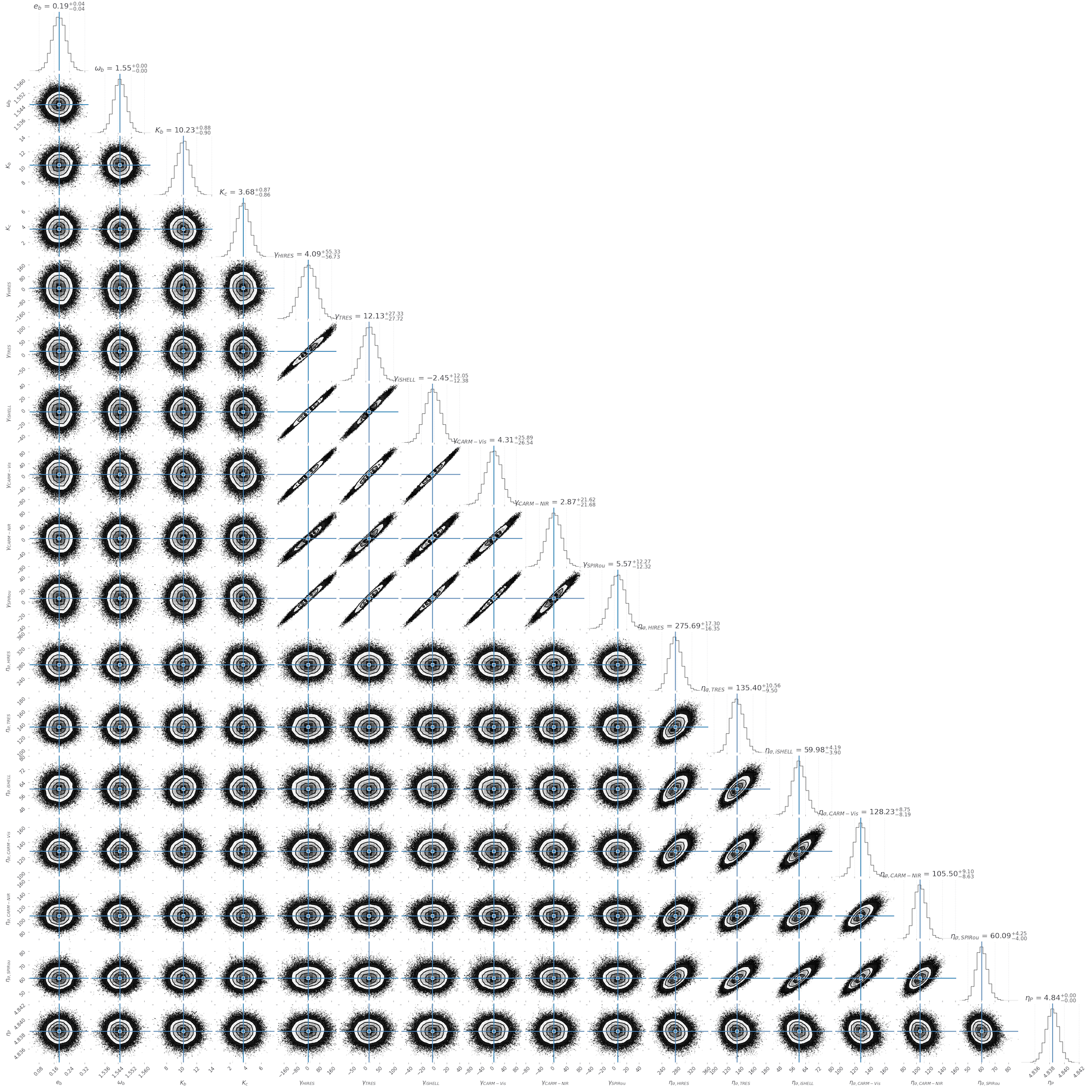}
    \caption{Posterior distributions for a two-planet fit to the RVs using $\mathbf{K_{J1}}$ (eq. \ref{eq:gp_j1}) to model the stellar activity.}
    \label{fig:corner_2planets_j1}
\end{figure}

\clearpage

\begin{figure}
    \centering
    \includegraphics[width=0.98\textwidth]{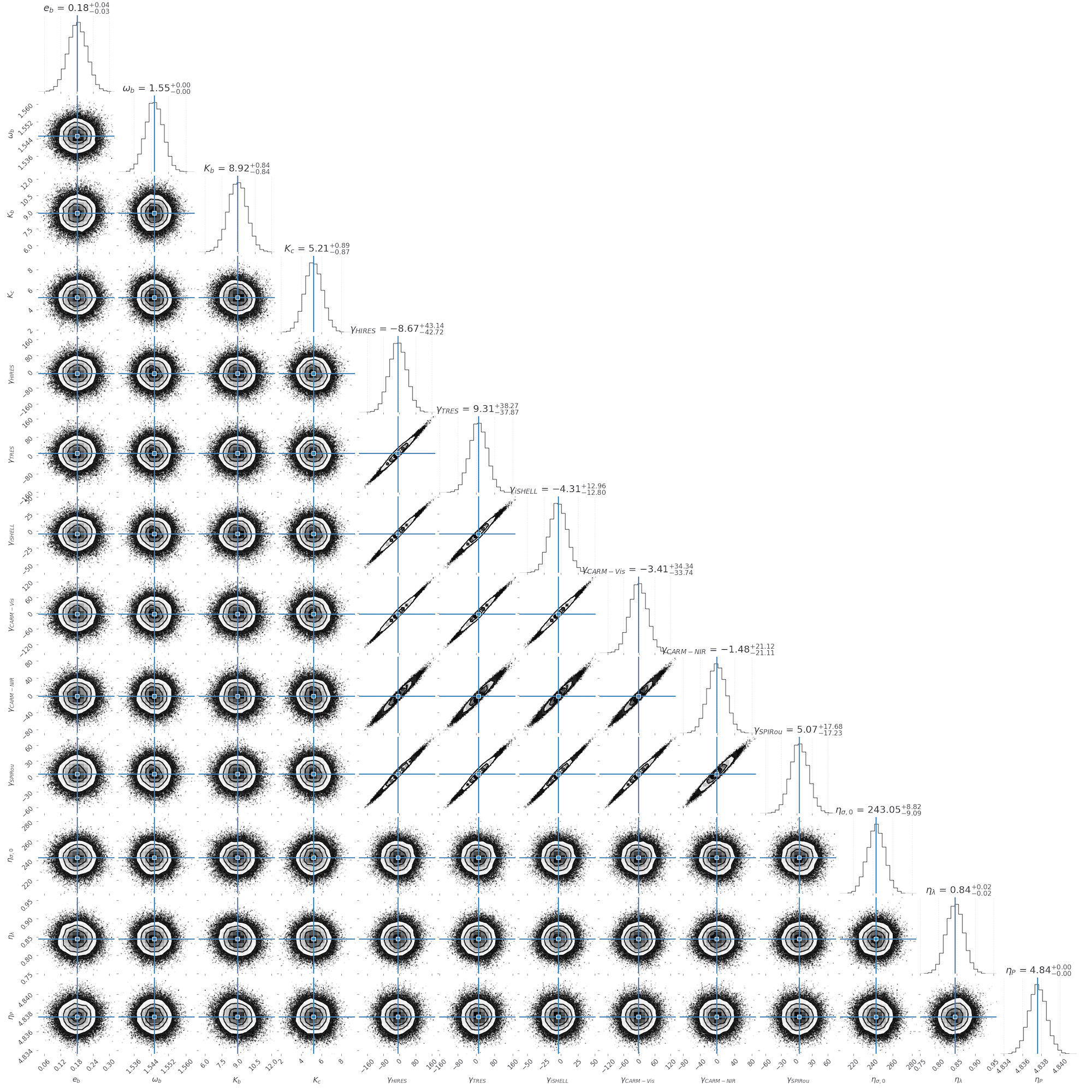}
    \caption{Same as fig. \ref{fig:corner_2planets_j1} but using $\mathbf{K_{J2}}$ to model the stellar activity.}
    \label{fig:corner_2planets_j2}
\end{figure}

\clearpage

\begin{figure}
    \centering
    \includegraphics[width=0.98\textwidth]{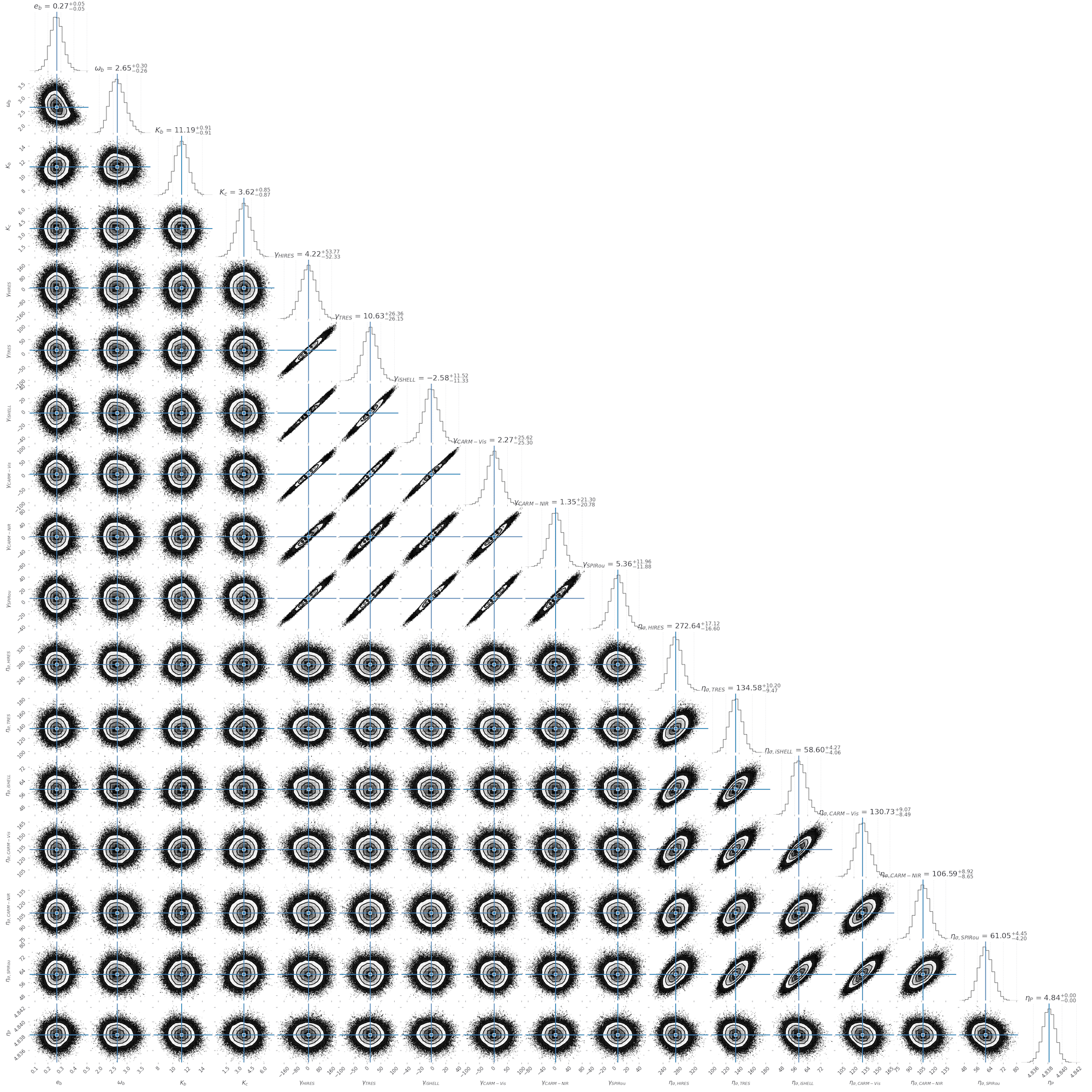}
    \caption{Same as fig. \ref{fig:corner_2planets_j1} but using less restrictive priors for $e_{b}$ and $\omega_{b}$ as well as $e_{c}$ and $\omega_{c}$.}
    \label{fig:corner_2planets_j1_ecc}
\end{figure}

\clearpage

\section{\texttt{optimize}} \label{app:optimize}

\texttt{optimize}\footnote{Documentation: https://optimize.readthedocs.io/en/latest/} is a generic, high-level optimization package in \textit{Python}, which generalizes the Bayesian-inspired classes used in \texttt{RadVel}. The primary container (\textit{Python} class) in \texttt{optimize} is referred to as an ``OptProblem''; primary attributes for this object are then helper-types to 1. construct the model, 2. compare the data and model with an objective function, and 3. perform the optimization and sample posterior distributions via Markov-Chain Monte-Carlo (MCMC) methods. Many attributes (such as initial parameters) are shared in multiple layers of this hierarchy for easier access and extension with appropriate methods to propagate changes to each. \texttt{Optimize} does not re-implement specific optimization algorithms, but rather is intended to be high-level wrapper around such routines (e.g., currently \texttt{scipy.optimize} and \texttt{emcee}).

\section{RV Measurements} \label{app:rvs_all}

\begin{longtable}[c]{|c|c|c|}
\caption{Nightly HARPS RVs analyzed in this work.}
    \\ \hline
    BJD & RV [m\,s$^{-1}$] & $\sigma_{\mathrm{RV}}$ [m\,s$^{-1}$] \\ \hline
    2452986.514817 & 73.91 & 4.81 \\
    2453157.898424 & 19.26 & 1.72 \\
    2453201.823450 & -32.8 & 2.56 \\
    2453468.892370 & -135.2 & 2.52 \\
    2453469.843534 & -170.56 & 2.92 \\
    2453499.868255 & -43.41 & 2.52 \\
    2453521.894368 & -345.97 & 6.36 \\
    2453551.803998 & -34.22 & 2.5 \\
    2453593.622139 & 137.84 & 3.33 \\
    2456568.510365 & -244.53 & 2.16 \\
    2456569.500104 & -32.99 & 1.8 \\
    2456570.565952 & 193.28 & 2.17 \\
    2456772.919271 & -54.34 & 1.64 \\
    2456773.918979 & -9.22 & 0.99 \\
    2456794.882288 & -119.81 & 2.53 \\
    2456795.885873 & 66.36 & 2.91 \\
    2456797.857541 & 20.03 & 3.0 \\
    2456844.806163 & 162.85 & 2.18 \\
    2456982.539577 & -65.03 & 1.8 \\
    2457223.648073 & 217.24 & 2.71 \\
    2457333.535731 & 94.85 & 1.29 \\
    2457493.891905 & 90.65 & 2.11 \\
    2457590.712986 & 73.08 & 1.4 \\
    2457904.813342 & -287.06 & 3.36 \\
    2457917.892902 & 161.13 & 1.97 \\
    2458035.528955 & -22.13 & 1.55 \\
    2458037.494745 & -41.59 & 2.39 \\
    2458206.872213 & -43.23 & 1.38 \\
    2458207.892023 & 33.78 & 1.56 \\
    2458208.884974 & 82.61 & 1.29 \\
    2458591.919473 & 120.82 & 1.42 \\
    2458594.904828 & 19.89 & 2.43 \\
    2458602.927758 & 9.22 & 1.47 \\
    2458605.916360 & -248.46 & 2.33 \\ \hline
\end{longtable}

\begin{longtable}[c]{|c|c|c|}
\caption{Nightly HIRES RVs analyzed in this work.}
    \\ \hline
    BJD & RV [m\,s$^{-1}$] & $\sigma_{\mathrm{RV}}$ [m\,s$^{-1}$] \\ \hline
    2453182.049556 & 58.09 & 4.01 \\
    2453195.938633 & -171.02 & 2.74 \\
    2453926.030459 & -270.15 & 3.02 \\
    2453926.979641 & 271.86 & 3.18 \\
    2453927.920257 & 68.35 & 2.13 \\
    2453931.079663 & -191.25 & 3.16 \\
    2453932.007322 & 256.77 & 2.76 \\
    2453932.978973 & 7.97 & 2.18 \\
    2453933.944291 & 162.29 & 2.36 \\
    2453934.930418 & -159.79 & 2.73 \\
    2453960.942379 & 257.92 & 2.98 \\
    2453962.030949 & 5.08 & 2.28 \\
    2454636.072465 & -9.82 & 2.53 \\
    2454688.871239 & -133.07 & 2.46 \\
    2454808.692607 & -83.96 & 3.06 \\
    2455015.018198 & -140.72 & 2.86 \\
    2455371.044475 & 100.3 & 1.9 \\
    2455727.981919 & 35.73 & 2.89 \\
    2456638.683238 & 325.16 & 4.18 \\
    2458645.065398 & -77.3 & 7.43 \\
    2459019.062251 & -121.05 & 2.81 \\
    2459025.094631 & -155.66 & 2.48 \\
    2459026.123675 & 399.32 & 3.35 \\
    2459028.098307 & 104.1 & 2.86 \\
    2459029.081905 & 14.31 & 2.53 \\
    2459030.078281 & -131.6 & 2.19 \\
    2459031.106844 & 414.91 & 3.41 \\
    2459032.105577 & -138.95 & 2.71 \\
    2459035.092001 & -76.31 & 2.26 \\
    2459036.101597 & 329.39 & 3.4 \\
    2459040.116030 & 3.72 & 2.12 \\
    2459041.117761 & 153.21 & 3.14 \\
    2459044.942379 & -0.97 & 2.22 \\
    2459051.921900 & 92.94 & 2.44 \\
    2459067.855784 & -21.57 & 3.09 \\
    2459068.895396 & -30.19 & 2.4 \\
    2459071.907618 & 43.81 & 3.37 \\
    2459072.852897 & -8.9 & 2.74 \\
    2459077.821906 & -44.12 & 2.71 \\
    2459078.813958 & -31.15 & 2.45 \\
    2459086.849156 & -147.83 & 3.51 \\
    2459087.868995 & 0.97 & 2.58 \\
    2459088.849805 & 139.3 & 2.57 \\
    2459089.824376 & -38.53 & 3.28 \\
    2459094.815792 & -127.27 & 2.67 \\
    2459097.855698 & -16.26 & 2.6 \\
    2459099.761856 & -139.97 & 3.05 \\
    2459101.904682 & -82.58 & 2.51 \\
    2459114.829534 & -6.09 & 2.08 \\
    2459115.914980 & -92.15 & 2.61 \\
    2459117.857924 & 48.2 & 2.29 \\
    2459121.710044 & 11.92 & 2.38 \\
    2459122.709275 & 58.74 & 2.34 \\
    2459123.703587 & 20.0 & 2.6 \\
    2459151.692854 & 40.18 & 2.41 \\
    2459153.753326 & 95.26 & 2.73 \\
    2459181.680326 & 42.79 & 2.6 \\
    2459187.682871 & 54.26 & 2.47 \\
    2459188.689611 & -6.89 & 2.6 \\
    2459189.684349 & -81.15 & 2.86 \\ \hline
\end{longtable}

\begin{longtable}[c]{|c|c|c|}
\caption{Nightly NIRSPEC RVs analyzed in this work.}
    \\ \hline
    BJD & RV [m\,s$^{-1}$] & $\sigma_{\mathrm{RV}}$ [m\,s$^{-1}$] \\ \hline
    2453522.56 & 48.5 & 50.0 \\
    2453523.55 & -19.5 & 50.0 \\
    2453596.37 & 111.5 & 50.0 \\
    2453597.38 & 19.5 & 50.0 \\
    2453669.19 & 113.5 & 50.0 \\
    2453670.2 & 262.5 & 50.0 \\
    2453928.5 & -74.5 & 50.0 \\
    2453929.45 & -148.5 & 50.0 \\
    2453930.46 & -93.5 & 50.0 \\
    2453931.4 & -24.5 & 50.0 \\
    2454308.43 & 123.5 & 50.0 \\
    2454309.41 & -166.5 & 50.0 \\
    2454311.4 & 20.5 & 50.0 \\
    2454312.36 & -157.5 & 50.0 \\ \hline
\end{longtable}

\begin{longtable}[c]{|c|c|c|}
\caption{Nightly CSHELL RVs analyzed in this work.}
    \\ \hline
    BJD & RV [m\,s$^{-1}$] & $\sigma_{\mathrm{RV}}$ [m\,s$^{-1}$] \\ \hline
    2455455.85303 & 16.79 & 38.76 \\
    2455479.800206 & 41.04 & 26.96 \\
    2455480.768983 & 143.4 & 56.64 \\
    2455482.756068 & 136.64 & 22.11 \\
    2455523.7002290003 & 0.0 & 20.78 \\
    2455752.0995830004 & 43.59 & 27.48 \\
    2455755.0752310003 & 141.67 & 44.57 \\
    2455758.967657 & -17.29 & 31.52 \\
    2455791.812748 & 123.61 & 54.8 \\
    2455793.8316450003 & 234.74 & 10.87 \\
    2456844.9609169997 & -35.3 & 15.47 \\
    2456917.743247 & -36.65 & 21.07 \\
    2457275.8463990004 & -115.88 & 26.14 \\
    2457551.12512 & -102.38 & 51.72 \\
    2457555.058318 & -150.06 & 20.78 \\
    2457564.0072459998 & -198.34 & 30.4 \\
    2457570.063826 & 82.72 & 36.9 \\
    2457587.023049 & -245.01 & 18.85 \\
    2457598.976608 & -90.04 & 25.01 \\
    2457618.963993 & 8.92 & 19.22 \\
    2457619.947171 & -72.55 & 21.1 \\ \hline
\end{longtable}

\begin{longtable}[c]{|c|c|c|}
\caption{Nightly TRES RVs analyzed in this work.}
    \\ \hline
    BJD & RV [m\,s$^{-1}$] & $\sigma_{\mathrm{RV}}$ [m\,s$^{-1}$] \\ \hline
    2456573.68979 & -128.0 & 11.6 \\
    2456574.640606 & -45.2 & 11.8 \\
    2456575.669816 & 183.9 & 14.4 \\
    2456576.660338 & -186.6 & 10.3 \\
    2456577.632961 & 203.9 & 12.3 \\
    2456578.625675 & -68.1 & 11.9 \\
    2456579.634451 & -64.0 & 13.8 \\
    2456580.634669 & 173.3 & 13.7 \\
    2456581.610641 & -117.8 & 11.3 \\
    2456582.623658 & 101.1 & 21.4 \\
    2456583.619796 & -35.2 & 9.7 \\
    2456584.624069 & -18.7 & 11.0 \\
    2456585.596255 & 59.0 & 11.7 \\
    2456586.617077 & -22.2 & 12.7 \\
    2456587.611581 & 0.0 & 11.7 \\
    2456588.591012 & -33.7 & 11.6 \\
    2456589.602899 & 45.0 & 11.8 \\
    2456590.622533 & -64.4 & 12.1 \\
    2456605.581655 & -250.0 & 13.7 \\
    2456606.565507 & 225.6 & 8.4 \\
    2456607.563214 & -172.0 & 12.1 \\
    2456608.58807 & -11.5 & 10.6 \\
    2456609.587085 & 211.2 & 13.7 \\
    2456610.565465 & -233.4 & 15.0 \\
    2456611.571286 & 213.0 & 9.2 \\
    2456615.560722 & -193.0 & 13.6 \\
    2456616.572477 & 176.9 & 13.8 \\
    2456622.557557 & -102.5 & 19.3 \\
    2456624.552961 & 85.6 & 46.8 \\
    2456625.563003 & 38.7 & 24.7 \\
    2458646.96613 & 49.9 & 24.2 \\
    2458647.964611 & 34.3 & 22.0 \\
    2458648.961007 & -44.6 & 17.8 \\
    2458649.958409 & -41.2 & 18.9 \\
    2458650.97301 & 164.3 & 16.2 \\
    2458651.966405 & 116.1 & 17.5 \\
    2458652.981466 & -33.8 & 24.0 \\
    2458657.953559 & 13.6 & 15.7 \\
    2458658.932441 & -107.5 & 17.0 \\
    2458659.912405 & 84.9 & 19.9 \\
    2458665.924079 & 130.6 & 17.8 \\
    2458674.932143 & 198.8 & 14.2 \\
    2458677.882169 & -99.1 & 20.1 \\
    2458685.884899 & 96.1 & 35.5 \\
    2458689.846153 & 110.1 & 14.7 \\
    2458693.842764 & -74.1 & 24.1 \\
    2458730.740018 & 17.1 & 23.4 \\
    2458731.739653 & -35.0 & 23.9 \\
    2458738.70382 & 145.2 & 32.6 \\
    2458742.696035 & -86.4 & 40.1 \\
    2458744.700854 & 234.4 & 25.3 \\
    2458745.730511 & -65.9 & 40.3 \\
    2458758.650170 & 42.1 & 25.3 \\
    2458759.694882 & 114.5 & 28.3 \\
    2458761.645681 & -117.3 & 17.2 \\
    2458762.677499 & 144.6 & 21.8 \\
    2458767.656488 & 107.5 & 33.1 \\
    2458768.653311 & 83.7 & 30.6 \\
    2458769.635106 & 15.1 & 22.5 \\
    2458770.623329 & -112.3 & 18.2 \\
    2458771.612582 & -144.9 & 21.1 \\
    2458772.628816 & 120.2 & 22.4 \\
    2458773.638865 & 91.1 & 25.1 \\
    2458774.591042 & -51.3 & 25.3 \\
    2458775.598544 & -135.9 & 24.1 \\
    2458779.573710 & -86.3 & 24.1 \\
    2458780.566456 & -148.0 & 20.2 \\
    2458782.572973 & 29.7 & 19.0 \\
    2458783.561459 & 123.9 & 30.0 \\
    2458784.56098 & -121.1 & 21.7 \\
    2458786.591881 & -45.4 & 22.6 \\
    2458787.627174 & -2.0 & 29.9 \\
    2458788.570062 & 211.2 & 22.7 \\ \hline
\end{longtable}

\begin{longtable}[c]{|c|c|c|}
\caption{Nightly iSHELL RVs analyzed in this work.}       \\ \hline
    BJD & RV [m\,s$^{-1}$] & $\sigma_{\mathrm{RV}}$ [m\,s$^{-1}$] \\ \hline
    2457684.759584 & 76.97 & 4.94 \\
    2457698.745971 & 47.96 & 3.05 \\
    2457699.710324 & 47.8 & 14.07 \\
    2457850.129559 & 87.59 & 7.24 \\
    2457856.130267 & -33.33 & 3.41 \\
    2457923.120317 & -1.31 & 4.77 \\
    2457931.026094 & -15.53 & 1.59 \\
    2457940.000525 & -7.44 & 0.26 \\
    2457982.918015 & -2.11 & 11.34 \\
    2457983.911491 & -25.53 & 31.67 \\
    2457984.906727 & 76.12 & 7.17 \\
    2458046.688896 & -53.89 & 6.97 \\
    2458047.677872 & 20.45 & 9.09 \\
    2458048.684528 & 1.5 & 7.03 \\
    2458049.677166 & -68.12 & 11.49 \\
    2458660.089282 & 8.42 & 3.31 \\
    2458666.92506 & 42.5 & 2.21 \\
    2458675.084455 & 71.1 & 4.86 \\
    2458739.930751 & 85.7 & 4.12 \\
    2458760.71097 & -48.38 & 4.51 \\
    2458761.729422 & -96.54 & 3.37 \\
    2458762.730214 & 16.08 & 6.35 \\
    2458763.779074 & -14.43 & 3.17 \\
    2458764.766543 & 18.96 & 5.06 \\
    2458765.71105 & -39.79 & 7.94 \\
    2458795.700288 & -59.42 & 3.14 \\
    2458796.706964 & -0.86 & 5.37 \\
    2458798.695629 & 32.0 & 4.13 \\
    2458799.69492 & -22.28 & 9.62 \\
    2459069.985765 & 67.29 & 11.02 \\
    2459071.979622 & 24.53 & 3.5 \\
    2459086.915335 & -5.96 & 4.43 \\
    2459087.917716 & 68.09 & 13.17 \\
    2459088.900423 & 39.07 & 5.6 \\
    2459089.900047 & 20.73 & 7.19 \\
    2459090.897927 & 50.16 & 4.41 \\
    2459115.810803 & -7.73 & 5.78 \\
    2459117.805977 & -23.08 & 3.53 \\
    2459118.807183 & -12.72 & 4.15 \\
    2459119.813063 & 3.98 & 5.34 \\
    2459120.804824 & -28.18 & 4.32 \\
    2459122.88224 & 3.77 & 4.62 \\
    2459123.807924 & -16.94 & 5.09 \\
    2459143.789273 & -35.3 & 5.73 \\
    2459145.789894 & 0.86 & 9.38 \\
    2459147.782831 & -7.73 & 4.04 \\ \hline
\end{longtable}

\begin{longtable}[c]{|c|c|c|}
\caption{Nightly IRD RVs analyzed in this work.}
    \\ \hline
    BJD & RV [m\,s$^{-1}$] & $\sigma_{\mathrm{RV}}$ [m\,s$^{-1}$] \\ \hline
    2458650.116682 & -36.37 & 2.06 \\
    2458653.123778 & 5.82 & 4.63 \\
    2458654.116977 & -62.9 & 1.88 \\
    2458655.126277 & 42.79 & 1.75 \\
    2458679.945743 & 1.82 & 5.4 \\
    2458771.845784 & -1.82 & 3.99 \\ \hline
\end{longtable}

\begin{longtable}[c]{|c|c|c|}
\caption{Nightly CARMENES-NIR RVs analyzed in this work.}
    \\ \hline
    BJD & RV [m\,s$^{-1}$] & $\sigma_{\mathrm{RV}}$ [m\,s$^{-1}$] \\ \hline
    2458678.567915 & 92.38 & 46.13 \\
    2458679.537705 & 315.17 & 54.39 \\
    2458680.53478 & 247.69 & 65.59 \\
    2458684.568375 & 437.36 & 75.05 \\
    2458686.548525 & 275.14 & 49.79 \\
    2458687.578395 & 201.29 & 45.37 \\
    2458688.584370 & 86.47 & 46.79 \\
    2458690.55771 & 180.81 & 23.17 \\
    2458691.505425 & 276.31 & 30.81 \\
    2458693.54918 & -19.49 & 18.84 \\
    2458694.59554 & 164.36 & 19.65 \\
    2458695.53909 & -3.88 & 25.06 \\
    2458696.523080 & 41.45 & 15.91 \\
    2458698.51838 & -112.18 & 13.87 \\
    2458699.48421 & 43.11 & 18.82 \\
    2458700.47701 & -12.46 & 23.43 \\
    2458701.471955 & -38.69 & 30.48 \\
    2458702.50203 & -267.32 & 55.65 \\
    2458704.48929 & -7.34 & 25.96 \\
    2458706.498265 & -60.66 & 24.28 \\
    2458711.444815 & -64.13 & 25.69 \\
    2458712.45243 & -208.9 & 30.92 \\
    2458714.507735 & -1.04 & 29.74 \\
    2458715.454715 & 104.33 & 19.43 \\
    2458718.45001 & 10.36 & 23.92 \\
    2458723.46396 & -10.17 & 29.09 \\
    2458724.419910 & -4.56 & 29.82 \\
    2458727.47503 & -162.58 & 33.76 \\
    2458742.39457 & -78.27 & 23.94 \\
    2458743.368875 & 105.08 & 32.28 \\
    2458744.35226 & 96.0 & 17.59 \\
    2458745.395085 & -0.64 & 29.71 \\
    2458755.39838 & -10.29 & 41.43 \\
    2458757.375210 & -54.66 & 41.67 \\
    2458759.369515 & 121.73 & 27.43 \\
    2458760.30872 & 10.36 & 28.45 \\
    2458761.35095 & -120.87 & 25.35 \\
    2458763.336130 & 26.53 & 27.49 \\
    2458765.32907 & -24.44 & 46.17 \\
    2458766.329605 & -141.11 & 33.1 \\
    2459049.54651 & 177.28 & 83.51 \\
    2459050.5627 & 326.43 & 75.53 \\
    2459051.54861 & 164.54 & 58.18 \\
    2459059.517 & 128.17 & 38.35 \\
    2459060.56397 & 0.64 & 45.11 \\
    2459061.51774 & -1.64 & 26.11 \\
    2459067.51658 & -22.7 & 35.8 \\
    2459070.49213 & -161.95 & 47.51 \\
    2459076.47969 & 53.98 & 28.45 \\
    2459078.48314 & -53.38 & 47.45 \\
    2459079.51194 & 84.71 & 32.93 \\
    2459081.44324 & -7.63 & 44.14 \\
    2459085.4860900003 & -66.12 & 30.86 \\
    2459086.4707 & -5.4 & 35.27 \\
    2459087.44822 & 43.18 & 42.15 \\
    2459095.462660 & 13.58 & 28.27 \\
    2459098.44795 & 15.8 & 34.37 \\
    2459099.41347 & 0.86 & 35.8 \\
    2459148.30047 & -76.46 & 68.62 \\
    2459154.2969 & 25.34 & 66.01 \\
    2459161.27302 & -70.52 & 61.17 \\
    2459170.25901 & -35.74 & 60.54 \\ \hline
\end{longtable}

\begin{longtable}[c]{|c|c|c|}
\caption{Nightly CARMENES-VIS RVs analyzed in this work.}
    \\ \hline
    BJD & RV [m\,s$^{-1}$] & $\sigma_{\mathrm{RV}}$ [m\,s$^{-1}$] \\ \hline
    2458678.568125 & -69.66 & 29.45 \\
    2458679.537290 & 241.65 & 40.59 \\
    2458680.53527 & 79.92 & 15.96 \\
    2458684.568335 & 232.51 & 25.58 \\
    2458686.54842 & 177.2 & 16.04 \\
    2458687.578910 & -17.01 & 20.79 \\
    2458688.584365 & -96.42 & 19.21 \\
    2458690.5572350 & 114.83 & 11.18 \\
    2458691.505425 & 155.65 & 13.93 \\
    2458693.549395 & -100.78 & 8.52 \\
    2458694.59568 & 135.7 & 6.79 \\
    2458695.53932 & 39.93 & 10.69 \\
    2458696.522970 & 57.79 & 6.7 \\
    2458698.518795 & -108.57 & 4.86 \\
    2458699.484225 & 81.68 & 8.55 \\
    2458700.47709 & 10.01 & 6.31 \\
    2458701.472325 & -1.34 & 7.9 \\
    2458702.502 & -163.13 & 16.45 \\
    2458704.489635 & 79.4 & 7.75 \\
    2458706.498955 & -53.41 & 13.14 \\
    2458711.445175 & -68.26 & 9.3 \\
    2458712.452575 & -145.53 & 8.27 \\
    2458714.507445 & 74.89 & 11.13 \\
    2458715.454290 & 119.79 & 6.96 \\
    2458718.449865 & -8.15 & 7.88 \\
    2458723.46386 & 10.82 & 9.14 \\
    2458724.42014 & 32.17 & 9.24 \\
    2458727.47467 & -120.64 & 15.58 \\
    2458742.394525 & -111.51 & 6.47 \\
    2458743.36923 & 109.79 & 10.05 \\
    2458744.35274 & 44.66 & 6.63 \\
    2458745.394785 & -8.64 & 10.77 \\
    2458755.399070 & -52.31 & 12.96 \\
    2458757.37524 & -40.49 & 11.8 \\
    2458759.369535 & 173.63 & 10.1 \\
    2458760.30896 & -50.26 & 9.67 \\
    2458761.350815 & -127.71 & 8.52 \\
    2458763.33635 & 21.39 & 6.78 \\
    2458765.32879 & -83.71 & 9.85 \\
    2458766.32998 & -150.2 & 7.98 \\
    2459049.54617 & -9.06 & 23.94 \\
    2459050.56301 & 261.02 & 21.89 \\
    2459051.54856 & -21.26 & 27.7 \\
    2459059.51739 & 21.86 & 12.02 \\
    2459060.56385 & 56.21 & 15.25 \\
    2459061.51767 & 3.43 & 9.96 \\
    2459067.51635 & -61.79 & 14.71 \\
    2459070.49224 & -120.99 & 17.32 \\
    2459076.47973 & 74.87 & 9.62 \\
    2459078.48285 & -66.09 & 12.86 \\
    2459079.51202 & 159.74 & 17.82 \\
    2459081.4433 & 18.72 & 16.94 \\
    2459085.48611 & -103.96 & 11.19 \\
    2459086.47053 & 0.0 & 11.41 \\
    2459087.44817 & -22.76 & 10.14 \\
    2459095.46273 & -17.99 & 15.28 \\
    2459098.44771 & 36.94 & 14.87 \\
    2459099.41364 & 1.26 & 13.44 \\
    2459113.41579 & 77.58 & 40.76 \\
    2459148.3007 & -199.06 & 41.82 \\
    2459154.29727 & 80.14 & 23.03 \\
    2459161.27325 & -25.96 & 17.31 \\
    2459170.25874 & -52.59 & 15.02 \\ \hline
\end{longtable}

\begin{longtable}[c]{|c|c|c|}
\caption{Nightly MINERVA-Australis RVs analyzed in this work.}
    \\ \hline
    BJD & RV [m\,s$^{-1}$] & $\sigma_{\mathrm{RV}}$ [m\,s$^{-1}$] \\ \hline
    2458683.150648 & 89.06 & 38.22 \\
    2458684.165833 & 208.9 & 10.54 \\
    2458711.967436 & -67.42 & 12.5 \\
    2458716.998287 & -61.47 & 10.95 \\
    2458719.003218 & 175.13 & 22.44 \\
    2458720.062263 & 81.08 & 6.24 \\
    2458725.959649 & 32.34 & 104.73 \\
    2458738.040897 & -55.32 & 7.36 \\
    2458739.963764 & 235.89 & 6.39 \\
    2458740.989093 & -125.63 & 5.43 \\
    2458741.999722 & -128.95 & 8.06 \\
    2458743.006845 & 0.0 & 5.15 \\
    2458792.965243 & -13.18 & 9.52 \\ \hline
\end{longtable}

\begin{longtable}[c]{|c|c|c|}
\caption{Nightly CHIRON RVs analyzed in this work.}
    \\ \hline
    BJD & RV [m\,s$^{-1}$] & $\sigma_{\mathrm{RV}}$ [m\,s$^{-1}$] \\ \hline
    2458740.72096 & 1.48 & 45.24 \\
    2458741.713716 & 149.09 & 46.54 \\
    2458742.711554 & -54.11 & 37.64 \\
    2458762.643643 & 112.47 & 56.45 \\
    2458763.64337 & -44.57 & 36.34 \\
    2458764.629147 & 72.97 & 42.89 \\
    2458765.631185 & -310.79 & 58.25 \\
    2458766.614416 & -169.88 & 44.79 \\
    2458795.574927 & 51.27 & 68.8 \\
    2458796.570968 & -1.48 & 71.87 \\
    2458797.56992 & -11.17 & 37.07 \\
    2458798.597397 & 359.32 & 52.94 \\ \hline
\end{longtable}

\begin{longtable}[c]{|c|c|c|}
\caption{Nightly SPIRou RVs analyzed in this work.}
    \\ \hline
    BJD & RV [m\,s$^{-1}$] & $\sigma_{\mathrm{RV}}$ [m\,s$^{-1}$] \\ \hline
    2458744.8212 & 59.5 & 5.0 \\
    2458750.7542 & -18.2 & 5.0 \\
    2458751.7453 & -52.4 & 5.0 \\
    2458752.7898 & 27.9 & 5.0 \\
    2458758.7288 & 51.3 & 5.0 \\
    2458759.8053 & 69.2 & 5.0 \\
    2458760.7278 & -29.6 & 5.0 \\
    2458761.7305 & -87.2 & 5.0 \\
    2458762.7315 & -2.9 & 5.0 \\
    2458764.7571 & 34.5 & 5.0 \\
    2458765.7694 & -39.8 & 5.0 \\
    2458769.7438 & 22.1 & 5.0 \\
    2458770.7407 & -41.6 & 5.0 \\
    2458771.7212 & -55.8 & 5.0 \\
    2458772.7416 & 19.4 & 5.0 \\
    2458787.7155 & 28.3 & 5.0 \\
    2458788.7045 & 41.3 & 5.0 \\
    2458789.7367 & -32.6 & 5.0 \\
    2458790.701 & -90.1 & 5.0 \\
    2458791.6983 & 2.1 & 5.0 \\
    2458792.6976 & 19.1 & 5.0 \\
    2458796.6859 & -13.5 & 5.0 \\
    2458797.7098 & 19.7 & 5.0 \\
    2458798.6873 & 27.0 & 5.0 \\
    2458799.6883 & -10.3 & 5.0 \\
    2458800.6896 & -46.2 & 5.0 \\
    2458801.6873 & 0.0 & 5.0 \\ \hline
\end{longtable}

\section{Fitting the Full RV Dataset} \label{app:full_fit}

\begin{figure*}
    \centering
    \includegraphics[width=0.98\textwidth]{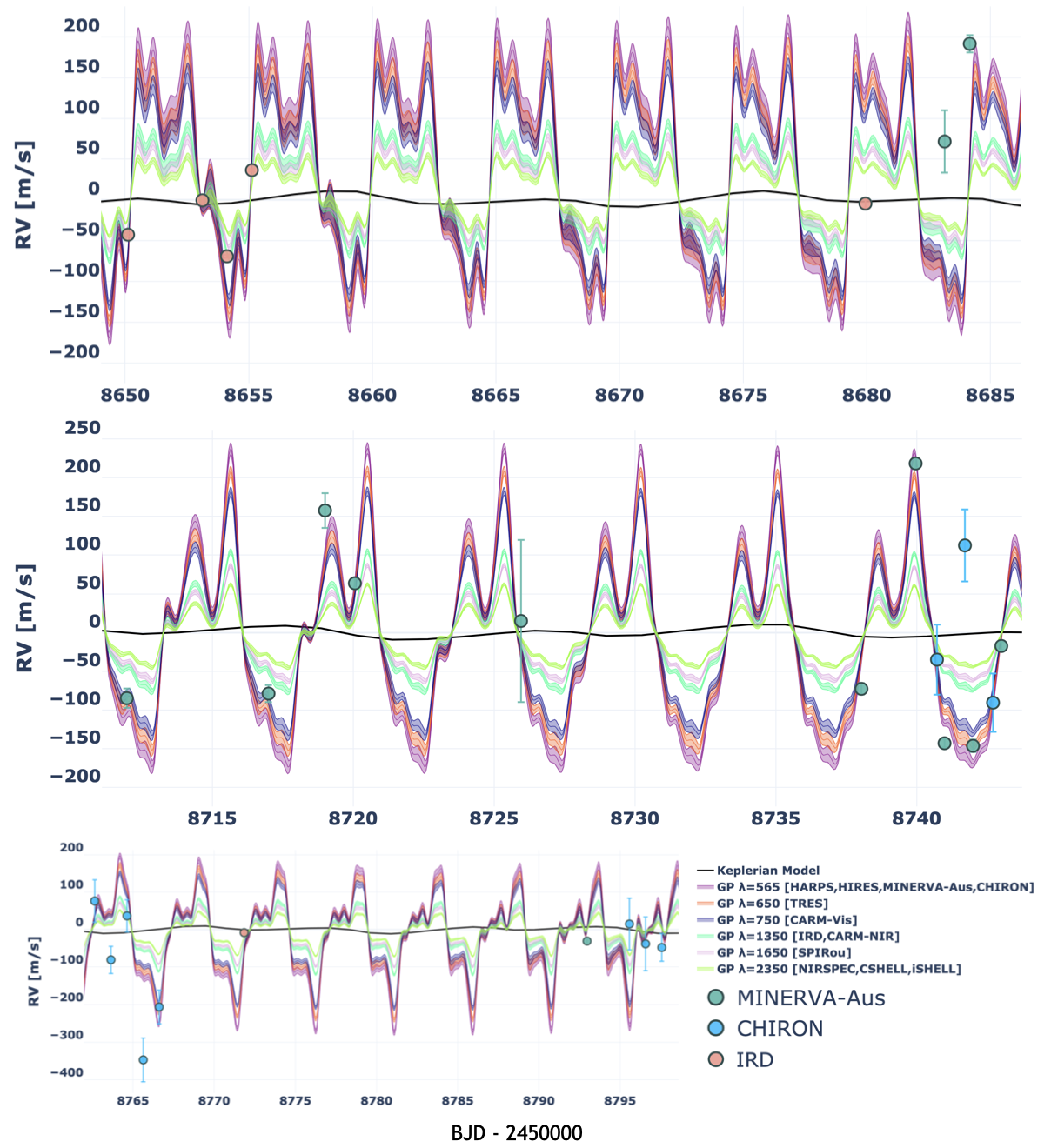}
     \caption{Here we show a subset of the 2019 RVs using kernel $\mathbf{K_{J2}}$ (eq. \ref{eq:gp_j1}) to model the stellar activity and including the full RV dataset. Although we do not include the MINERVA-Australis or IRD RVs in our primary fits in section \ref{sec:rv_fitting}, we find they are generally consistent with our stellar activity model. We do not show the phased CHIRON RVs due to their larger residuals.}
     \label{fig:rvs_all_ird_minervaaus}
\end{figure*}

\begin{figure*}
     \centering
     \includegraphics[width=0.98\textwidth]{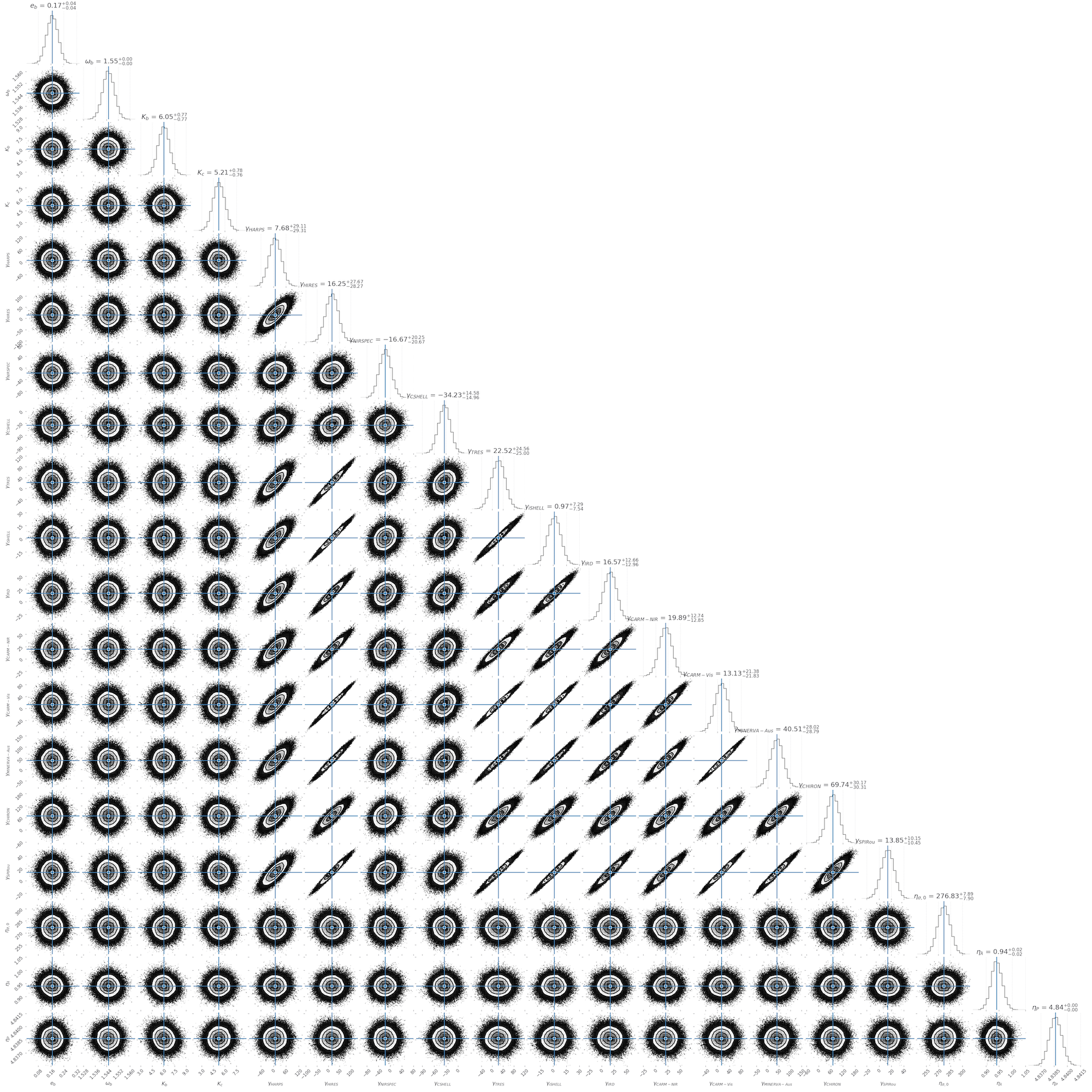}
     \caption{Posterior distributions for a two-planet fit to the full RV dataset using $\mathbf{K_{J2}}$ to model the stellar activity. Blue lines correspond to the 50$^{\textrm{th}}$ percentile of the distribution. Upper and lower uncertainties correspond to the $15.9^{th}$ and $84.1^{st}$ percentiles, respectively. The semi-amplitude $K_{b}$ is $\approx$ 30\% smaller than the subset of 2019-2020 data yields (Table \ref{tab:pars_results_2planets}), however $K_{c}$ is relatively unchanged.}
     \label{fig:corner_2planets_all_data}
 \end{figure*}

Here we present fits to the full radial velocity dataset (see section \ref{sec:data}). Although the baseline of the full dataset is nearly 17 years (first epoch in Dec. 2003), the uncertainties for the period and time of transit for AU Mic b and c are small enough to be fixed (see Table \ref{tab:pars_priors}). We only use kernel $\mathbf{K_{J2}}$ (eq. \ref{eq:gp_j2}) to model the stellar activity as we do not seek to fit for per-spectrograph activity amplitudes for datasets with $\lesssim$ 10 measurements. A first-order estimation for the secular acceleration \citep{2013ApJ...764..131C} of AU Mic is negligible given the precision of our measurements and baseline ($\Delta$ RV $<3$ cm\,s$^{-1}$), so no long-term linear or quadratic trend is used. The posteriors are shown in figure \ref{fig:corner_2planets_all_data}. The GPs and Keplerian model are shown in figure \ref{fig:rvs_all_ird_minervaaus}.

\begin{figure*}
    \centering
    \includegraphics[width=0.98\textwidth]{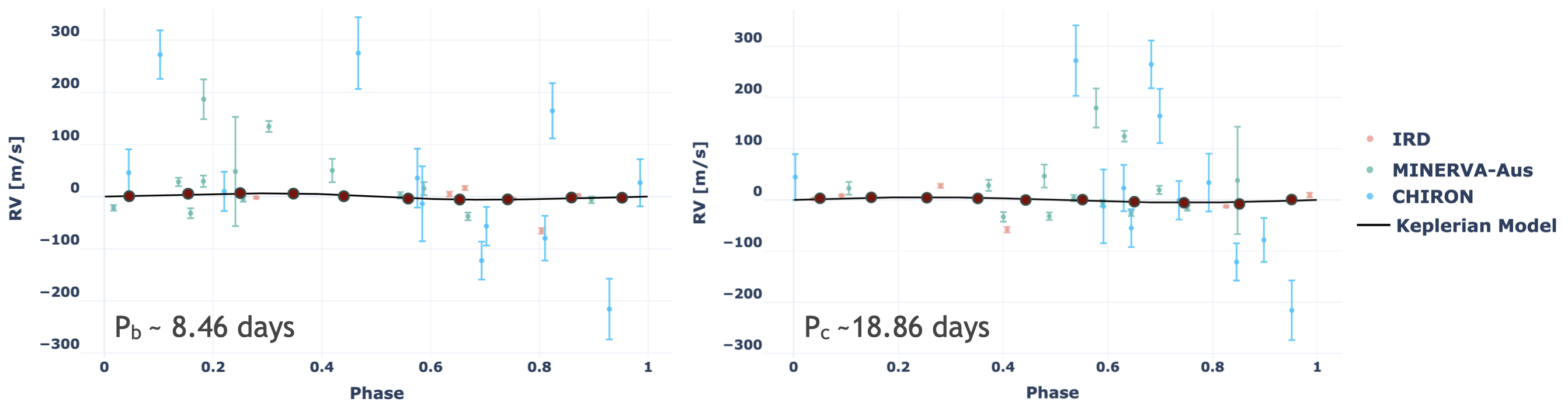}
     \caption{Same as fig. \ref{fig:rvs_all_ird_minervaaus}, but showing the phased RVs.}
     \label{fig:rvs_all_ird_minervaaus_phased}
\end{figure*}

\end{document}